%% file: main.tex
\documentclass[a4paper,fleqn,usenatbib]{mnras}

\usepackage{mathptmx}
\usepackage{siunitx}
\usepackage{cleveref}

\usepackage{caption}
\usepackage{multicol}
\usepackage{subfig}
\usepackage{comment}
\usepackage{rotating}
\usepackage[T1]{fontenc}
\usepackage{ae,aecompl}
\usepackage[export]{adjustbox}
\usepackage{fancyhdr}

\usepackage{graphicx}	% Including figure files
\usepackage{amsmath}	% Advanced maths commands
\usepackage{amssymb}	% Extra maths symbols
\usepackage{geometry}
\usepackage{pdflscape}

\title[Unveiling the TeV gamma-ray nature of EHBLs]{A new hard X-ray selected sample of extreme high-energy peaked BL Lac objects and their TeV gamma-ray properties}

\author[L. Foffano et al.]{
L. Foffano$^{1,2}$\thanks{E-mail: luca.foffano@phd.unipd.it},
E. Prandini$^{1,2}$,
A. Franceschini$^{1,3},$
and S. Paiano $^{2,3}$\\
$^{1}$University of Padova, Physics and Astronomy department, via Marzolo 8, 35131 Padova, Italy\\
$^{2}$Istituto Nazionale di Fisica Nucleare (INFN), via Marzolo 8, 35131 Padova, Italy\\
$^{3}$Istituto Nazionale di Astrofisica (INAF), vicolo dell'osservatorio 3, 35131 Padova, Italy\\\vspace{-20pt}
}
\date{\vspace{-5pt}}

\pubyear{2019}

\usepackage[usenames, dvipsnames]{color}

\begin{document}
\label{firstpage}
\pagerange{\pageref{firstpage}--\pageref{lastpage}}
\maketitle

%\linenumbers

\begin{abstract}
Extreme high-energy peaked BL Lac objects (EHBLs) are an emerging class of blazars with exceptional spectral properties. 
The non-thermal emission of the relativistic jet  peaks in the spectral energy distribution (SED) plot with the synchrotron emission in X-rays and with the gamma-ray emission in the TeV range or above. 
These high photon energies may represent a challenge for the  standard modeling of these sources. They are important for the implications on the indirect measurements of the extragalactic background light, the intergalactic magnetic field estimate, and the possible origin of extragalactic high-energy neutrinos. 
In this paper, we perform a comparative study of the multi-wavelength spectra of 32  EHBL objects detected by the \emph{Swift}-BAT telescope in the hard X-ray band and by the \emph{Fermi}-LAT telescope in the \mbox{high-energy} \mbox{gamma-ray band}.
The source sample presents uniform spectral properties in the \mbox{broad-band} SEDs, except for the TeV gamma-ray band where an interesting bimodality seems to emerge. This suggests that the EHBL class is not homogeneous, and a possible sub-classification of the EHBLs may be unveiled.
Furthermore, in order to increase the number of EHBLs and settle their statistics, we discuss the potential detectability of the 14 currently TeV gamma-ray undetected sources in our sample by the Cherenkov telescopes.

\end{abstract}

\begin{keywords}
 BL Lacertae objects: general - galaxies: active - gamma-rays: galaxies - X-rays: general \vspace{-15pt}

\end{keywords}

%%%%%%%%%%%%%%%%%%%%%%%%%%%%%%%%%%%%%%%%%%%%%%%%%%

%%%%%%%%%%%%%%%%% BODY OF PAPER %%%%%%%%%%%%%%%%%%

\section{Introduction}
Blazars are active galactic nuclei (AGN) predominantly characterized by  non-thermal emission covering the entire electromagnetic spectrum from radio up to gamma rays. The observational properties of these objects are interpreted such that the relativistic jet is closely aligned with the line-of-sight of the observer. Their spectral energy distribution (SED) in the $\nu \,\text{F}_\nu (\nu)$ presentation (where $\text{F}_\nu$ is the flux density at the frequency $\nu$) typically consists of two main components. 
The origin of the first hump is well established and commonly interpreted as synchrotron radiation emitted by relativistic electrons moving in the magnetic field of the jet. 
On the other hand, the nature of the second component peaking at higher energies is currently under debate. It is generally supposed to be produced by inverse Compton (IC) scattering of low-energy target photons \citep{IC} that may come from the same electrons producing the synchrotron hump in the so-called Synchrotron-Self-Compton (SSC) model (e.g. \citealt{Tavecchio:SSC}), or alternatively from external photon fields that are up-scattered by IC process in the so-called External Compton case \citep{EC}.
This second hump could also result from a combination of leptonic and hadronic processes \citep{aharonian2000, Murase}.
In fact, part of the jet power may be used to accelerate also relativistic protons. When they reach sufficiently high energies that p-$\gamma$ pion  production processes take place in a significantly magnetized environment, the consequently produced electromagnetic cascades may contribute to the high-energy hump in addition to proton, muon, and pion synchrotron radiation \citep{Mannheim:1993, Boettcher2010}.
Another contribution to the high-energy hump could be provided by the so-called hadronic cascade scenario. This model assumes that the observed high-energy radiation is produced in the intergalactic space through photo-hadronic reactions by ultra-high energy cosmic rays (UHECRs) beamed by the blazar jet  up to energies of $10^{19-20}$ eV  \citep{Tavecchio-cascade-scenario}.

Blazars are
historically subdivided in two main categories: Flat Spectrum Radio Quasars (FSRQs) and BL~Lac~objects. The former are characterized by their radio spectral index almost null at a few GHz and their strong emission lines in the optical spectrum. Conversely, in BL Lac objects  these lines are faint or not present at all. 

These two categories are plausibly part of a phenomenological sequence that blazars may be following \citep{fossati98}, based on the anti-correlation between the bolometric luminosity and the peak energy of their SED humps. 
Assuming that the acceleration mechanism is similar for all the blazars, 
this so-called ``blazar sequence'' was soon interpreted as due to the different radiative cooling of the emitting electrons in different sources \citep{Ghisellini:1998,blazarsequence08,blazarsequence17}.

The FSRQ are the `redder'' blazars (i.e. with lower peak frequencies): their synchrotron peak is usually located between $10^{11}$ and $10^{15}$ Hz, while the high-energy one ranges from $10^{21}$ to $10^{24}$ Hz. Their high bolometric luminosity is probably correlated to a rich environment around the jet that make the second hump of the SED being dominated by external radiation to the cooling. 

Beyond FSRQs, the sequence shows the presence of four different subclasses of BL Lac objects, classified depending on the energy of the synchrotron peak. Here the seed radiation for the IC process is supposed to be provided by the internally produced synchrotron photons: this implies larger typical electron energies and a smaller Compton dominance, meaning that the synchrotron peak shifts toward higher energies, and the ratio between the luminosity of this peak and the high-energy one increases.
According to the frequency of the synchrotron peak $\nu_{\text{peak}}^{\text{sync}}$, the four classes are composed by the low peaked BL Lac objects (LBL, with $\nu_{\text{peak}}^{\text{sync}}<10^{14}$ Hz), then by the intermediate peaked BL Lac objects (IBL, with $\nu_{\text{peak}}^{\text{sync}}$ between $10^{14}$ and $10^{15}$ Hz), the high peaked BL Lac objects (HBL, $\nu_{\text{peak}}^{\text{sync}}$ between $10^{15}$ and $10^{17}$ Hz), and finally by the extreme high peaked BL Lac objects (EHBLs). 

\vspace{-7pt}

\subsection*{Extreme high peaked BL Lac objects}
The EHBLs \citep{Costamante:2001pu} form an emerging class of BL Lac objects with extreme properties. They are characterized by a synchrotron emission that peaks at exceptionally high energies in the medium and hard X-ray band. 
\mbox{Hereafter}, we adopt the definition of EHBLs as sources having their synchrotron peak frequency exceeding $10^{17}$ Hz. This represents an arbitrary value commonly used in literature that reconciles  the definition of blazars with $\nu_{\text{peak}}^{\text{sync}}$ > 1 keV adopted by \citet{Costamante:2001pu} with the frequency-based subdivision of blazars in LSP/ISP/HSP proposed by \citet{fermi-bright-blazars}. 

In the standard SSC model view, the synchrotron peak located in the X-ray band pushes the second hump close to the very-high-energy gamma-ray band (VHE, energies above 100 GeV). 
However, recent observations revealed that in some objects the second peak is actually shifted at even higher energies - above the TeV gamma-ray region - and this makes their SEDs a challenge for the standard only leptonic SSC model. In that scenario, in fact, the decreasing scattering cross section with energy in the Klein-Nishina regime would inevitably lead to rather soft SSC spectra at TeV energies, that goes against the observations \citep{tavecchio2009}.
 
The SSC model can still be used to fit the emission of these sources, but the results tend to accommodate the observed SED of these objects at the price of using particularly extreme model parameters (\citealt{Tavecchio:2009zb}, and a collection of some of the most updated results on EHBL modeling can be found e.g. in \citealt{cerruti2015}), generally requiring either high values of the minimum Lorentz factor $\gamma_\text{min}$ of the electron distribution or invoking unrealistically large Doppler factors $\delta$  \citep{tavecchio2009}. In particular, high values of $\gamma_\text{min}$ might occur in very specific conditions: for example when electrons are injected  with a narrow energy distribution into the emission region and their subsequent cooling is inefficient, or in the case of stochastic turbulence that may be responsible for the electron acceleration in blazar jets (see e.g. \citealt{Asano:2013}). On the other hand, the request for large Doppler factors $\delta$ would imply a fast flow of the plasma or an extremely small viewing angle, and generally this goes against radio observations of the movement of knots inside the jets or against the statistics of observed blazars (see e.g. \citealt{Lister:2013}).

Many different alternative scenarios within the leptonic model have been proposed. Some works, for example, interpret the high values of $\gamma_\text{min}$ in a time-dependent one-zone model with extremely hard Maxwellian particle distributions \citep{tev_maxwellian_distribution, Lefa2011} or with a low-energy cut-off of the electron distribution at VHE \citep{Katarzynski2006}.
Other works suggest that such very hard VHE gamma-ray spectra might be reproduced by models including external Compton up-scattering of ambient photon fields \citep{Lefa2011}. However, the latter are commonly thought to dominate in powerful FSRQs rather than in HBLs, where no strong emission from the accretion disk or the environment has been detected yet.
Finally, in the case of 1ES~0229+200, that is one of the most studied TeV gamma-ray detected EHBLs, the hard TeV spectrum was successfully interpreted also in the previously mentioned intergalactic cascades scenario \citep{Murase}.

Interestingly, EHBLs do not show high and rapid variability, as other blazars like HBLs do (e.g. the low variability of 1ES 0229+200 on the long-term lightcurve in VHE gamma~rays of \citealt{Cologna:2015nia}). This effect may be related to the low flux of these sources, but leptonic models predict large flux variations on short timescales that have never been observed on them.
The hard VHE gamma-ray spectra and the absence of rapid flux variability 
make EHBLs interesting candidates for hadronic and  lepto-hadronic emission models, that can well reproduce their observed SED (e.g. in \citealt{cerruti2015}).

Within the blazar sequence context, EHBLs are the less luminous blazars but with the highest Doppler boosting factors. This suggests that they could be one of the most efficient and extreme accelerators in the Universe: the hadronic contribution - in addition to the main leptonic emission mechanism - could let them to produce and accelerate \mbox{UHECRs}. This topic has received enhanced focus by the scientific community thanks to the recently found first correlation between a neutrino event and a flaring blazar \citep{icecube17}. In the last years, the possible presence of hadronic processes in the blazar jets led some authors to look for a correlation between the direction of the neutrino events and the presence of nearby blazars. These works resulted in the determination of a hint of correlation between some HBL objects and some neutrino events  \citep{Padovani:2016}, suggesting that the most extremely high-peaked BL~Lac~objects may show an even higher correlation with the neutrino events \citep{Resconi:2016}.

\vspace{-5pt}

\subsection*{Importance of the TeV gamma-ray band}
The challenging interpretation of the SED of EHBLs is an important puzzle in the blazar context, and the VHE gamma-ray band is fundamental to study their enigmatic spectral properties.

The EHBL study in the VHE gamma-ray band is mainly performed by the current generation of major Imaging Atmospheric Cherenkov Telescopes (IACTs): H.E.S.S., MAGIC, and VERITAS. These telescopes detected more than 200 sources in the TeV gamma-ray band in the last fifteen years, with about 70 sources identified and classified as blazars, the majority of them being HBLs. Given the relatively small field of view (3.5 to 5 degrees), IACTs operate mainly in pointing mode and (except for the Galactic survey of the H.E.S.S. telescope in \citealt{HESS_galactic_survey}) cannot provide wide surveys of the whole sky in VHE gamma rays. On the other hand, due to their low flux in this energy band, EHBLs generally need large integration time to be detected. Thus, since no automatic procedures exist to produce lists of candidates directly observable by the IACTs, only a few of them have been observed and characterized in the VHE gamma-ray regime. 
Among them, some objects show an extremely hard spectrum in this energy band. The archetypal EHBL is 1ES~0229+200 - the source with the highest synchrotron peak ever found \citep{Kaufmann:2011-1es0229}. Its archival SED has been enriched by several multi-wavelength (MWL) observational campaigns during the last years, and for this reason from now on we adopt it as our reference EHBL. 
Such hard spectra similar to 1ES~0229+200 one have been reported on few other sources, like for example 1ES~0347-121 \citep{0347discovery}, RGB J0710+591 \citep{0710discovery}, and 1ES 1101-232 \citep{1101discovery}. 
Such objects have been named ``hard-TeV blazars'' by \citet{Nustar_EHBLs} due to the high energy peak position located above about 10 TeV. For this reason, they may represent a sub-category of the EHBL class and their relation with ``ordinary'' blazars has to be further investigated.

EHBLs with such a hard VHE gamma-ray spectrum as that observed in 1ES~0229+200 are also important probes for testing models of the extragalactic background light (EBL): gamma rays with energy above tens of GeV may be absorbed along cosmological distances due to the interaction with the diffuse extragalactic background light in the intergalactic space, where electron-positron pairs are produced by the $\gamma$-$\gamma$ interaction \citep{Hauser:2001}. This effect is an increasing function of the photon energy and the distance of the emitting source. 
Since EHBLs are expected to have their second hump peaking in the VHE gamma-ray band, where the imprint of the EBL absorption in the TeV spectrum can be informative, the study of their spectral properties could help in the analysis of EBL features, particularly at long infrared wave-lenghts. However, the absorpion of TeV photons through the EBL interaction makes the EHBL  difficult to be detected by Cherenkov telescopes: the higher is their redshift, the lower is the TeV flux we can observe.

Additionally, these objects are also considered good astrophysical probes for constraining extragalactic magnetic fields. In fact, it is assumed that the observed galactic magnetic fields result from the amplification of much weaker seed fields that may have extragalactic and cosmological origin, and whose nature is largely unknown.
The VHE gamma-ray photons produced by these objects are supposed to interact with EBL depositing electron-positron pairs in the intergalactic space and, if the extragalactic magnetic fields are strong enough, they may be able to deviate electron and positron trajectories, producing an observable extended emission around the initial point source. This approach has been tested on 1ES~0229+200 and 1ES~0347-121 by \citet{EGMF}.

\vspace{-10pt}

\subsection*{Selection method}

One of the first works where an EHBL selection method was proposed is \cite{costamante2002}.
Their main assumption was that the seed photons for the TeV IC emission are located in the optical and radio bands. The authors considered sources with - at a given X-ray flux - high radio and optical fluxes in order to extract new possible candidates. More recently, \cite{bonnoli2015} compiled a list of EHBL candidates looking for sources with high X-ray to radio flux ratio, and dominance of thermal radiation  from the host galaxy in the optical spectral range. While \cite{costamante2002} basically looked for candidates with high TeV flux asking for dense seed photon fields, in  \cite{bonnoli2015} the authors selected sources with higher X-ray to radio flux aiming at identifying hard TeV spectra candidates.

In this work, instead, we aim both at identifying new EHBL sources and studying their broad-band emission. In fact, since no catalog of EHBLs has never been produced yet, first of all we aim at increasing the number of sources classified as EHBLs using the definition based on the synchrotron peak position. Secondly, given the observational evidence of different TeV spectral features (see \Cref{sec:TEVdetected}), we aim at enlarging the statistics of the TeV population in order to identify a possible sub-classification inside the EHBL class. 
The driving idea is that, since EHBLs SED peaks are located at higher energies with respect to HBLs ones, their synchrotron peak should be well detected in hard X-rays, and their IC hump should be rather faint in HE gamma rays  in order to peak in the TeV gamma-ray band. To perform this selection, we identify sources with a high frequency of the synchrotron peak and high flux in the hard X-ray band. Moreover, we make use of the large integration time  by the \emph{Fermi}-LAT telescope to improve our data in HE gamma rays.

\vspace{30pt}

\noindent
The structure of this paper is the following. In \Cref{sec:2} we describe the selection method  and provide the final sample of EHBL sources, comparing their observational properties in \Cref{sec:results}. In \Cref{sec:TEVdetected} we provide a discussion and interpretation of the TeV gamma-ray detected sources in our sample, and in \Cref{sec:TEVundetected} we show our expectations about the currently TeV gamma-ray undetected objects. Finally, we report in \Cref{sec:concl} the conclusions of this work and future prospects.

\vspace{-5pt}

\section{Source selection} \label{sec:2}

\subsection{Final sample}
For the selection of EHBL candidates we searched for sources with a firm detection in the hard X-ray energy range using the \emph{Swift}-BAT 105-months catalog \citep{BATcatalog105}. This is an all-sky  survey   in   the   ultra-hard   X-ray   band   (14-195 keV) provided by the Burst Alert Telescope (BAT) instrument \citep{BATinstrument} on board of the \emph{Neil Gehrels Swift} satellite \citep{swiftsatellite}, consisting in a large coded-mask telescope optimized to detect transient gamma-ray bursts. The \emph{Swift}-BAT telescope has a wide field of view of about $ 60 \times 100$ degrees, that resulted in a catalog of the brightest sources in this energy range. This is - up to now - the most sensitive and uniform hard X-ray all-sky survey, reaching a sensitivity of $8.40\cdot 10^{-12}$ erg $\text{cm}^{-2}$ $\text{s}^{-1}$ over 50\% of the sky and $7.24\cdot 10^{-12}$ erg $\text{cm}^{-2}$ $\text{s}^{-1}$ over 90\% of the sky.

Among the 1632 sources of the \emph{Swift}-BAT 105-months catalog, all the known TeV emitting EHBLs,  namely 1ES~0229+200, 1ES~0347-121, RGB~J0710+591, and 1ES 1101-232 are present, having their synchrotron peak located in the hard X-ray band.
The \emph{Swift}-BAT catalog, once properly exploited, can therefore allow us to obtain the \mbox{up-to-now} most complete and representative sample of EHBLs with bright flux in the hard X-ray band according to the \emph{Swift}-BAT sensitivity. 

To find at least a significant fraction of these sources,  at the present stage we have confined our analysis to the 158 objects that are classified as ``Beamed AGN (Blazar/FSRQ)'' (class number 80 of the \emph{Swift}-BAT 105-months catalog), whose classification was confirmed among previously published catalogs of AGNs and blazars. The study of other sources of the catalog could provide new EHBL candidates. For example, this analysis may be applied to the 114 ``unknown AGN'' (class 70) sources of the catalog, or even to the 129 completely ``unknown class'' objects (classes 10-11-12). However, such analysis would need a refined approach to validate the nature of each source, and it will be studied in a future work.

In order to study the nature 
of these 158 sources in the HE gamma-ray band, we considered data of the pair-conversion Large Area Telescope (LAT) instrument on board the \emph{Fermi} satellite, that is an all-sky survey in the energy range from 20 MeV to more than $300$ GeV \citep{Fermitelescope}. In particular, we adopted the \emph{Fermi}-LAT 3LAC catalog \citep{3LACcatalog}. This is the third  catalog  of  AGNs detected  by  the \emph{Fermi}-LAT telescope after  four  years  of operation in which a detailed analysis of the sources was performed. 

Our selection procedure was mainly performed by cross-matching the \emph{Swift}-BAT 105-months catalog with the \emph{Fermi}-LAT 3LAC catalog. The final sample is composed of 32 blazars with synchrotron peak above the frequency $\nu_{\text{peak}}^{\text{sync}} > 10^{17}$ Hz that are detected in both the catalogs, and is presented in \Cref{tab:sourcelist}. More details on the selection method are reported in \Cref{appendix:selection}. 

For each of these candidates, we updated the HE gamma-ray data with a new ten-years \emph{Fermi}-LAT analysis (see \Cref{appendix:fermi}). The results are reported in \Cref{tab:sourcelist} and discussed in \Cref{sec:results}.

\vspace{-10pt}

\subsection{Broad-band SEDs}

For each of our final EHBL candidates, in \Cref{fig:sed_superposition}  in \Cref{appendix:figures} we compiled the available MWL (not simultaneous) archival data from the ASI Science Data Center (SSDC) serve (including our ten-years \emph{Fermi}-LAT analysis). This let us to perform a visual inspection of the SEDs to check the extreme nature of the sources, comparing them with the data of the reference EHBL 1ES~0229+200. 

According to the main idea driving our selection method, in the HE gamma-ray band these sources show the rising part of the second bump that finally peaks in the TeV band. Thus, this criterium by which the EHBLs may exhibit a rather faint detection in HE gamma rays could be a key feature in the selection and characterization of EHBLs. A more detailed discussion of this point is reported in \Cref{sec:results}.

The shift at higher energies of the two peaks in the SED provide access to the optical radiation of the host galaxy for low-redshift objects in our sample. 
This is an important feature of the so far studied EHBLs: the majority of our candidates (81\%) has a known redshift value, and the sources with unknown redshift may be good candidates to be addressed by specific optical campaigns.

Since the synchrotron peak position is at the basis of the definition of EHBL, this value plays an important role in classifying these sources.
In order to consider the best values of $\nu_{\text{peak}}^{\text{sync}}$, we performed a log-parabolic fit of the synchrotron peak in the X-ray band using for each source the available  \emph{Swift}-XRT, \emph{Beppo}-SAX, and \emph{Swift}-BAT 105-months archival data. We obtained overall good quality fits, and all our results are in general compatible with the values reported in the 2WHSP catalog \citep{2whsp}. In some cases (e.g. in 1ES 1959+650) there is some tension with respect to the 2WHSP catalog that we attribute to the high variability of the sources and to the fact that these estimations are all based on not-simultaneous archival data. However, in the following analysis we adopt our estimation of the synchrotron peak position also because we are able to provide a statistical error. The results are reported in \Cref{tab:sourcelist}.

It is worth to note that in our final sample 5 over 6 already known TeV EHBL emitters reported in \citet{Nustar_EHBLs} are present (the only one missing is 1ES~0414+009 because not detected by \emph{Swift}-BAT 105-months). In that paper, they performed a detailed study of the synchrotron peak of these sources thanks to simultaneous \emph{Swift}-XRT and NuSTAR \citep{NuSTARsatellite} observations. 
All their estimations are, however, compatible with our previous results (please see \Cref{fig:comparison_synchropeaks} for the comparison).

Finally, to check if our 32 objects are already detected in the TeV gamma-ray band, we performed a cross-match between the main sample, the TeGeV catalog \citep{tegev}, and the TeVCat catalog\footnote{http://tevcat.uchicago.edu}. We found that 18 sources out of 32 are already known as TeV gamma-ray emitters.
This result underlines that, in addition to the well known hard-TeV sources, there are other VHE gamma-ray emitting blazars that are by definition EHBLs even if their behavior differs from that of 1ES~0229+200.

%%%%%%%%%%%%%%%%%%%%%%%%%%%%%%%%%%%%%%%%%%%%%%%%%%%%%%%%%%%%%%%%%%%%%%%%%%%%%%%%%%%%%%%%%%%%%%%%%%

\input{table_big.tex}
\newpage

\begin{figure}
\centering
\hspace*{-20pt}
\includegraphics[width=0.48\textwidth]{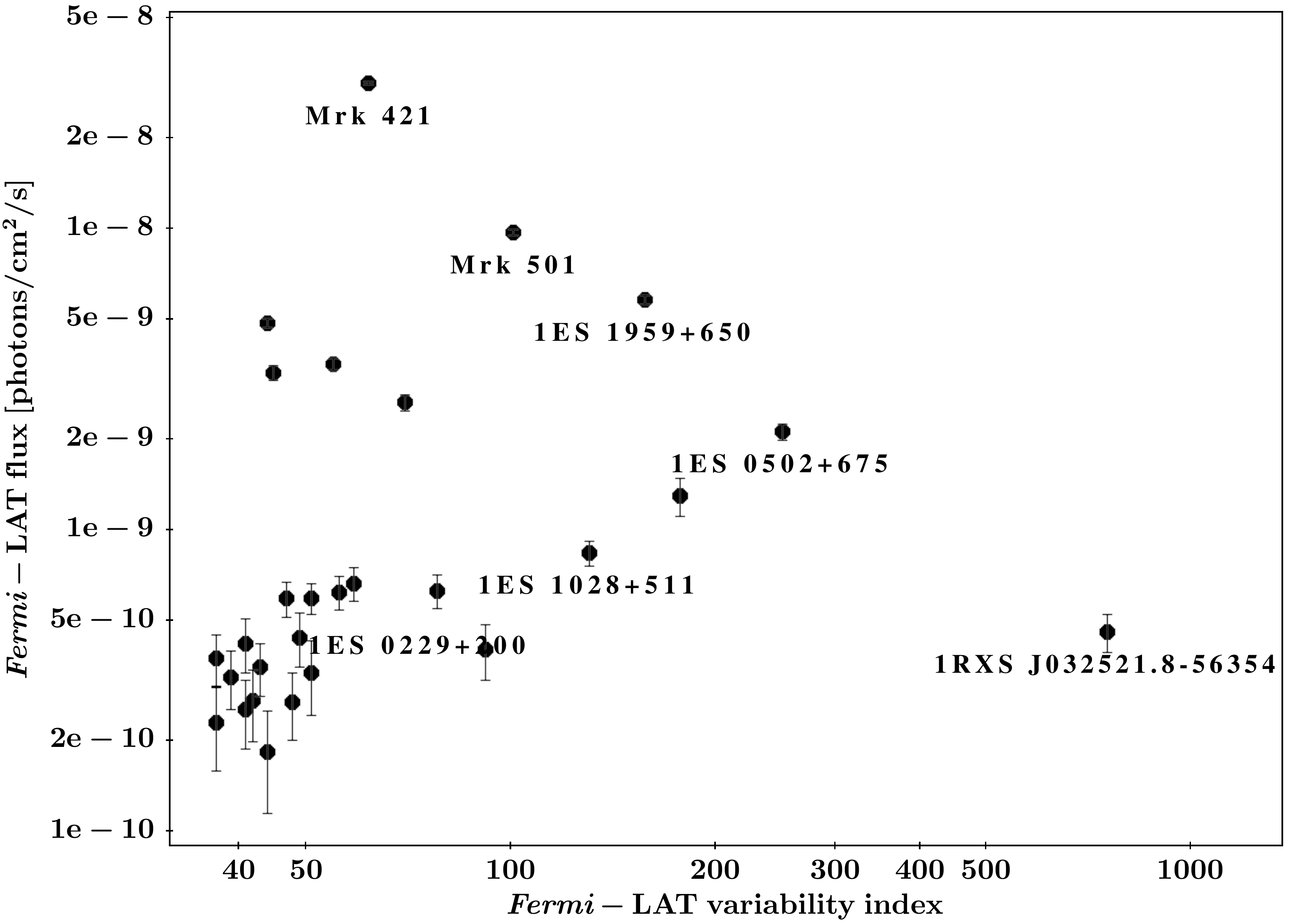}
\caption{Distribution of the \emph{Fermi}-LAT 3FGL flux versus the corresponding variability index of the sources in the main sample (logarithmic scale).}
\label{fig:flux_vs_variability}
\end{figure}

\section{Results} \label{sec:results}

In \Cref{tab:sourcelist} we report the results of our ten-years \mbox{\emph{Fermi}-LAT} analysis and we list the main properties extracted from other catalogs for all the 32 sources of the final sample.\\

\noindent
\textbf{\emph{Swift}-BAT fluxes.} The \emph{Swift}-BAT total flux in the 14-195\,keV band shows that 31 out of 32 sources (97\%) generally have an X-ray flux of the same order of 1ES~0229+200.  
Generally they show a flux of the synchrotron peak around $10^{-11}$~\text{erg}~$\text{cm}^{-2}$~$\text{s}^{-1}$, probably  related to the  sensitivity limit of the \emph{Swift}-BAT 105-months catalog. We notice that the two Markarians (Mrk 421 and Mrk 501 from now on) make an exception and are particularly bright also in this energy band.

The following column $\mathtt{\Delta\,\text{\texttt{flux}}_{\text{\texttt{BAT}}}}$ describes the variability of the hard X-ray flux of each source, calculated as difference between the highest and the lowest flux measured by \emph{Swift}-BAT between December 2004 and August 2013 (see \citealt{BATcatalog105}). Almost all sources show an average stable flux in this energy band, very  similar with respect to that of 1ES~0229+200. \\

\noindent
\textbf{\emph{Fermi}-LAT variability indices.} We report also the \emph{Fermi}-LAT variability indices as reported in the \emph{Fermi}-LAT 3FGL catalog. We find that 27 out of 32 sources (84\%) show values below the threshold of 72.44 and are considered not variable in the HE gamma-ray band \citep{fermi3fgl}. In \Cref{fig:flux_vs_variability} we show a qualitative plot relating these two observables available from the \emph{Fermi}-LAT 3FGL catalog: some interesting sources e.g. 1ES~1959+650 and Mrk 501 do not follow the mean low variability behavior.\\

\noindent
\textbf{\emph{Fermi}-LAT flux.} In \Cref{tab:sourcelist} we report also our new estimation of the total flux in the 1-300 GeV band, showing in parentheses the ratio of the flux of each source with respect to the reference 1ES~0229+200. All sources show total fluxes of the same order of that of 1ES~0229+200, except for 1ES~1959+650 and the two Markarians that are particularly bright. \\

%%%%%%%%%%%%%%%%%%%%%%%%%%%%%%%%%%%%%%%%%%%%%%%%%%%%%%%%%%%%%%%%%%%%%%%%%%%%%%%%%%%%%%%%%%%%%%%%

%\pagestyle{empty}
\begin{figure*}
\centering
\subfloat[][Superimposition of the MWL SEDs of the 13 already TeV gamma-ray detected sources with publicly available TeV data. The fluxes have been rescaled to the 1ES~0229+200 flux of $3.34 \cdot 10^{-12}$ erg $\text{cm}^{-2}$ $\text{s}^{-1}$ at $1.7 \cdot10^{17}$ Hz.]
{\includegraphics[width=0.99\textwidth]{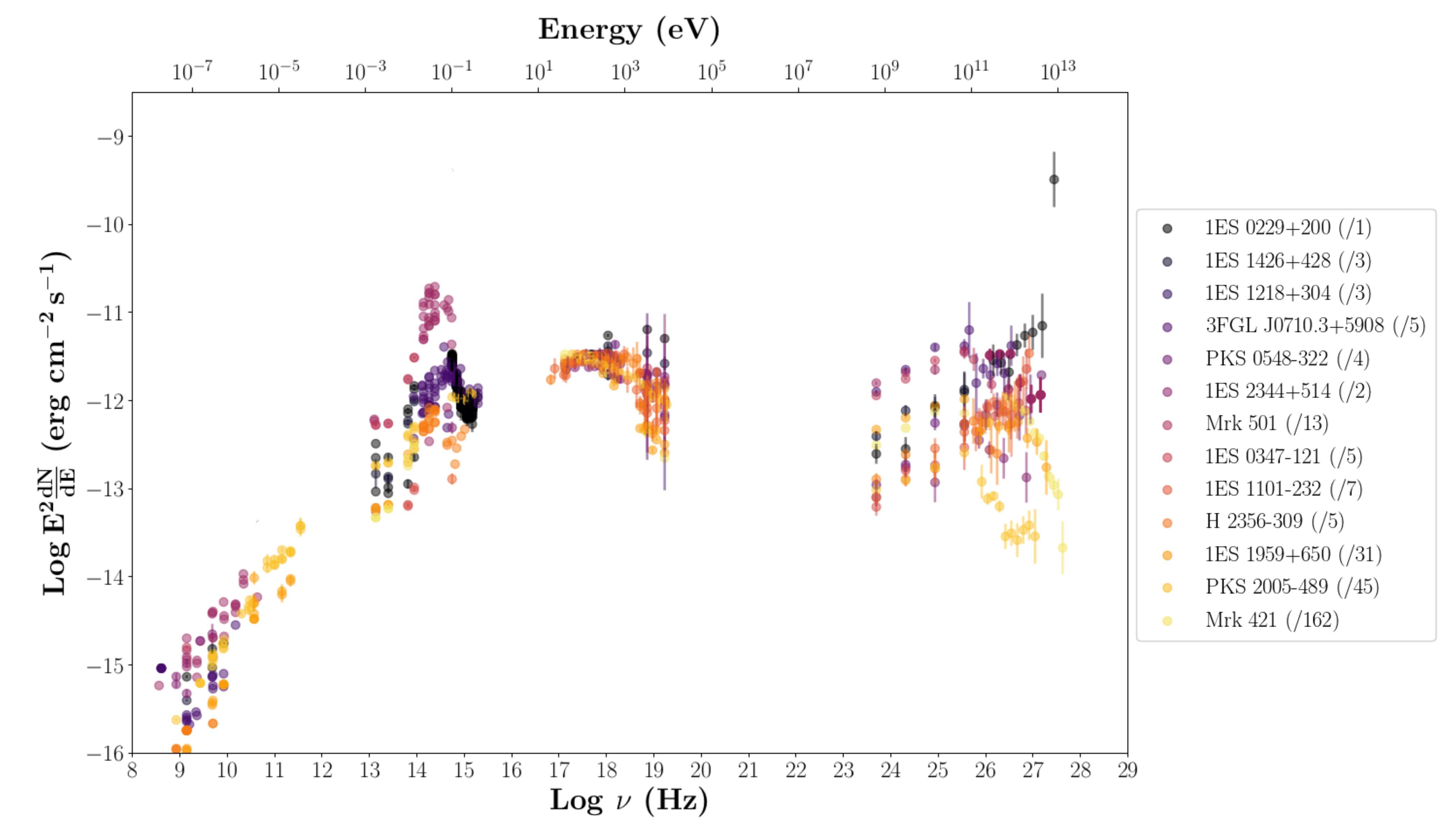}\label{fig:SEDsuperpositionALL}}\\
\subfloat[][Focus on the HE-VHE gamma ray range of the previous figure but with fluxes rescaled to the 1ES~0229+200 flux at 147 GeV with $1.93 \cdot 10^{-12}$ erg $\text{cm}^{-2}$ $\text{s}^{-1}$.]{\includegraphics[width=0.99\textwidth]{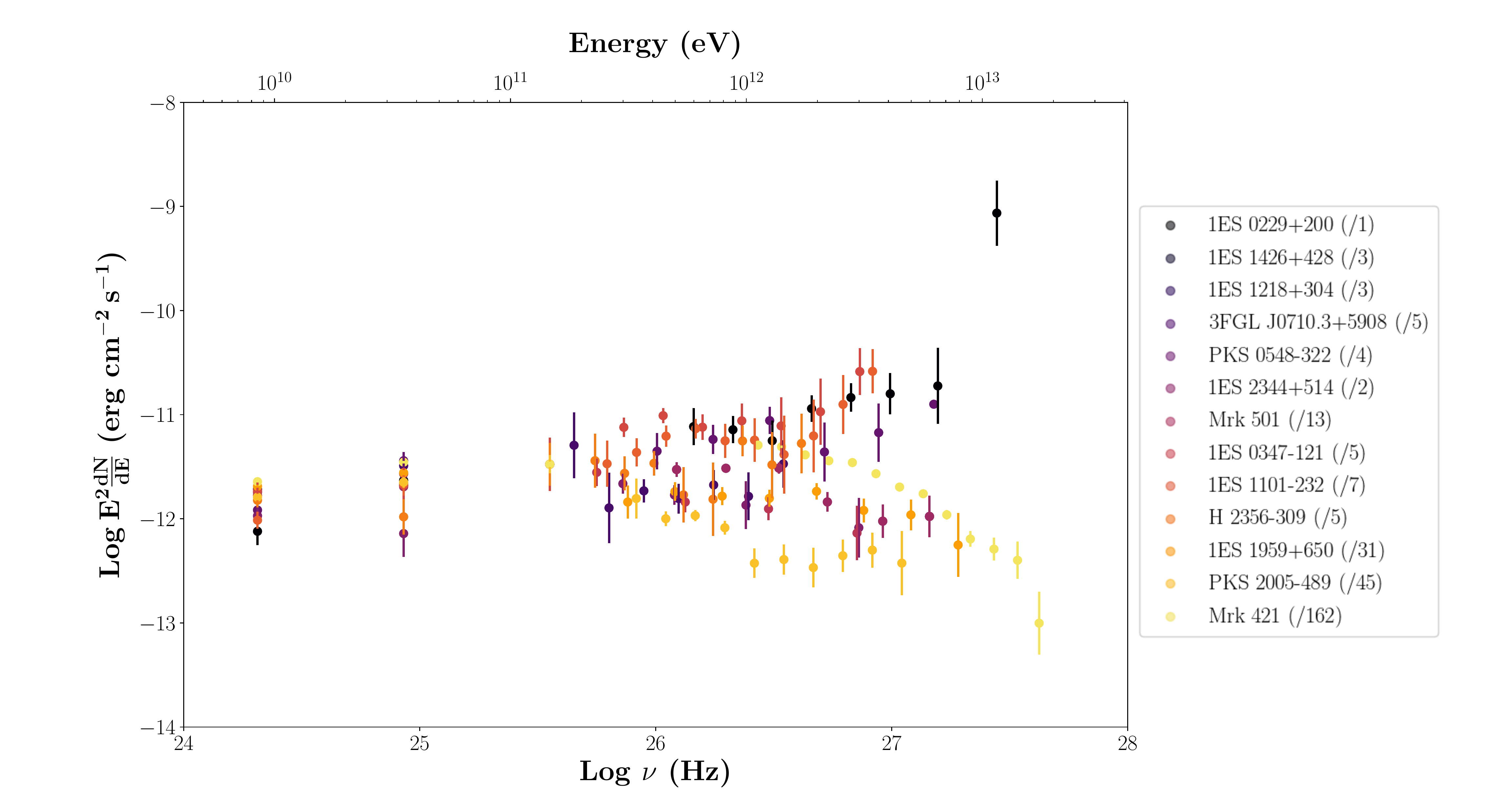}\label{fig:SEDsuperpositionHE}}
\caption{The color table that follows up-to-bottom \Cref{tab:tevdetected}, meaning that the color gradually goes from dark brown when $\nu_{\text{peak}}^{\text{sync}}$ of the source is high to light yellow when its $\nu_{\text{peak}}^{\text{sync}}$ is low. In parentheses we report the ratio of the spectra of each source with respect to the normalization chosen in the 1ES~0229+200 spectrum. The plotted data are already corrected for EBL absorption with the model by \citet{Franceschini17} to show the intrinsic spectrum of the source. }
\end{figure*}

\vspace{10pt}
%%%%%%%%%%%%%%%%%%%%%%%%%%%%%%%%%%%%%%%%%%%%%%%%%%%%%%%%%%%%%%%%%%%%%%%%%%%%%%%%%%%%%%%%%%%%%%%%

\noindent
\textbf{\emph{Fermi}-LAT spectral index.} In the following column we present the spectral 
index resulting from our analysis and in parentheses the compatibility $\lambda=\frac{|A-B|}{\sqrt[2]{\sigma_A^2+\sigma_B^2}}$ of the spectral index of each source (A) with that measured in 1ES~0229+200 (B). 
The \emph{Fermi}-LAT spectral indices are generally very hard (lower than 2.0) and well compatible with the value $-1.74 \pm 0.13$ of 1ES~0229+200. \\

\noindent
\textbf{\emph{Fermi}-LAT TS values.}
The \emph{Fermi}-LAT TS values are distributed over a wide range of values. Generally they are around few hundreds, but some very bright objects in HE gamma rays like 1ES~1959+650 and the two Markarians  show TS values up to hundred thousands. On the other hand, some other sources like 1RXS~J225146.9-320614 (TS=47) and 1ES~0927+500 (TS=49) show low TS values also after ten years of data. This can explain their absence in the 3FGL catalog.

Since we are looking for sources that should not show high signal in HE gamma rays due to hard spectrum of the IC bump that peaks above this energy band, we have to remember that some of the brightest sources  may show mainly thanks to their high flux rather than  to their extreme spectral properties. For example, the two Markarians are generally considered very bright HSP sources, even though in some cases they show strong variability and sometimes Mrk~501 behaves like an EHBL during some flaring episodes \citep{Pian98_mrk501, mkr501-flare2012}. We will study this relation in \Cref{sec:TEVdetected}.\\

\input{tev_table.tex}

\noindent
\textbf{Synchrotron peak frequency.} The logarithm of the frequency of the synchrotron peak $\nu_{\text{peak}}^{\text{sync}}$ (see \Cref{tab:sourcelist} for more details) ranges from 17 to 18.3. Thus, all the sources in our sample can be classified as EHBL following the definition based on the synchrotron peak frequency.

%%%%%%%%%%%%%%%%%%%%%%%%%%%%%%%%%%%%%%%%%%%%%%%%%%%%%%%%%%%%%%%%%%%%%%%%%%%%%%%%%%%%%%%%%%%%%%%%%%%

\section{TeV gamma-ray detected sources} \label{sec:TEVdetected}

The analysis of \Cref{sec:results} reveals essentially a compatibility between the main observables of the EHBLs and our reference source, 1ES~0229+200. This effect may be to some extent expected because we are dealing with sources that - in most of the cases - are at the limit of the sensitivity of our instruments. On the other hand, some sources with very bright and variable flux in these bands (e.g. 1ES~1959+650 and the two Markarians) seem to deviate from the mean values of the spectral parameters that the other EHBL candidates show.

In order to find distinguishable features between our candidates, we now try to further investigate the properties of the TeV gamma-ray band for the sources.

\subsection{Broad-band spectra comparison}

To better understand the relation between the main spectral observables of the sources at different wavelengths, we report in \Cref{fig:SEDsuperpositionALL} a comparison of the MWL SEDs of all the TeV gamma-ray detected sources with data from the SSDC website (TeV data are publicly  available only for 13 out of 18 sources, please see \Cref{tab:tevdetected} and \Cref{appendix:tevdata} for details and references). In this figure, all fluxes have been rescaled to the 1ES~0229+200 flux of $3.34 \cdot 10^{-12}$ erg $\text{cm}^{-2}$ $\text{s}^{-1}$ at $1.7 \cdot10^{17}$ Hz and the scaling factor for each source is reported in parentheses in the legend. All sources have been deabsorbed by EBL effect at their own redshift with the model by \citet{Franceschini17} to obtain the intrinsic spectrum of the source. We chose a color scale that follows up-to-bottom \Cref{tab:tevdetected}, meaning that the color gradually goes from dark brown when the $\nu_{\text{peak}}^{\text{sync}}$ of the source is high to light yellow when its $\nu_{\text{peak}}^{\text{sync}}$ is low.

Some common features arise in \Cref{fig:SEDsuperpositionALL}: the synchrotron peak is located in the $10^{17}$-$10^{18}$ Hz range as expected, the optical wavelengths generally show the presence of the galaxy, and the flux in radio and hard X-rays is scattered due to different slopes and brightness in these bands. Interestingly, in HE gamma rays all sources are shaping the increasing flux of the second hump. Even though the slopes in \Cref{tab:tevdetected} are generally  well compatible between each other, there is a moderate dispersion of the differential energy flux in this band. This spread seems to be amplified in VHE gamma rays above about 100 GeV, where the SED points for some sources diverge.

This tendency in the TeV gamma rays, in addition to some indicators like the TeV detection that is not uniform through the sample, suggest us that the properties of the EHBL candidates in the TeV VHE are crucial.

\begin{figure*}
\centering
\vspace*{-1.5cm}
\subfloat[][1ES~0229+200.]{\includegraphics[width=0.5\textwidth]{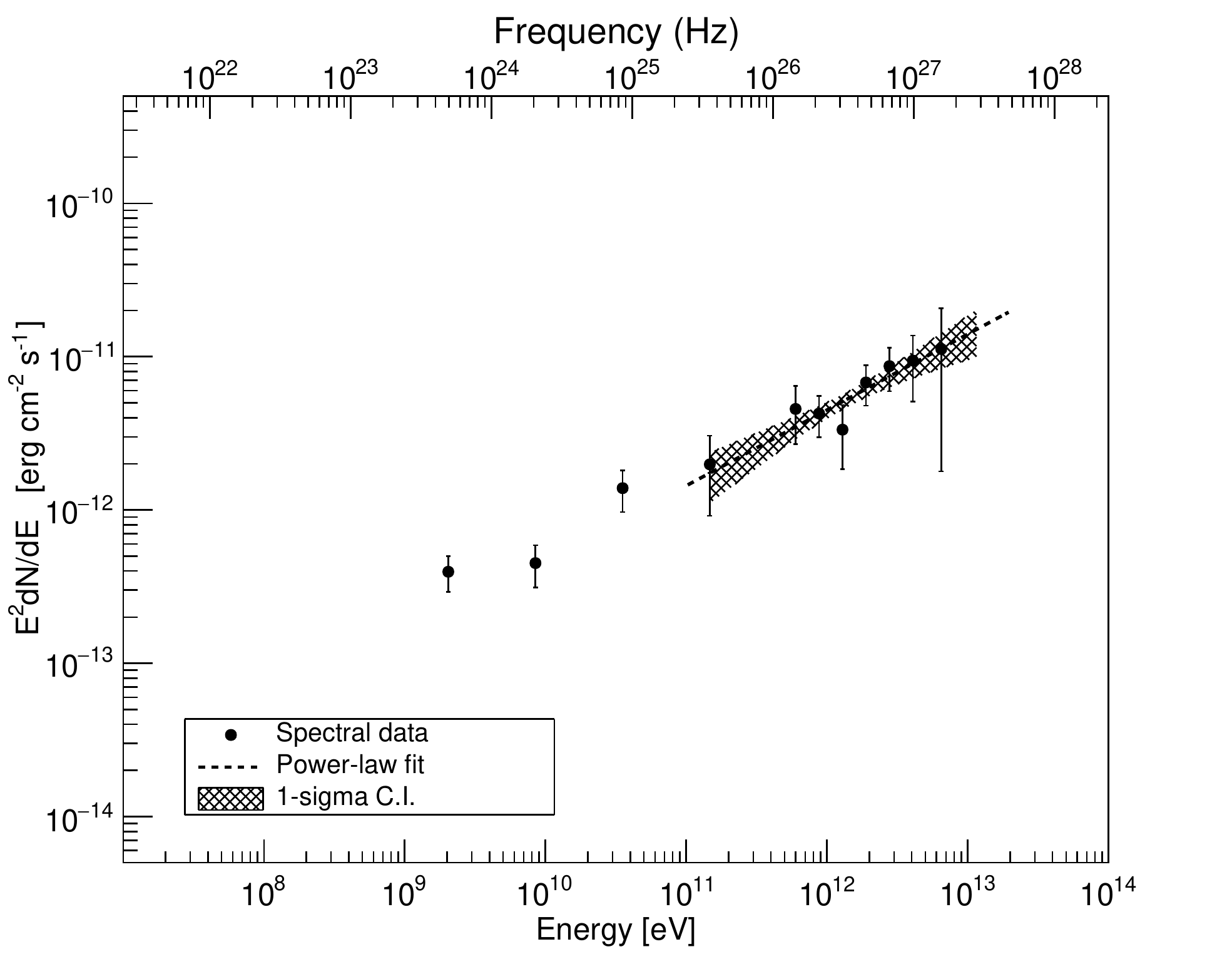}\label{fig:TeV-fits-a}}
\subfloat[][ Mrk 501.]{\includegraphics[width=0.5\textwidth]{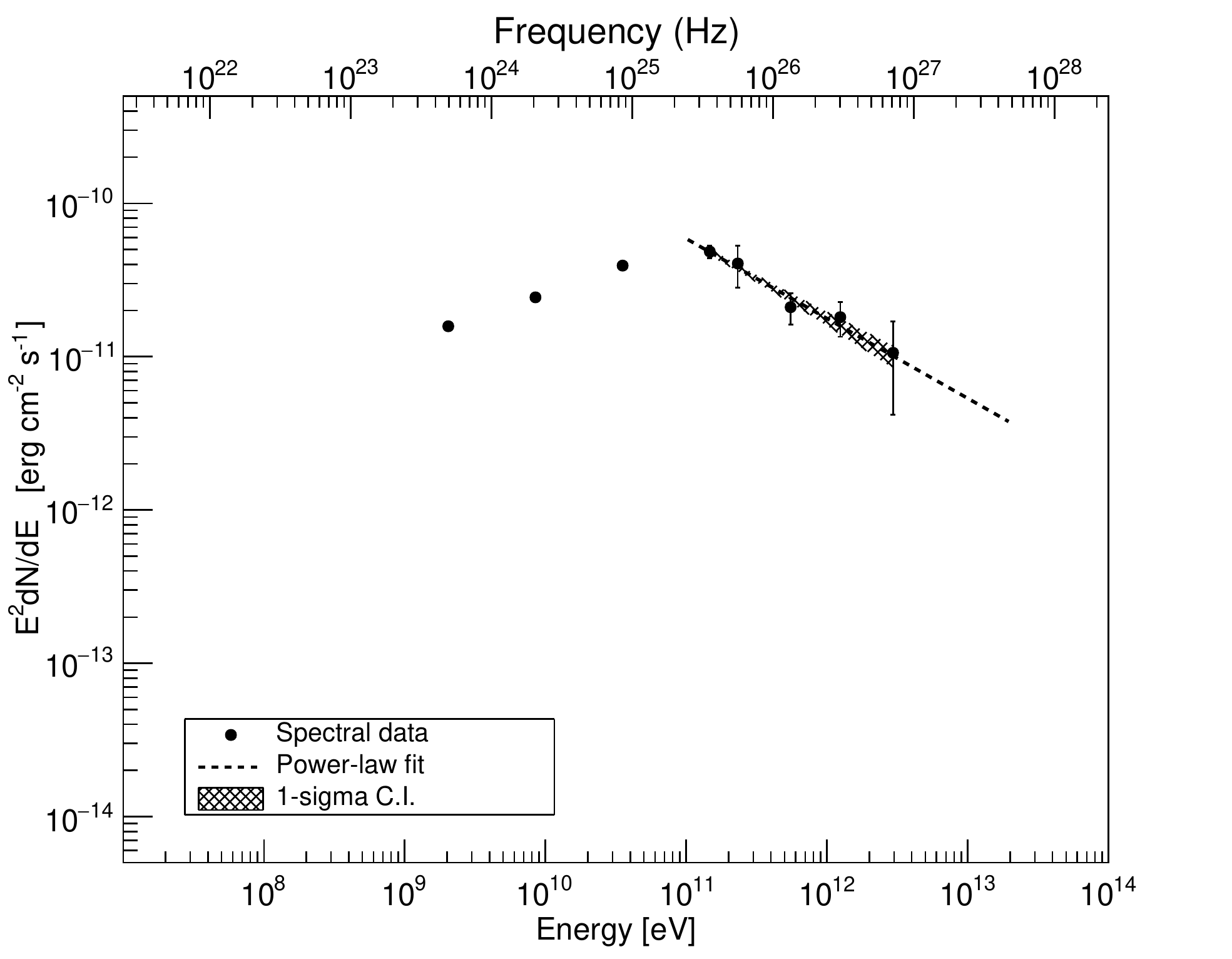}\label{fig:TeV-fits-b}}\\
\caption{TeV power-law fit of 1ES~0229+200 (a) and Mrk 501 (b) in the 100 GeV-10 TeV interval.
These two sources show opposite behaviors in the TeV regime: while 1ES~0229+200 is compatible with the previous  power-law  fit all along the GeV - TeV band, Mrk~501 shows curvature in the spectrum that is down-going in  hard TeV gamma-ray band.}
\label{fig:TeV-fits}
\vspace*{+1cm}
\end{figure*}

\thispagestyle{empty}

\subsection{The TeV behavior}

To better study this trend, we present in \Cref{fig:SEDsuperpositionHE} a zoom on the HE and VHE gamma-ray bands of the previous plot, normalizing now at the 1ES~0229+200 \emph{Fermi}-LAT point at 147 GeV with flux of $1.93 \cdot 10^{-12}$ erg $\text{cm}^{-2}$ $\text{s}^{-1}$. This figure shows that the sources maintain a similar slope in the low energy part of the HE gamma-ray spectrum, but they split in two populations when dealing with data above about 100 GeV. Remembering that all sources in our sample are compatible within the errors with the definition of EHBL (meaning the synchrotron peak position above $10^{17}$ Hz), we see here that some of them show their second  IC peak in the 0.1-1.0 TeV band and with decreasing flux above 1 TeV (e.g. Mrk~501 and Mrk~421), but some others (e.g. the already known hard-TeV EHBLs 1ES~0229+200 and 1ES~0347-121) show an increasing
flux up to the TeV regime, and without clear sign of peak before 10 TeV.

For this reason, we performed two linear fits of the IC data for the 13 sources for which we have HE and VHE gamma-ray data publicly available. We show in \Cref{fig:TeV-fits} an example of such interpolations on two sources 1ES~0229+200 and Mrk~501 where the differences are particularly visible. The overall results are reported in \Cref{tab:tevdetected}.\\

\textbf{HE peaking EHBLs.} 
First of all, we fitted all the HE and VHE gamma-ray data in the entire 100 MeV - 10 TeV range (the whole IC hump). In \Cref{tab:tevdetected}, the \texttt{HE-VHE $\chi^2$} column reports the reduced $\chi^2$-test we performed and shows that that only 5 out of 13 sources present good compatibility with a power-law fit over all this band, meaning that their spectrum is compatible with a continuously increasing IC flux up to the TeV band (as in \Cref{fig:TeV-fits-a}). These sources are the well known hard-TeV EHBLs like 1ES~0229+200, 1ES~0347-121, 3FGL~J0710.3+5908, 1ES~1426+428, and 1ES~1101-232.
Some other sources like the two Markarians, 1ES 1959+650, and 1ES 2344+514, present strong deviations from this linear model, meaning that the HE and the VHE spectra have different slopes and they are probably shaping the IC peak (as in \Cref{fig:TeV-fits-b}). For these sources, we performed a log-parabolic fit overall the IC hump in order to determine a rough estimation of the IC peak, reporting the results in the \texttt{IC peak} column: all sources with high \texttt{HE-VHE $\chi^2$} values show estimated IC peak in 0.1-1 TeV range, while the others clearly peak above their last TeV point (about above 10 TeV).\\

\textbf{TeV peaking EHBLs.} 
Moreover, we performed a power-law fit in the 0.1-10 TeV region (the interval where the TeV slopes seem to diverge) to see if the sources can be distinguished in their up-going or down-going TeV fluxes. In \Cref{tab:tevdetected}, the \texttt{TeV slope} column reports the value of the slope in this band, and the \texttt{TeV $\chi^2$} column indicates that all sources except for Mrk 421\footnote{This is expected because there is a very detailed spectrum available for this source, and the power-law fit is no more a good approximation even in a relatively small energy range.} are now well modeled by the power-law fit in this band.

\begin{figure*}
\centering
\includegraphics[width=\textwidth]{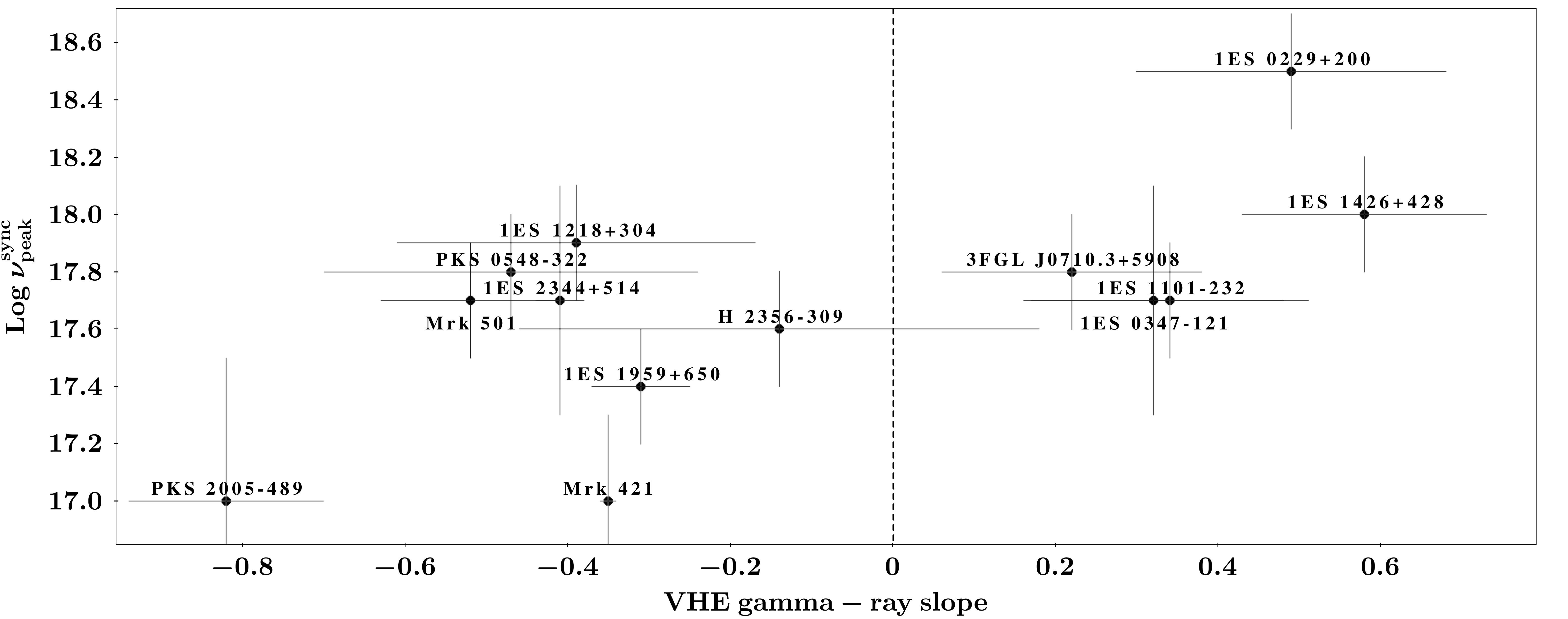}
\caption{Synchrotron peak position of the TeV gamma-ray detected sources   with respect to the distribution of their slopes in the 0.1-10 TeV range.}
\label{fig:TeV_slopes_vs_synchropeak}
\end{figure*}

\begin{figure*}
\centering
\includegraphics[width=\textwidth]{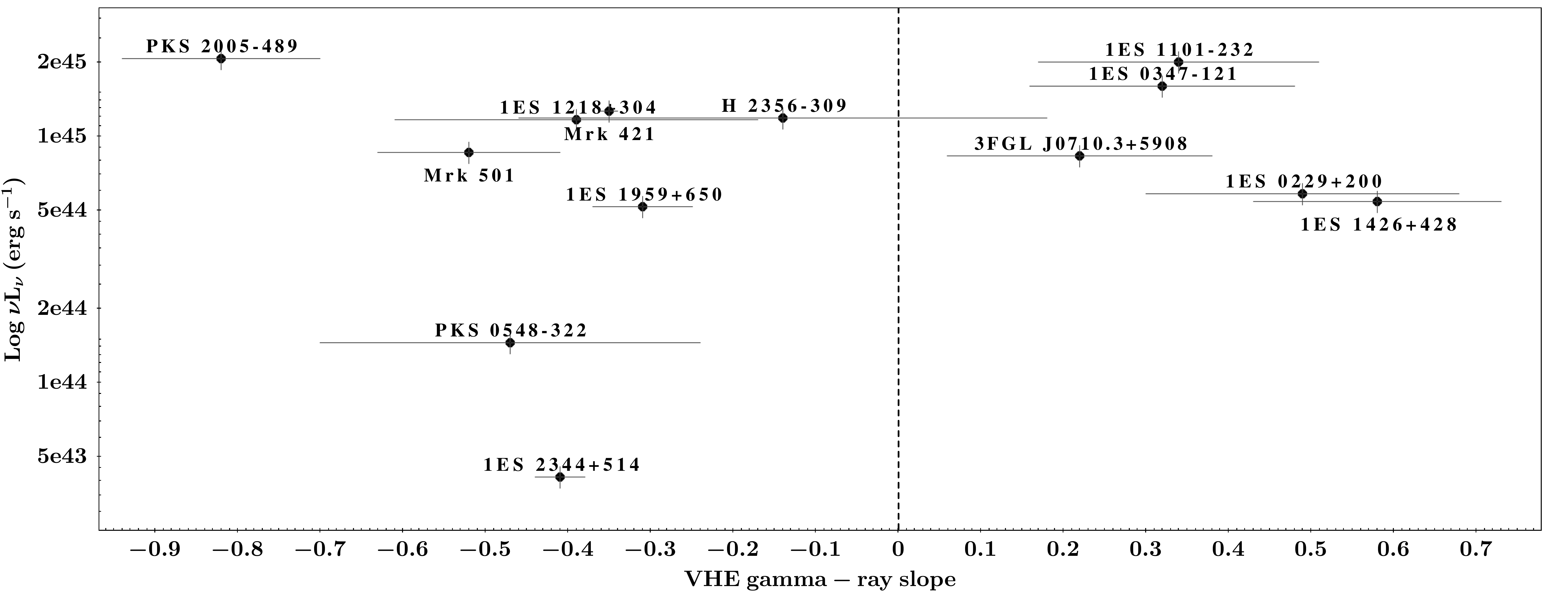}
\caption{Synchrotron peak luminosity of the TeV gamma-ray detected sources   with respect to the distribution of their slopes in the 0.1-10 TeV range.}
\label{fig:TeV_slopes_vs_synchropeak_LUM}
\end{figure*}

We show in \Cref{fig:TeV_slopes_vs_synchropeak} the distribution of the TeV slopes  with respect to the synchrotron peak position of the sources. We notice that the EHBLs seem to cluster in two main different groups: while all the already known hard-TeV EHBLs are located on the right hand side of the plot where the slope is positive, the majority of the other sources are located in the left hand side where the slope is negative and thus are already showing the descending part of the IC hump.

The source H~2356-309 may represent a transitional case because it is located, considering the errors, where the slope in this energy range is on average flat. This could suggest that the IC peak of this source is located in the region around few TeV, and that the source may be starting the descending part of the IC hump. However, to fully understand this feature we would need more precise TeV gamma-ray data.

Except for H~2356-309, the objects on the right of the plot and those on the left are largely incompatible in their TeV slopes, and they support the evidence for two distinct populations of EHBLs.\\

\noindent
\textbf{Synchrotron peak and TeV data.}
We can now see if there is a relation between the TeV slopes and the synchrotron peak positions of this sample of TeV detected sources. In \Cref{fig:TeV_slopes_vs_synchropeak} the sample splits in two similar groups with opposite TeV slope and same range of synchrotron peak frequencies, and objects with the same synchrotron peak position show completely different TeV behaviors. For example, Mrk~501 and 3FGL~J0710+5153 show the same synchrotron peak position, but the first one is down-going in the TeV regime, while the latter is up-going up to the TeV band. Even if it is important to remember that the estimation of $\nu_{\text{peak}}^{\text{sync}}$ is correlated with the variability of the source and the lack of simultaneous data, this fact seems implying that there is not a strong correlation between the synchrotron peak position and the TeV behavior of these sources. Thus, two parallel classes of objects may appear: independently on the synchrotron peak position, some EHBLs are already peaking in the 0.1-1 TeV band, some others increase their flux up to the TeV regime.

%\textbf{
In order to verify the total power emitted by the EHBL objects, in \Cref{fig:TeV_slopes_vs_synchropeak_LUM} we present the luminosity of the synchrotron peaks of the same sources versus their TeV \mbox{gamma-ray} slopes.  All the sources are distributed over a quite narrow range of synchrotron peak luminosities (especially the hard-TeV EHBLs on the right hand side) at about $10^{45}$ erg $s^{-1}$. Some sources like PKS 0548-322 and 1ES 2344-514 of the left-hand side of the plot show lower luminosity peak values.
%}

The question weather the two sub-classes of sources form two distinct populations or a continuum transition cannot be answered with this limited dataset. We need more statistics on TeV gamma-ray detected EHBLs to fully understand this point, and only additional TeV gamma-ray observations will help to disclose this topic.

\subsection{Are EHBLs a unique population?}

All these considerations suggest us that EHBLs are not a homogeneous class. In fact, considering that this sample is composed by sources that match the definition of EHBL based on the synchrotron peak location, the sample may be subdivided at least into two main sub-categories: EHBL sources peaking in the 0.1-1.0 TeV range (we will call them ``HBL-like'' EHBLs), and exceptional EHBLs peaking above 10 TeV (the known ``hard-TeV'' EHBLs). The former seem to represent a sort of continuation of the HBL class, meaning with a high synchrotron peak above $10^{17}$ Hz but the IC peak located in a region compatible with HBL sources and with decreasing slope in the TeV gamma-ray band. %\textbf{
Most of these sources (like 1ES~1959+650, 1ES~2344+514, 1ES~1218+304, Mrk 501, Mrk 421) are also known to be showing moderate flux variability or even frequent flaring activity.

On the other hand, the latter show an IC hump peaking well above the 10-TeV band  and represent the sample of hard-TeV blazars proposed also by \citet{Nustar_EHBLs}.
Looking also at \Cref{fig:TeV_slopes_vs_synchropeak_LUM}, these hard-TeV blazars seem to be a standalone category characterized by high total emitted power, rather stable flux, and hard TeV spectra.

Another important category that could emerge with a more detailed statistics of EHBL sources in the VHE gamma-ray regime could be represented by objects like H~2356-309. Such source, with its flat \texttt{TeV slope} value, is probably peaking at few TeV and might represent a ``transitional'' type of EHBLs with spectral properties between the HBL-like and the hard-TeV EHBLs.

Interestingly, all these results regarding the sub-classification of EHBLs seem not to be strongly correlated with the synchrotron peak position.

\input{tevundetected_table.tex}

\vspace{10pt}

It is important to note that in this paper we can provide only a first limited sample of EHBLs, and that our suggestion should be tested on richer samples of sources. The natural expectation is that, increasing the statistics of the EHBL population, the appearance of a bimodality - or that of a continuum transition - might become more evident.

This study is one example in which the historical classification of EHBLs based on the synchrotron peak position turns out to be difficult to be applied. In fact, the estimation of the peak position is generally challenging because it is correlated with the state of the source, and the different estimations may easily led to wrong classifications of these EHBLs as HBL sources, or vice versa. This is the reason for which, during the selection procedure, we preferred to keep considering in the analysis all the EHBL candidates also if their $\nu_{\text{peak}}^{\text{sync}}$ was not strictly over the value of $10^{17}$ Hz.
Additionally, most of the HBL-like EHBLs showed frequent flaring episodes, during which also their synchrotron and IC peak positions changed.
For this reason, only simultaneous X-ray campaigns would help in order to constrain the peak positions of such blazars and to provide new candidates to the sample to be further investigated with the Cherenkov telescopes.
In fact, it is worth to note that all the spectral features characterizing EHBLs (by definition  based on the synchrotron peak location) generally do not help in distinguishing sources of these two categories, and only the observation at TeV gamma rays of these sources is the key factor in this classification. 
 
%%%%%%%%%%%%%%%%%%%%%%%%%%%%%%%%%%%%%%%%%%%%%%%%%%%%%%%%%%%%%%%%%%%%%%%%%%%%%%%%%%%%%%%%%%%%%%%%%%%

%%%%%%%%%%%%%%%%%%%%%%%%%%%%%%%%%%%%%%%%%%%%%%%%%%%%%%%%%%%%%%%%%%%%%%%%%%%%%%%%%%%%%%%%%%%%%%%%%%%%%%%%%%%%%%%

\section{TeV gamma-ray undetected sources} \label{sec:TEVundetected}

Since the TeV gamma-ray characterization of EHBLs seems to be a key element in the study of this population, the currently undetected TeV gamma-ray  sources lack of fundamental informations to classify them in the different categories that we presented in the previous section.
However, we can extrapolate the updated HE gamma-ray points to evaluate their possible detectability in the TeV gamma-ray band.

\subsection{Spectral extrapolations to the VHE} \label{subsec:extrapolations}

In order to estimate the detectability of the currently VHE gamma rays undetected targets, we extrapolate each HE spectrum in \Cref{fig:PWL-extrapolation}.\\

\noindent
\textbf{Power-law fit.}
For the extrapolation, we assume a power-law function with exponential cut-off as
\[
\hspace{50pt}\frac{dN}{dE}  =  N_0  \Big( \frac{E}{E_0} \Big)^{-\Gamma} \cdot \, \exp{ \Big( -  \frac{E}{E_{\text{cut-off}}} \Big)}
\] 
where $N_0$ is the normalization factor, $E_0$ is the pivot energy, $\Gamma$ is the photon index. These parameters were estimated from the fit of the spectral points of each target in the \mbox{\emph{Fermi}-LAT} range above 1 GeV.   
The obtained spectra are then corrected for EBL absorption with the model by \cite{Franceschini17}  to show the intrinsic spectrum of the source.

Power-law with exponential cut-off functions are very often used to model gamma-ray spectra from both galactic and extragalactic accelerators, as detailed in  \citet{2017APh....88...38R}.

In our case, the choice of this function is driven by the observational evidence (reported also in \citet{Nustar_EHBLs}) that the hard-TeV EHBLs show power-law spectra up to the deep TeV range, with the IC peak not detected by the current TeV gamma-ray instruments. On the other hand, in the case of HBL-like EHBLs, the IC hump peaks in the HE to VHE gamma-ray band, and might be detected by the Cherenkov instruments. For this reason, the exponential cut-off in our function describes the possible peak of the IC hump in the case the selected object presents a HBL-like TeV behaviour.\\

\begin{figure}
\centering
\hspace*{-10pt}
\includegraphics[width=0.99\columnwidth]{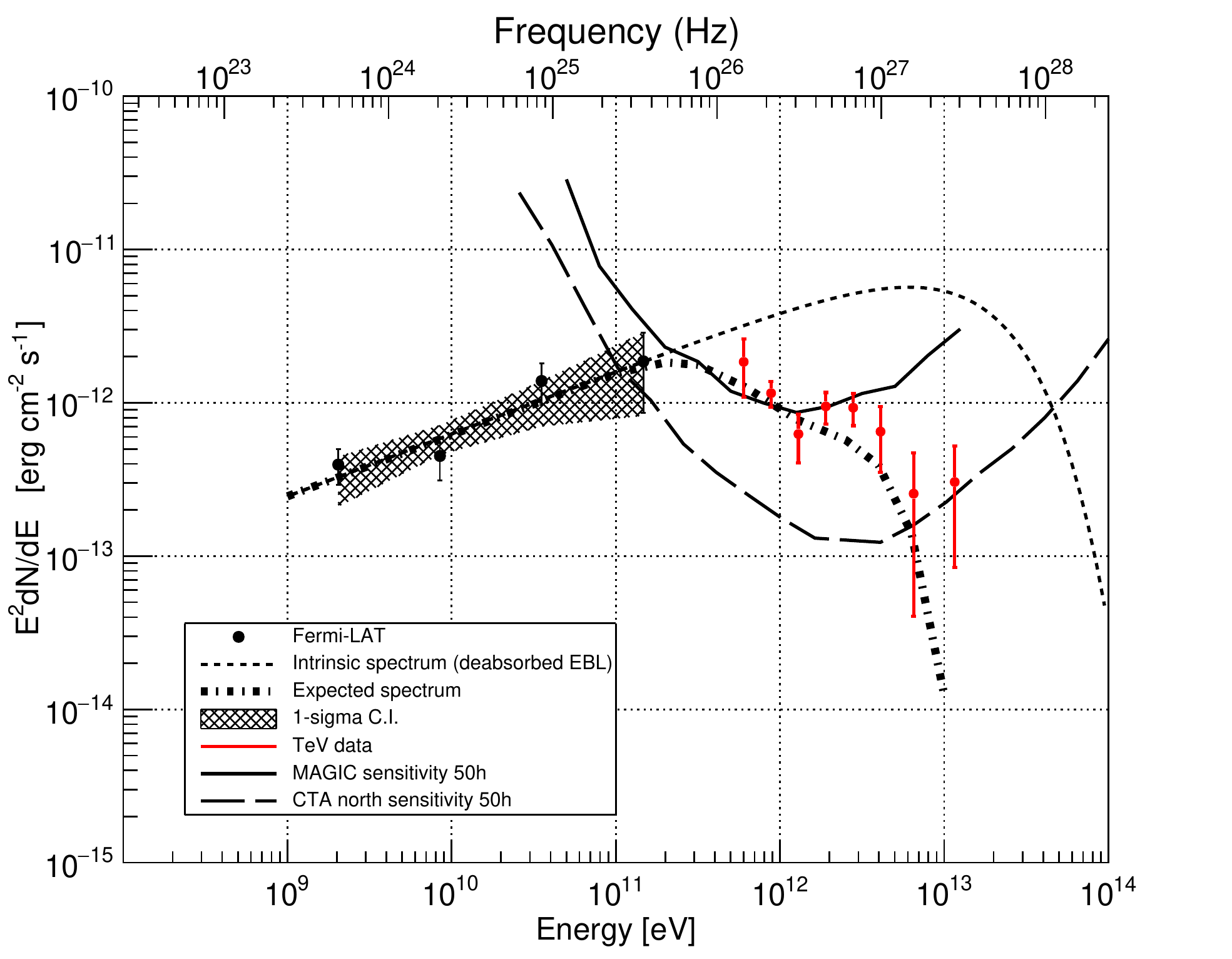}
\caption{Extrapolation at the VHE of the \emph{Fermi}-LAT points obtained for 1ES~0229+200 with $E_{\text{cut-off}}=12$ TeV. We report also the real TeV data (not EBL de-absorbed) of \citet{0229_hess_points} to show that they match well our extrapolation only if considering the right value of the $E_{\text{cut-off}}$.}
\label{fig:1ES0229+200_extrapolation_with_TeV_data}
\end{figure}

\noindent
\textbf{The cut-off at high energies. }
In \Cref{fig:1ES0229+200_extrapolation_with_TeV_data} we show an example of our extrapolation on the 1ES~0229+200, showing both the extrapolation and the TeV data obtained by the SSDC website  \citep{0229_hess_points}. Considering in this plot $E_{\text{cut-off}}=12$ TeV as reported in \citet{Nustar_EHBLs}, as expected we see that our extrapolation matches the real TeV data (not EBL de-absorbed). Thus, the value of $E_{\text{cut-off}}$ strongly affects the TeV detectability of EHBLs. In \citet{Nustar_EHBLs}, the authors show that in all the considered known TeV gamma-ray emitting EHBLs, the IC peak lies above 2\,TeV. However, in our previous analysis in \Cref{sec:TEVdetected} we have seen that some sources like H 2356-309 may peak around 1 TeV. Thus, since for the TeV undetected sources we are missing this information, in our analysis we will adopt  $E_{\text{cut-off}}=$1\,TeV as a conservative cut-off energy in their spectral extrapolations.\\

\noindent
\textbf{Cherenkov visibility and observability. }
In each of these plots the differential sensitivities of the current generation of IACTs are also shown: 23 sources are observable from the northern hemisphere (we are considering visibility from La Palma, below 50 degree in zenith) and they are compared to CTA-North and MAGIC sensitivity (here we use MAGIC sensitivity to VERITAS one because of the lower energy threshold of the former), while for the 9 sources that are better observable from the southern hemisphere we show the CTA-South and H.E.S.S..

Some sources in our sample have already been observed by the VERITAS collaboration \citep{Veritas-2masx} and by the H.E.S.S. collaboration \citep{hess-ul}, however without detection. In \Cref{tab:tevundetected} we report their time exposure, the significance, and the resulting upper-limit values.
All upper-limits provided by the VERITAS collaboration and by the H.E.S.S. collaboration are in good agreement with our spectral extrapolations in \Cref{fig:PWL-extrapolation}.
This fact underlines that these low VHE flux EHBLs may need longer exposures to be detected, and will be good targets for the forthcoming CTA telescopes.\\

\noindent
\textbf{Implications of the redshift. }
The known redshift values of our sources are limited between 0.03 and 0.36. Since we are dealing with photons up to several TeV, in this redshift range the EBL absorption is expected to affect significantly the observability of our sources. This effect is evaluable in our figures: while the ones with redshift below 0.1 are not significantly affected by EBL absorption in the HE and VHE bands up to some hundreds of GeV (see for example \Cref{fig:PKS0352-686-PWL-extrapolation} in relation with other sources, considering the different spectral index and absolute flux), sources with higher redshift (z > 0.25-0.30) suffer a non negligible effect that decreases significantly the observed radiation in the same band (see for example \Cref{fig:1ES1028+511-PWL-extrapolation}). This EBL absorption affects the flux of these sources at Earth, and their detection for redshifts higher than about 0.5 would need a large amount of integration time from the current IACTs and, in some cases, also by the forthcoming CTA observatory. \\

\noindent
\textbf{Hint of HE curvature. }
In order to check the hypothesis of power-law behavior, for all sources we performed a curvature test on the \emph{Fermi}-LAT data with a $\chi$-square ratio test on the power-law model used for extrapolations in \Cref{fig:PWL-extrapolation}. We report the results in \Cref{tab:tevdetected} and \Cref{tab:tevundetected}. For 29 out of 32 sources (88\%) no signs of curvature seem to be present in the second peak at high energies and the computed $\chi$-square ratio test is of the order of 1. On the other hand, an hint of curvature is present in the \emph{Fermi}-LAT SEDs of RBS~259, 1ES~1028+511, and RX~J0324.6+3410. It is worth to note that an hint of curvature already in the \emph{Fermi}-LAT data means that IC peak is probably  located well below the energies of our reference 1ES~0229+200. This probably means that they are probably good candidates as HBL-like EHBLs, rather than hard-TeV EHBLs. \\

\noindent
\textbf{Results}
In our EHBL list, some currently TeV gamma-ray undetected sources show particularly favorable spectral features that could make them well detectable also by the current generation of IACTs. In particular, considering the combination of known redshift, hard spectra in HE gamma rays, and good extrapolations to the VHE (see  \Cref{fig:PWL-extrapolation}), some sources like PKS~0352-686, 1RXS~J225146.9-320614, BZB~J1417+2543, and BZB~J0244-5819, may be good targets also for the MAGIC, the H.E.S.S., and VERITAS telescopes, likely detectable in less than 50h of observations.

%%%%%%%%%%%%%%%%%%%%%%%%%%%%%%%%%%%%%%%%%%%%%%%%%%%%%%%%%%%%%%%%%%%%%%%%%%%%%%%%%%%%%%%%%%%%%%%%%%%%%%%%%%%%%%%

\section{Conclusions} \label{sec:concl}

In this paper we aim at providing a sample of extremely high-peaked BL Lac (EHBL) objects and studying their broad-band spectral features.

Since the SED of these sources in the $\nu \,F_{\nu}(\nu)$ plot reveals a synchrotron peak located in hard X-rays, we adopt the \emph{Swift}-BAT 105-months hard X-ray catalog to build our reference source sample and to eventually achieve a statistical flux completeness. We complement the SED data of  these luminous objects in the hard X-ray band with archival radio, UV, optical, and soft X-ray data.  For the HE gamma-ray band, we use the \emph{Fermi}-LAT 3LAC catalog, and update the results with a new ten-years \emph{Fermi}-LAT analysis of our source sample. 

Our work results in a sample of 32 EHBL objects out of 158 of the reference  \emph{Swift}-BAT 105-months blazar sample. Most of them are characterized by spectral properties similar to those of the archetypal EHBL 1ES~0229+200 at energies below about 100 GeV. The broad-band SEDs of all objects show a synchrotron peak located at frequencies higher than $10^{17}$ Hz, confirming their EHBL nature.
Since the synchrotron peak position is particularly relevant for the classification of EHBLs, we have provided a new estimation of the synchrotron peak frequency for all sources using all the available X-ray and hard X-ray archival data.

Even though all sources in the sample generally show comparable spectral observables, some discrepancies appear in their TeV gamma-ray spectra. Some indicators like the flux variability in hard X-rays and HE gamma rays, the curvature of their HE gamma-ray spectra, and the TeV detection, are not homogeneous in the sample. This fact inspired us to study the available VHE spectra of the TeV gamma-ray detected sources.

%%%%%%%%%%%%%%%%%%%%%%%%%%%%%%%%%%%%%%%%%%%%%%%%%%%%%%%%%%%%%%%%%%%%%%%%%%%%%%%%%%%%%%5

We found that the sources seem to be subdivided at least into two main groups: ``HBL-like'' EHBLs with VHE gamma-ray spectra peaking in the 0.1-1.0 TeV range, and ``hard-TeV'' EHBLs peaking above about 10 TeV. The former are probably an extrapolation of the HBL class to sources with a high synchrotron peak above $10^{17}$ Hz, but with their IC hump  peaking already in the near TeV gamma-ray band. 
These sources are characterized by moderate to high flux variability, and in some cases even notable flaring activity.
Conversely, the latter show a rather stable flux and an IC peak energy exceeding the 10 TeV threshold.

Richer samples of sources may provide more information on weather the two sub-classes of sources form two distinct populations or a continuum transition. 

For example, in our sample the source H~2356-309 might represent a ``transitional'' type of EHBL with IC hump peaking at few TeV and intermediate spectral features with respect to HBL-like and hard-TeV EHBLs.

The results obtained in this sample of objects confirm the features found in the literature with regards to the different IC peak locations of the EHBL sources and provide a first collection of such spectral differences in a unique sample of sources. Furthermore, we found that the TeV behavior of the EHBL class seems not to be strongly correlated with both the synchrotron peak position and luminosity. All this considerations suggest us that EHBLs may not be a homogeneous class: these differences might support different approaches to the modeling of such sources, whose physical interpretation will be covered in a future work.

It is important to note that, considering that this sample is composed by sources that match the definition of EHBL based on the synchrotron peak location, all the spectral features characterizing EHBLs are generally compatible between these two categories (according to the sensitivity of our instruments), and only the TeV characterization of these sources is the key factor in this classification.

The EHBLs in our source sample make a good sample to which TeV observational campaigns should be addressed.
In fact, the combination of the shift  to higher energies of the two humps  and the relatively low redshift of all sources in our sample (below $z=0.36$) makes the optical radiation by the host galaxy visible in the SEDs. For this reason, the majority of our candidates (84\%) shows a known redshift value, and the sources with unknown redshift are good candidates to be targeted by specific optical campaigns. 
Additionally, the low redshift of the candidates implies that the EBL absorption of TeV gamma rays expected from their hard VHE spectra is not too severe, and the detection by the IACTs may need a reasonable exposure time.
The combined effect of a low redshift and hard TeV spectra make these the ideal targets to constrain the EBL intensity at long IR wavelengths, that has never been observed.

In our EHBL list, some currently TeV gamma-ray undetected sources show particularly favorable spectral features that could make them well detectable also by the current generation of IACTs. An evolution in this field will certainly be played by the forthcoming CTA observatory, that with its improved sensitivity (almost an order of magnitude) will have a key-role in detecting new EHBLs. All these sources, both the ones already observed and those never observed by the current IACTs, are excellent targets for the CTA telescopes. 

\vspace{40pt}

\label{lastpage}

\section*{Acknowledgements}

We thank the journal referee for the insightful and valuable comments that helped us to improve the paper.
We would like to thank also Eugenio Bottacini, Giovanni Busetto, and Michele Doro for their helpful suggestions during the preparation of this work.

This research has made use of public \emph{Swift} and \emph{Fermi} data obtained from the High Energy Astrophysics Science Archive Research Center (HEASARC), provided by NASAs Goddard Space Flight Center through the Science Support Center. Part of this work is also based on archival data and on-line services provided by the ASI Science Data Center (ASDC).  This research has made use the TeVCat online source catalog (\url{http://tevcat.uchicago.edu}).

This publication makes use of data products from the Wide-field Infrared Survey Explorer, which is a joint project of the University of California, Los Angeles, and the Jet Propulsion Laboratory/California Institute of Technology, funded by the National Aeronautics and Space Administration.

This research made use of Enrico, a community-developed Python package to simplify \emph{Fermi}-LAT analysis \citep{enricosoftware}.

%%%%%%%%%%%%%%%%%%%%%%%%%%%%%%%%%%%%%%%%%%%%%%%%%%



\bibliographystyle{mnras}
\bibliography{bib_paper_n1}
%\nocite{*}

%%%%%%%%%%%%%%%%%%%%%%%%%%%%%%%%%%%%%%%%%%%%%%%%%%

%%%%%%%%%%%%%%%%% TABLE BIG %%%%%%%%%%%%%%%%%%%%%

\newpage

%%%%%%%%%%%%%%%%%%%%%%%%%%%%%%%%%%%%%%%%%%%%%%%%%%

%%%%%%%%%%%%%%%%% APPENDICES %%%%%%%%%%%%%%%%%%%%%

\appendix

\section{Source selection procedure} \label{appendix:selection}
\begin{figure}
\centering
\hspace*{-10pt}
\includegraphics[width=0.99\columnwidth]{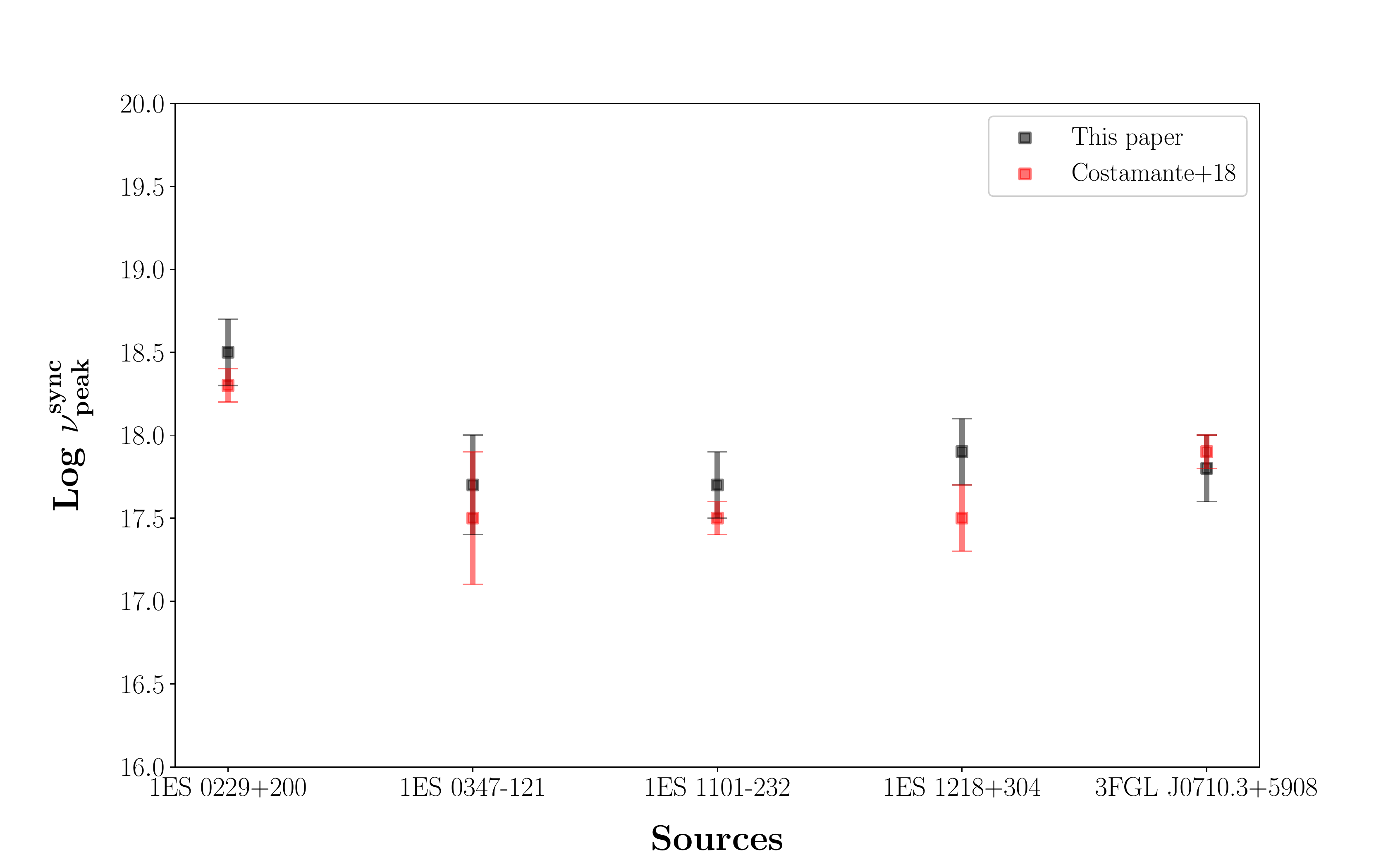}
\caption{Comparison of the synchrotron peaks estimated in this work and the ones in \citet{Nustar_EHBLs}.}
\label{fig:comparison_synchropeaks}
\end{figure}

The main step of the selection procedure was the cross-match between the \emph{Swift}-BAT 105-months catalog and the \emph{Fermi}-LAT 3LAC catalog (as described in \Cref{sec:2}). We used a search radius of 20 arcmin corresponding to the maximum value of the error box distribution of the BL Lac objects reported in the  \emph{Fermi}-LAT  3LAC catalog. This operation resulted in two sub-samples of 86 \emph{Fermi}-LAT 3LAC detected sources and 72 \emph{Fermi}-LAT 3LAC undetected sources.\\

\begin{figure}
\centering
\hspace*{-10pt}
\includegraphics[width=1.05\columnwidth]{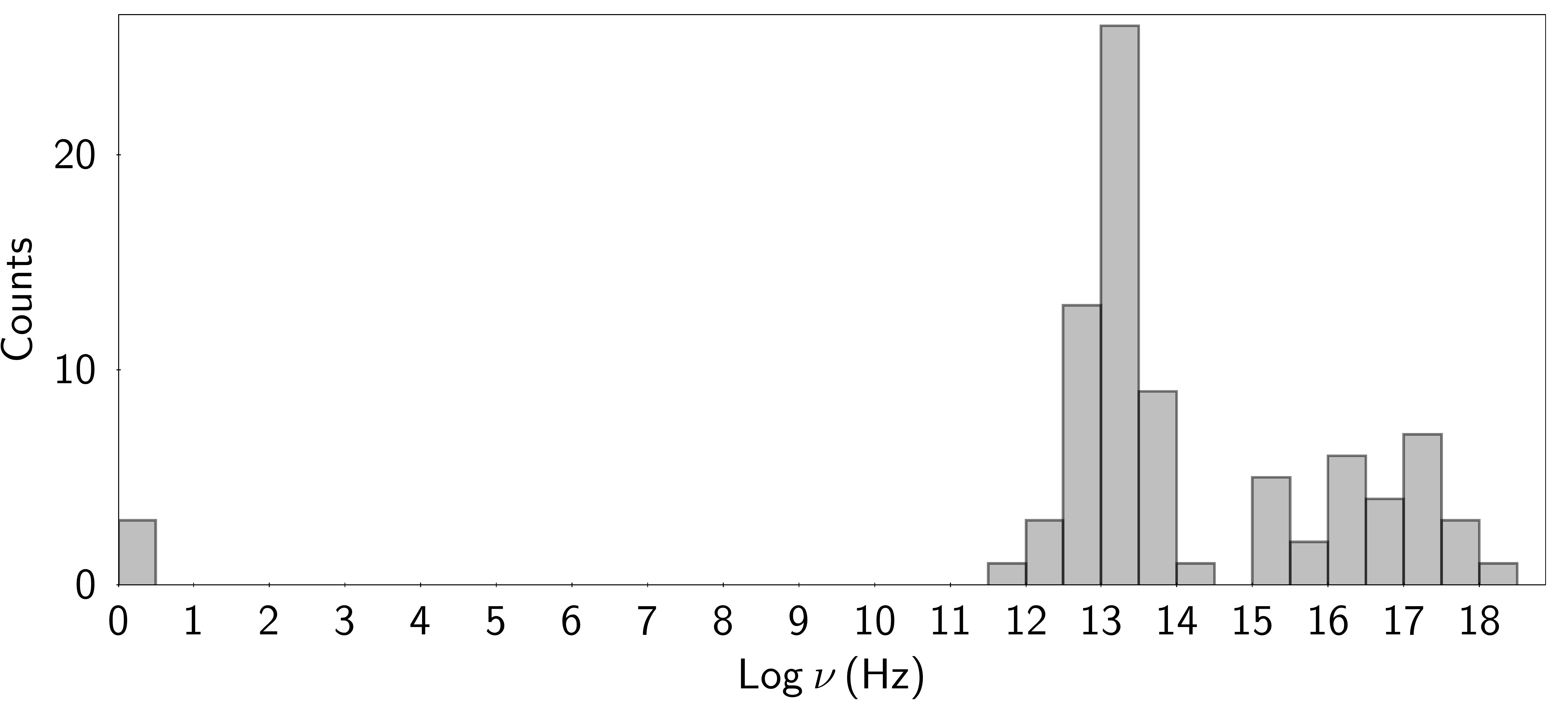}
\caption{Distribution of the synchrotron peak frequencies in the sample of 86 sources obtained cross-matching the \emph{Swift}-BAT beamed AGNs and the \emph{Fermi}-LAT 3LAC catalog.}
\label{fig:BAT105_vs_3LAC_histogram}
\end{figure}

\noindent
\textbf{\emph{Fermi}-LAT 3LAC detected sources. } Concerning the sample of 86 \emph{Fermi}-LAT 3LAC detected sources, we plot the distribution of their synchrotron peak frequencies in \Cref{fig:BAT105_vs_3LAC_histogram}. Except for the three sources with unknown synchrotron peak frequency, we notice that this sample is splitted in two main groups: the blazars with synchrotron peak around $10^{13}$ Hz, meaning a sample of LBL objects, and the blazars with synchrotron peak above $10^{15}$ Hz. We explain this fact as a selection effect due to the sensitivity of the BAT instrument around $10^{-12}$ erg $\text{cm}^{-2}$ $\text{s}^{-1}$ in the hard X-ray band. This implies that this sample is composed by a group of HBL/EHBL objects of which the BAT instrument detects the last part of the synchrotron peak, and by a sample of LBL sources of which the BAT instrument already detects the second peak. Following this idea, IBL objects with synchrotron peak frequency around $10^{14}$ Hz are excluded from this distribution because the BAT energy range in the SED of those sources is located in the region between the two peaks and the instrument is not sensitive enough to detect them.

For what concerns our analysis, we decided to keep all sources with synchrotron peak exceeding  at least $10^{15}$ Hz and not only $10^{17}$ Hz as definition of EHBL because the estimation of the synchrotron peak position is commonly affected by large errors due to the data selection used to fit the polynomial to the MWL data.
For this reason, in this selection of 86 \emph{Fermi}-LAT 3LAC detected sources, we selected all the sources with the synchrotron peak situated above $10^{15}$ Hz, excluding all ISP and LSP objects, and reducing the sample to 28 sources. In this sample we decided to add the three sources with undefined synchrotron peak: thus the sample increases to 31 objects.

Thus, with a cross-match using a search radius of 10 arcmin corresponding the maximum value of the error box distribution of the BL Lac sources reported in the \emph{Swift}-BAT 105 catalog, we checked in the 2WHSP catalog the synchrotron peak position of our previous sample of 31 \emph{Fermi}-LAT 3LAC detected objects, finding 28 out of 31 sources of which we report the updated value in \Cref{tab:sourcelist}.\\

\noindent
\textbf{Exceptions.}
For some sources in the sample, we performed a detailed search in literature to check their classification. In particular, we noticed that  
the source RX J0324.6+3410 was classified as Sy 1 galaxy by \citet{Motch:1997hz} and is now classified as NLSY1 galaxy (see \citealt{Zhou:2007bf} and \citealt{Healey:2007by}) in the \emph{Swift}-BAT 105-month catalog. 
Thus, this source is not interesting in our sample. \\

\noindent
\textbf{\emph{Fermi}-LAT 3LAC undetected sources. }
Coming now to the \emph{Fermi}-LAT 3LAC undetected sources (72 sources), we decided to
check if some new blazar of the 2WHSP catalog is present. We found three more sources PKS~0706-15, PKS~2300-18 and 1RXS~J225146.9-320614, that additionally show association in the \emph{Fermi}-LAT 3FGL catalog. After  a detailed search in literature, we noticed that the source PKS~2300-18 actually is not a blazar but is classified as Sy 1 galaxy by \citet{Healey:2007by}, and it is not interesting for our study. Finally, the other two sources these sources PKS~0706-15 and 1RXS~J225146.9-320614 have been added to the main sample, reaching a total final number of 32 sources.\\

%%%%%%%%%%%%%%%%%%%%%%%%%%%%%%%%%%%%%%%%%%%%%%%%%%%%%%%%%%%%%%%%%%%%%%%%%%%%%%%%%%%%%%%%
\vspace{-20pt}

\section{\emph{Fermi}-LAT analysis} \label{appendix:fermi}
For all the sources in our final sample we performed an updated analysis over ten years of operation of the \emph{Fermi}-LAT telescope using the data publicly available on the \emph{Fermi} Science Support Center. The results  are reported in \Cref{tab:sourcelist}.

\begin{figure*}
\centering
\includegraphics[width=0.9\textwidth]{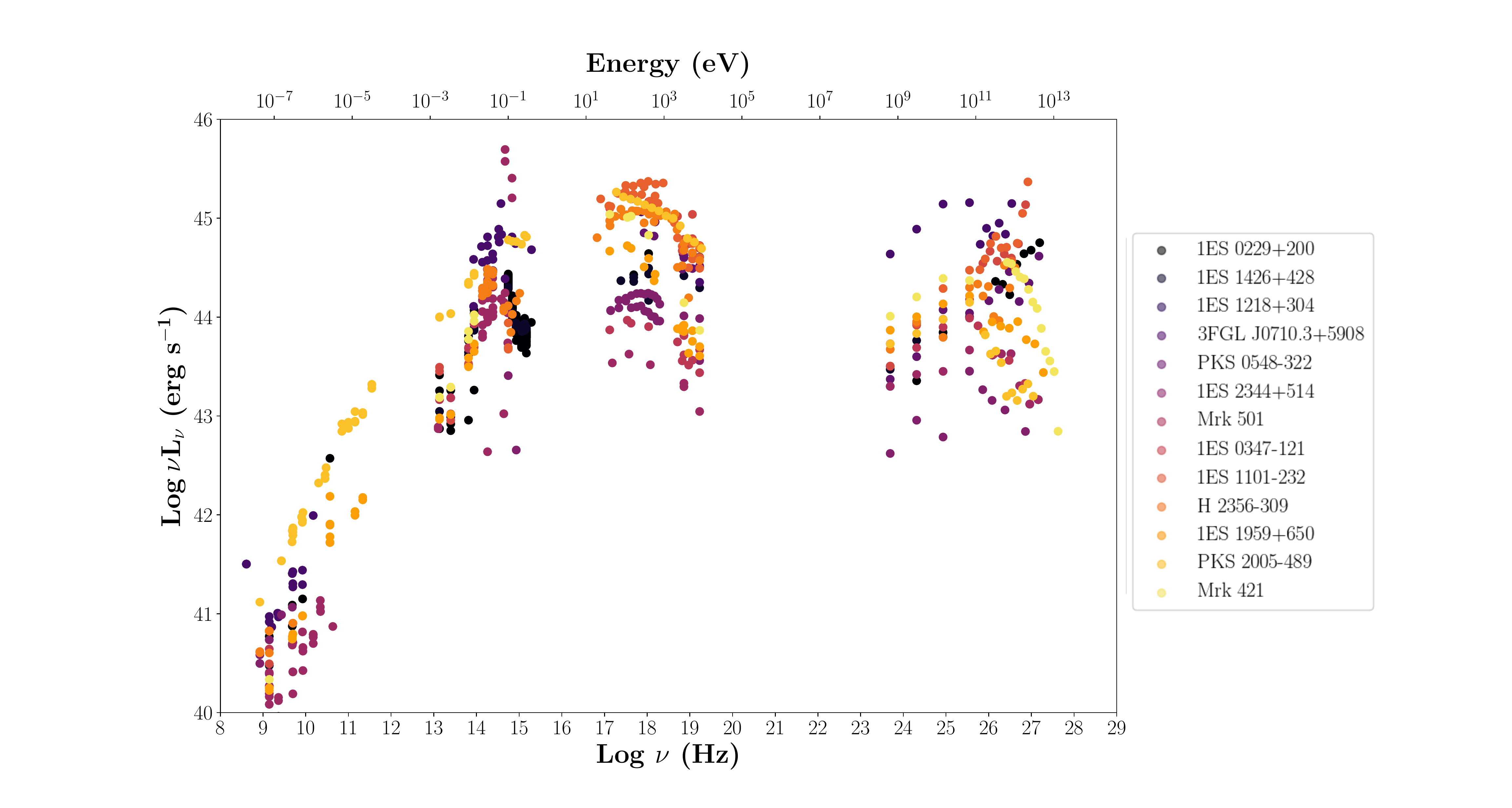}\label{fig:SEDsuperpositionALL-lum}
\caption{Superimposition of the MWL SEDs of the 13 already TeV gamma-ray detected sources with publicly available TeV data. The fluxes have been converted in luminosity by considering the redshift reported in the \emph{Swift}-BAT 105-months catalog. The plotted data are already corrected for EBL absorption with the model by \citet{Franceschini17} to show the intrinsic spectrum of the source. }
\end{figure*}

We analyzed them using the \emph{Fermi}-LAT Science Tools version \texttt{v10.r0.p5}, together with the Pass 8 instrument response functions, the corresponding \texttt{iso-P8R2-SOURCE-V6-v06} isotropic model and \texttt{gll-iem-v06} galactic diffuse background model. 
The event selection was based on Pass 8 reprocessed source (event type 3 and class 128) photons in the 1-300 GeV energy range, collected from 2008 August 4 (MET 239557417) to 2018 July 19 (MET 553654936) and coming from a 15$^\circ$ radius region of interest (ROI) centered at the nominal position of the source. The cut in energy above 1 GeV was set because, taking into account that the detector Point Spread Function improves with increasing energy and that the spectral slope of HSP sources is relatively hard, it helps to avoid contamination from nearby sources.\\
The events were selected and filtered through standard quality cuts. 
The target was modeled with \emph{PowerLaw2}:
\[
\hspace{70pt}\frac{dN}{dE}  =   \frac{  N_0 (\Gamma+1) \, E^{\Gamma} }{ E_{\text{max}}^{\Gamma+1}-E_{\text{min}}^{\Gamma+1}}
\] 
where $N_0$ is the normalization factor, $\Gamma$ is the photon index $E_{\text{max}}$  and $E_{\text{min}}$ are fixed parameters representing the range where the integral flux is calculated.\\
Besides the target and backgrounds, all the \emph{Fermi}-LAT 3FGL catalog point sources in the field were included in the model \citep{3FGLcatalog}. A standard binned analysis was then performed. The test statistic (TS) was used to verify the significance of excess signal of our sources. The TS is defined as \citep{TSmapsConfidenceRadius}
\[
\hspace{70pt}\text{TS}=-2 \; \ln \Bigg(  \frac{L_{\,null}}{L_{\,source}} \Bigg)
\]
where $L_{\,null}$ and $L_{\,source}$ are the likelihoods of observing a certain flux for a model respectively without the candidate source (the null hypothesis) and including the additional candidate source. This quantity allows to determine how much a source emerges from the background: a TS$\approx$25 is equivalent to a 5$\sigma$ detection \citep{FermiCat1}, and only cases with TS$>$25 are considered as a positive detection of point-like source.

\section{TeV gamma-ray data} \label{appendix:tevdata}
The TeV gamma-ray data were collected looking for data up to the highest available energy. We tried to avoid flaring states of the sources, but some of them are known variable sources and may present strong fluctuations in this band in different observational campaigns. 

We used the following TeV gamma-ray data: 
\begin{itemize}
\item 1ES~0229+200 in \citet{0229_hess_points} (MJD 53614),
\item 1ES~1426+428 in \citet{hegra_1426_points} (MJD 51179-51909),
\item 1ES~1959+650 in \citet{veritas_1959_points} (MJD 54417), 
\item PKS~0548-322 in \citet{0548-hess-discovery} (MJD 53279-54495), 
\item 3FGL~J0710.3+5908 in \citet{0710discovery} (MJD 54801-54891),   
\item Mrk~501 in \citet{mrk501_tev_magic_data} (MJD 54913), 
\item 1ES~2344+514 in  \citet{2344_tev_data} (MJD 54377-54476),  
\item 1ES~0347-121  in \citet{0347-hess-discovery} (MJD 53973-54090), 
\item 1ES~1101-232 \citet{1101discovery} (MJD 53111-53445),  
\item 1ES~1218+304 in \citet{1218-veritas-tevpoints} (MJD 54115-54180),  
\item H~2356-309 in \citet{h2356_discovery} (MJD 53530-53615), 
\item PKS 2005-489 in \citet{pks2005_tev_data}  (MJD 53171-54345),    
\item Mrk~421 in \citet{mrk421_tev_data} (MJD 53107-53114.9).
\end{itemize}

\vspace{-25pt}
\section{Figures} \label{appendix:figures}

%%%%%%%%%%%%%%%%%%%%%%%%%%%%%%%%%%%%%%%%%%%%%%%%%%

%\newpage

\input{mwlplots.tex}

\newpage

\begin{figure*}
\centering
\vspace{-2cm}
\subfloat[][PKS 0352-686 with $z=0.085$ (left) and PKS 0706-15 with $z=0.001$ (right).] {\includegraphics[width=0.52\textwidth]{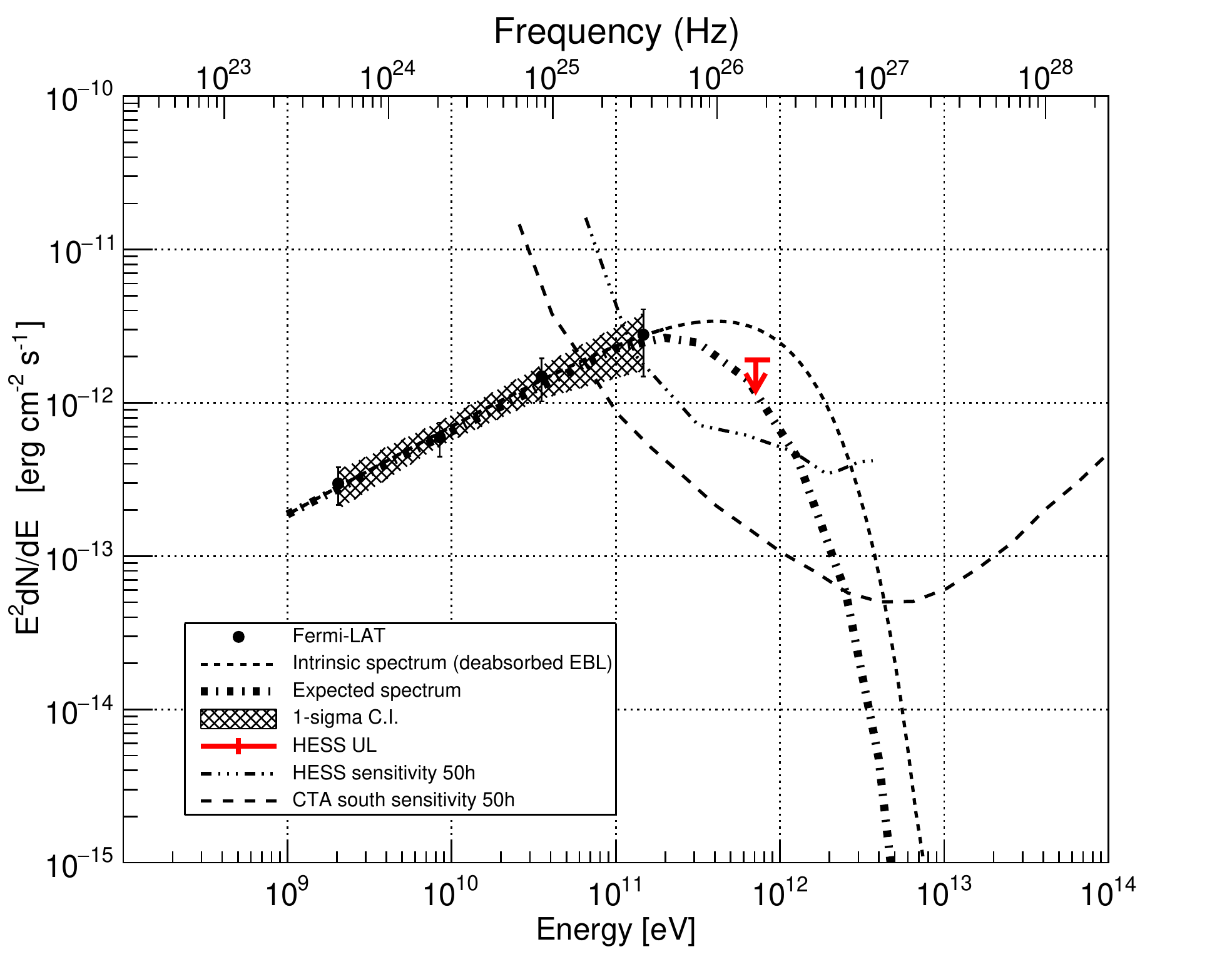}\label{fig:PKS0352-686-PWL-extrapolation} \includegraphics[width=0.52\textwidth]{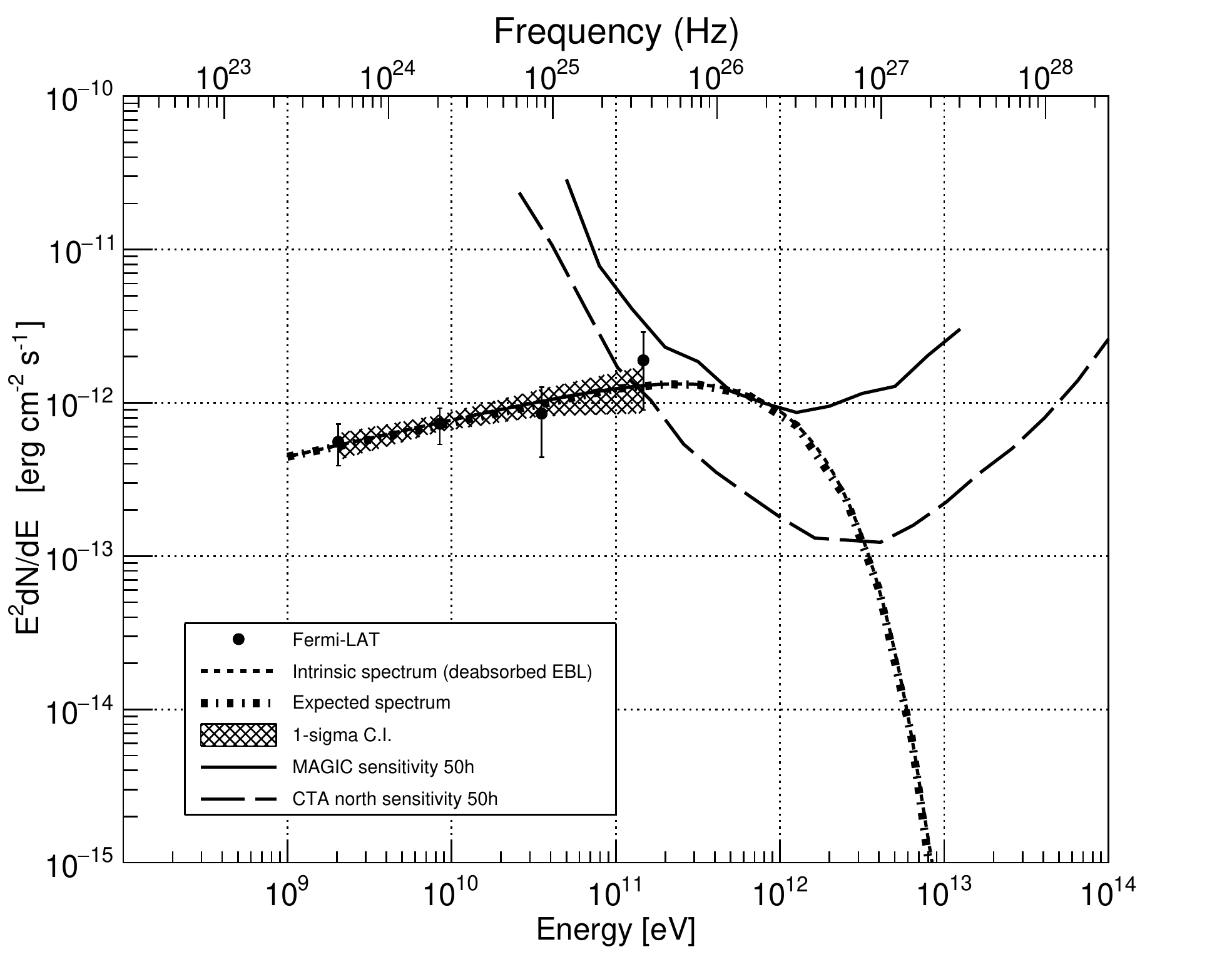}\label{fig:PKS0706-15-PWL-extrapolation}}\\
\subfloat[][1ES 0120+340 with $z=0.272$ (left) and RBS 259 with $z=0.001$ (right).] {\includegraphics[width=0.52\textwidth]{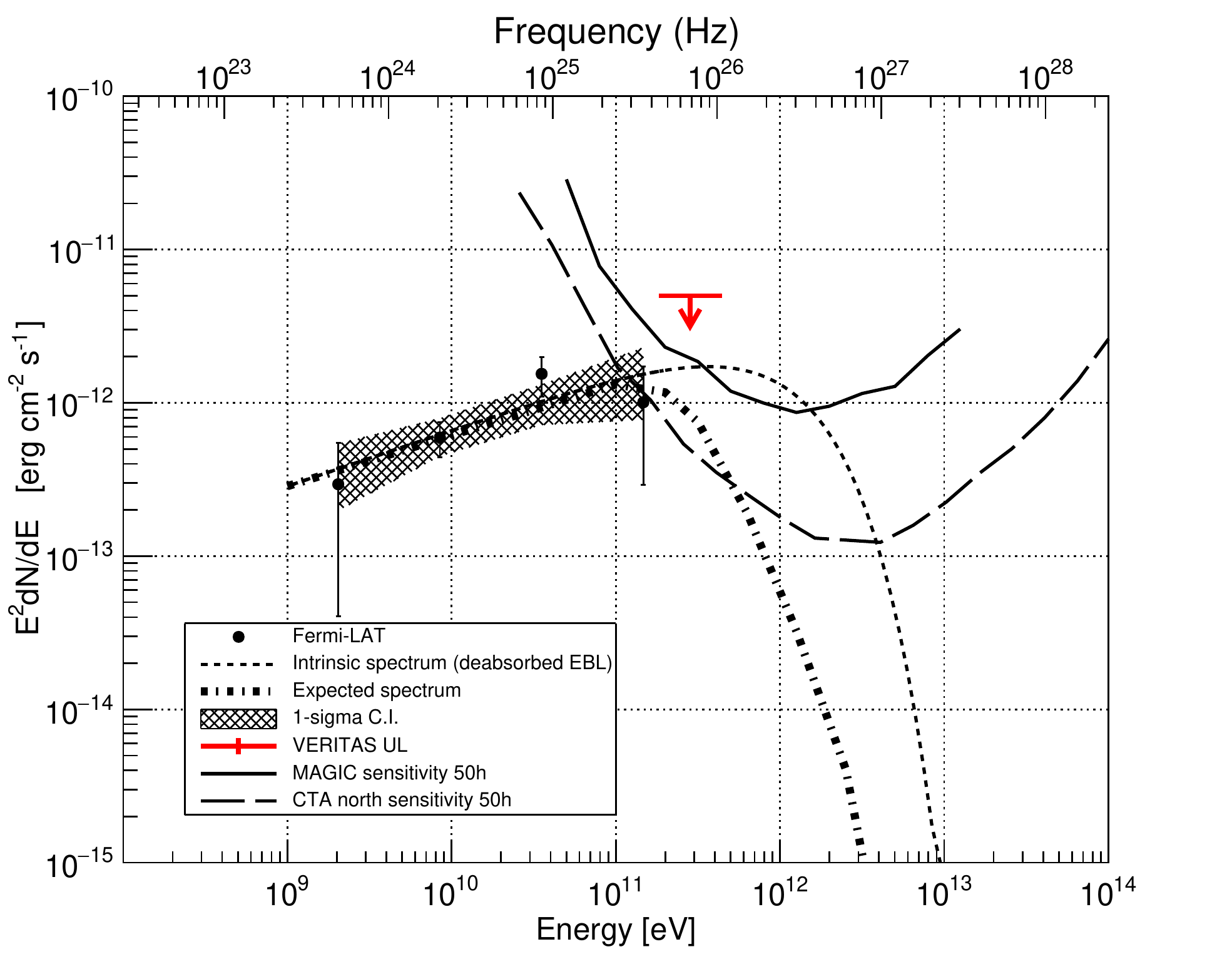}\label{fig:1ES0120+340-PWL-extrapolation} \includegraphics[width=0.52\textwidth]{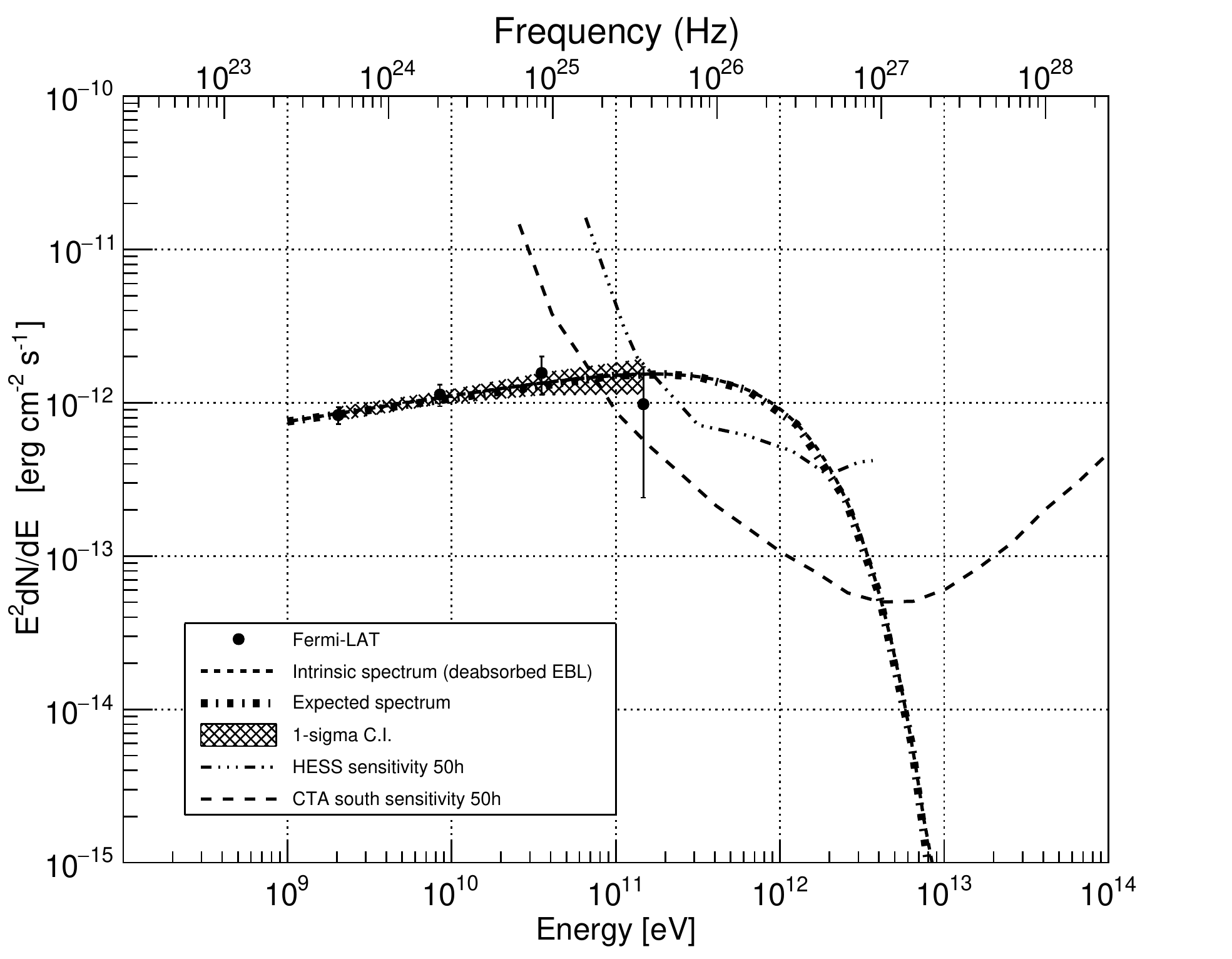}\label{fig:RBS259-PWL-extrapolation}}\\
\subfloat[][1RXS J225146.9-320614 with $z=0.246$ (left) and 1ES 0927+500  with $z=0.187$ (right).] {\includegraphics[width=0.52\textwidth]{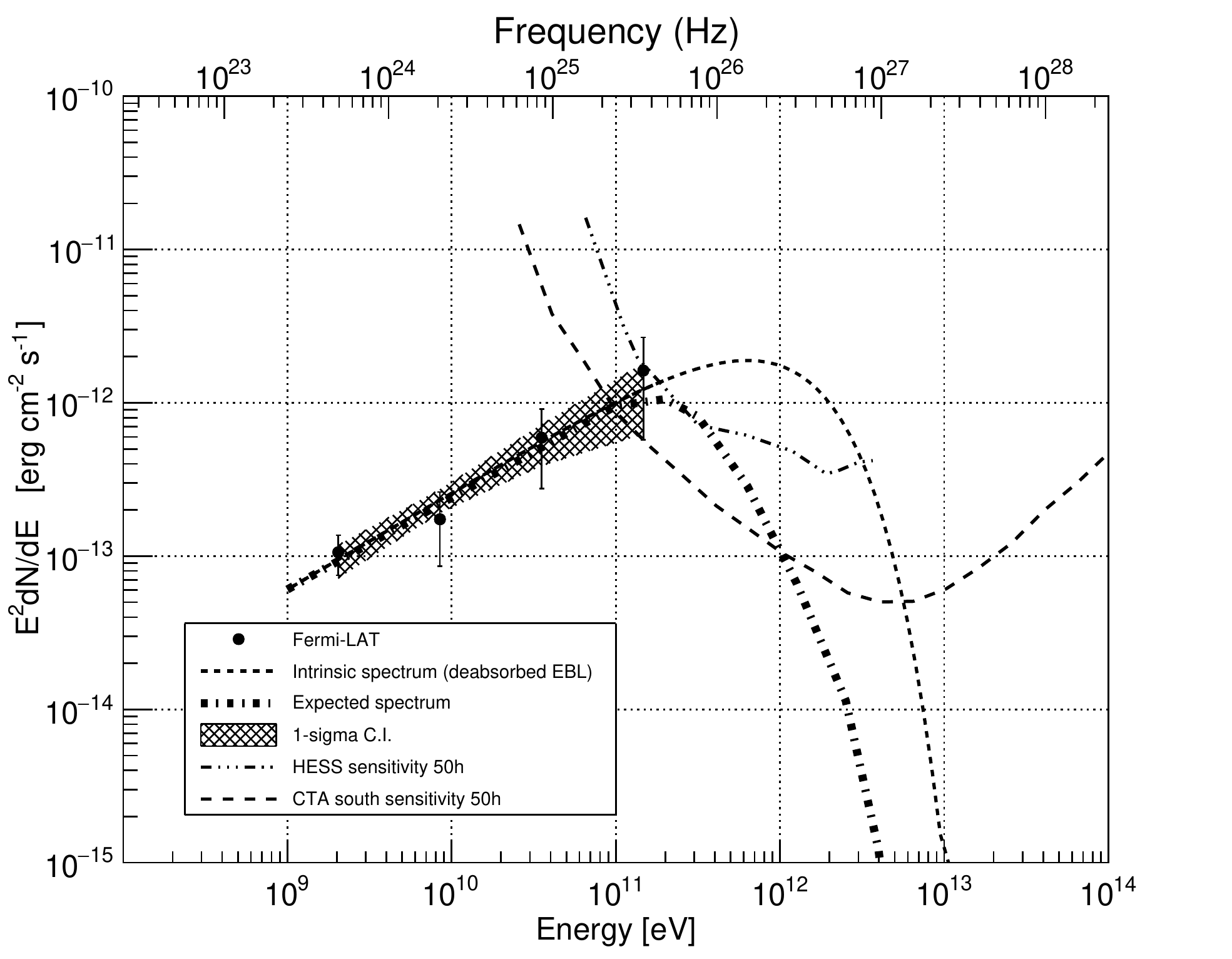}\label{fig:1RXSJ225146.9-320614-PWL-extrapolation} \includegraphics[width=0.52\textwidth]{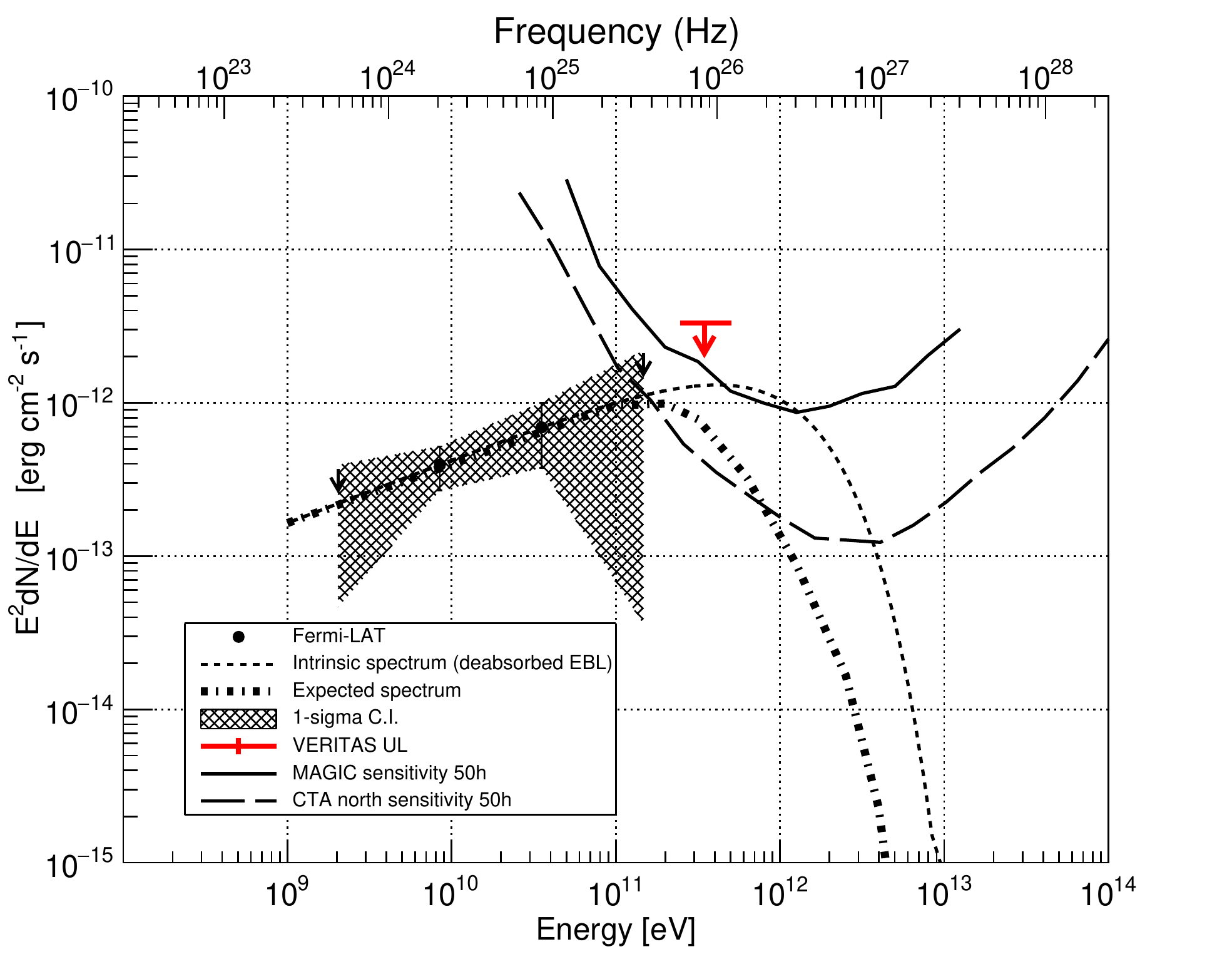}\label{fig:2MASSJ09303759+4950256-PWL-extrapolation}}\\
\end{figure*}

\begin{figure*}
\centering
\ContinuedFloat

\subfloat[][3FGL J0733.5+5153  with $z=0.001$ (left) and RBS 1895 with $z=0.194$ (right).] {\includegraphics[width=0.52\textwidth]{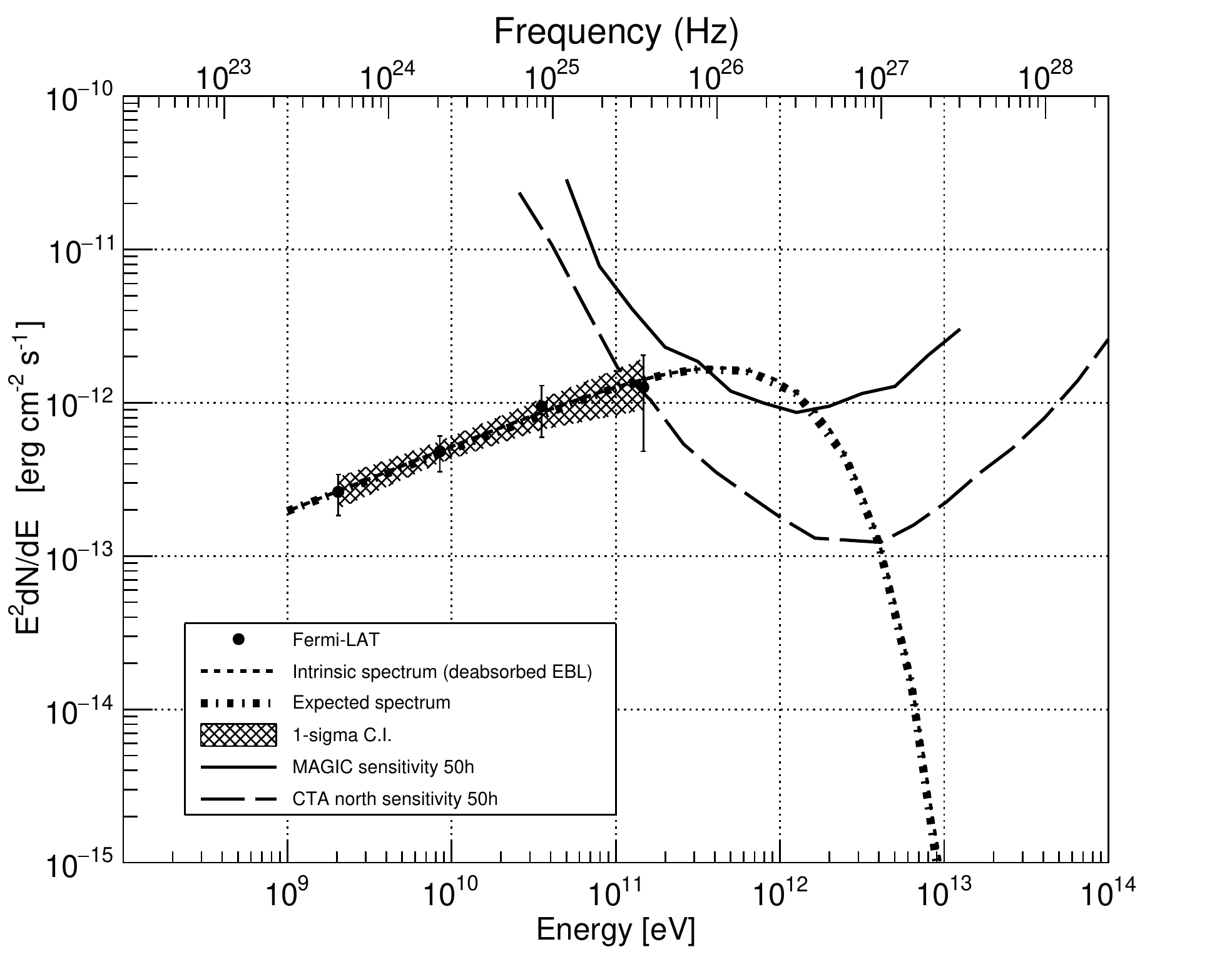}\label{fig:2MASXJ07332681+5153560-PWL-extrapolation} \includegraphics[width=0.52\textwidth]{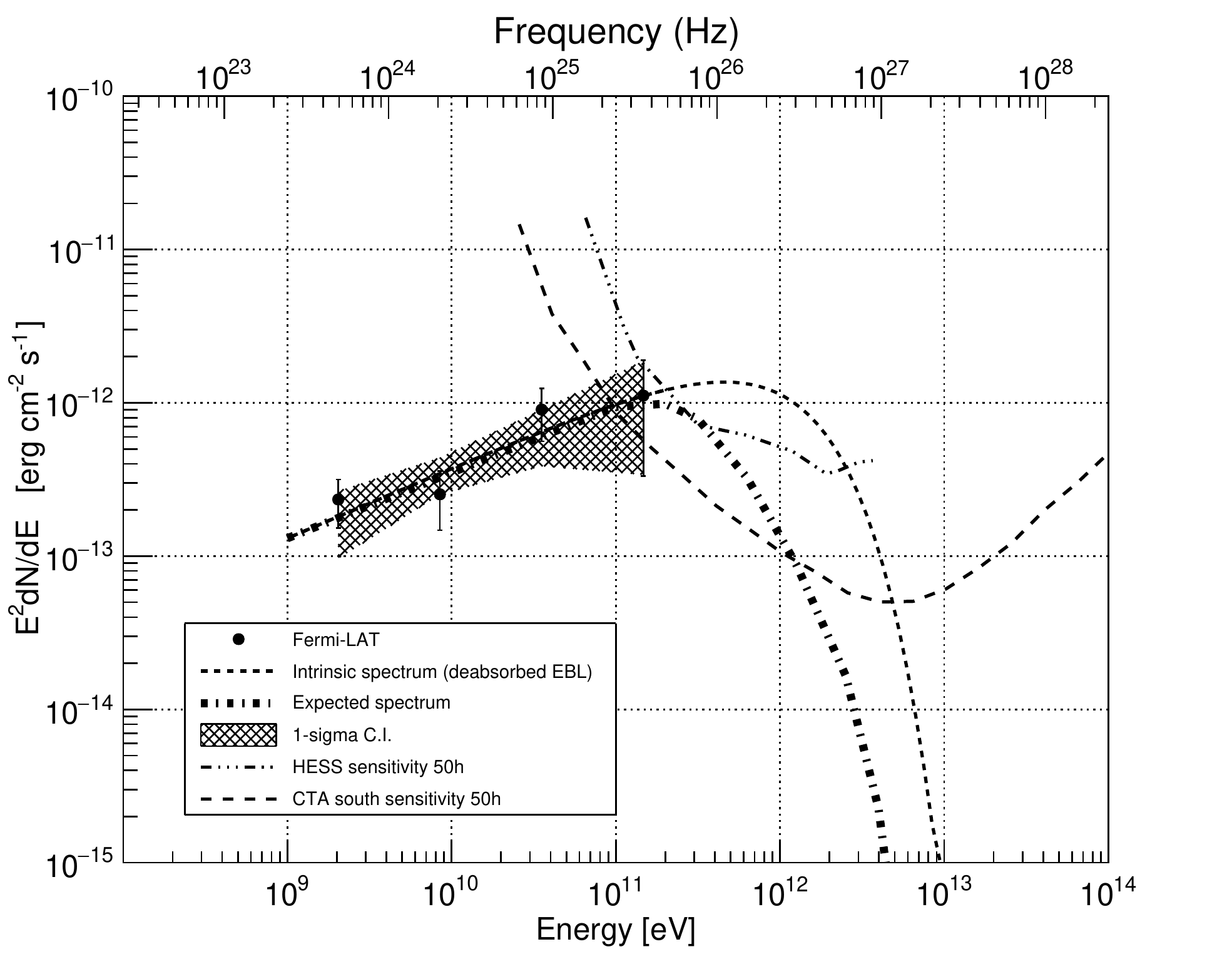}\label{fig:RBS1895-PWL-extrapolation}}\\
\subfloat[][BZB J1417+2543 with $z=0.23$ (left) and TXS 0637-128  with $z=0.001$ (right).] {\includegraphics[width=0.52\textwidth]{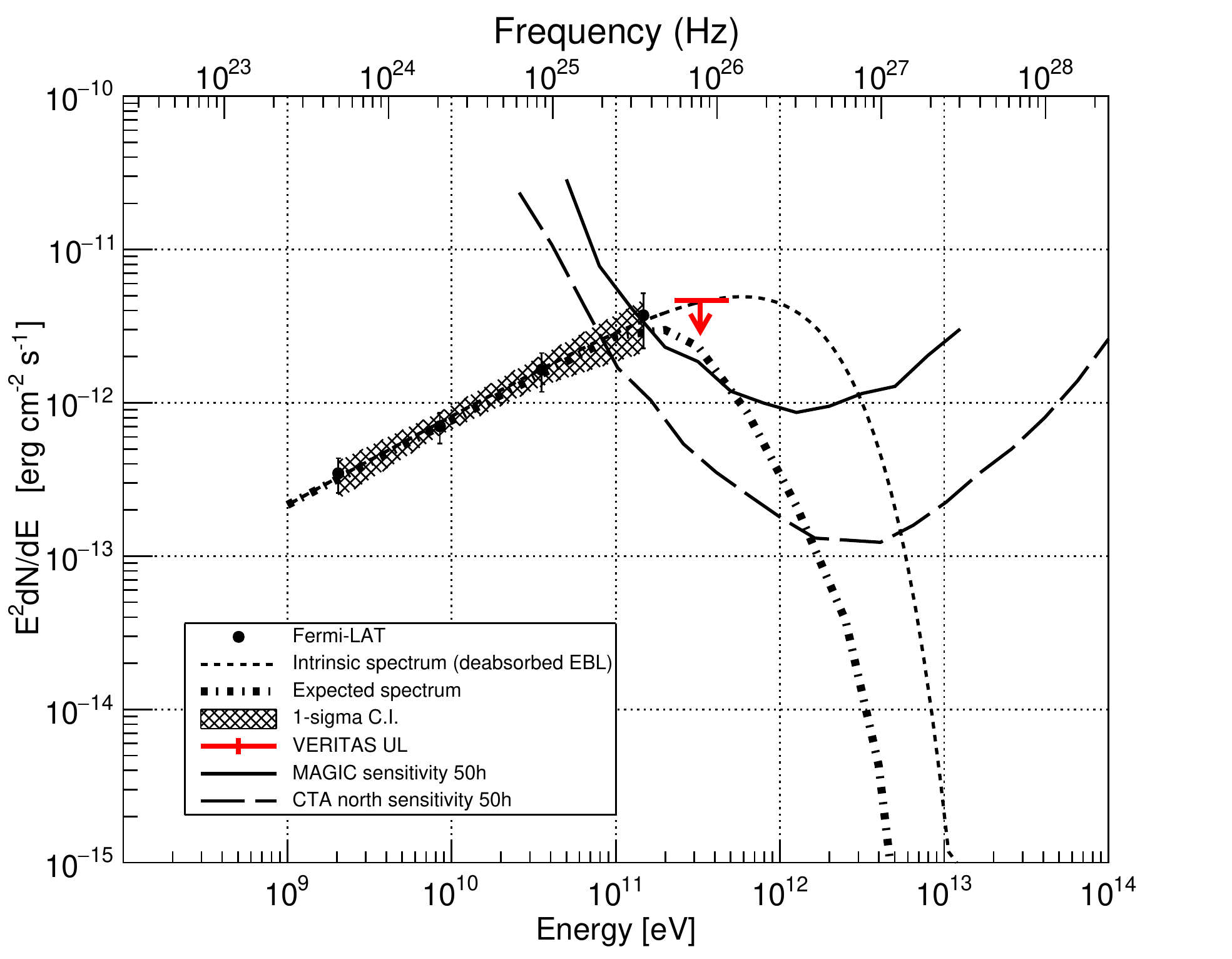}\label{fig:BZBJ1417+2543-PWL-extrapolation} \includegraphics[width=0.52\textwidth]{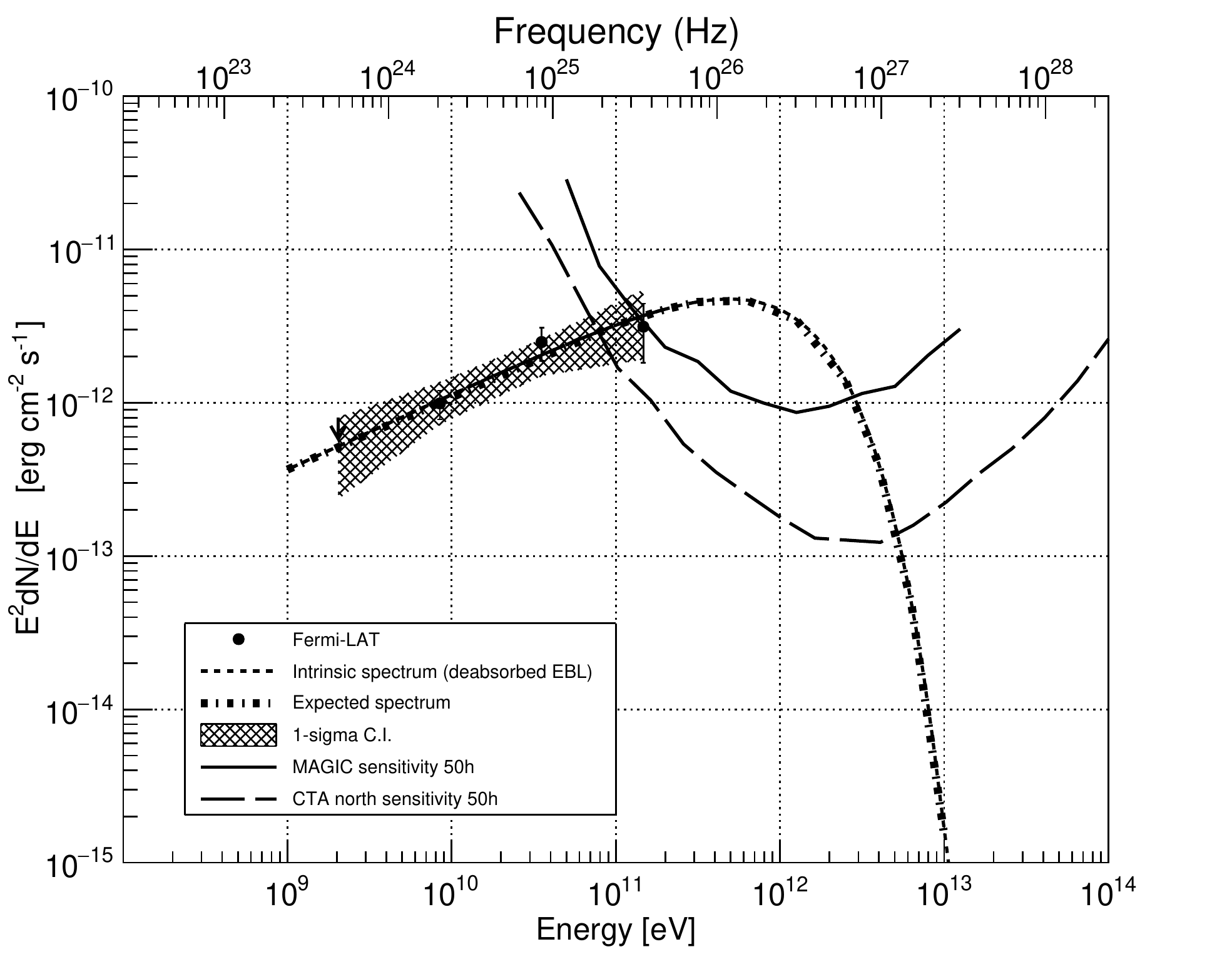}\label{fig:PMNJ0640-1253-PWL-extrapolation}}\\
\subfloat[][1RXS J021417.8+514457 with $z=0.049$ (left) and 1ES 1028+511 with $z=0.36$ (right).] {\includegraphics[width=0.52\textwidth]{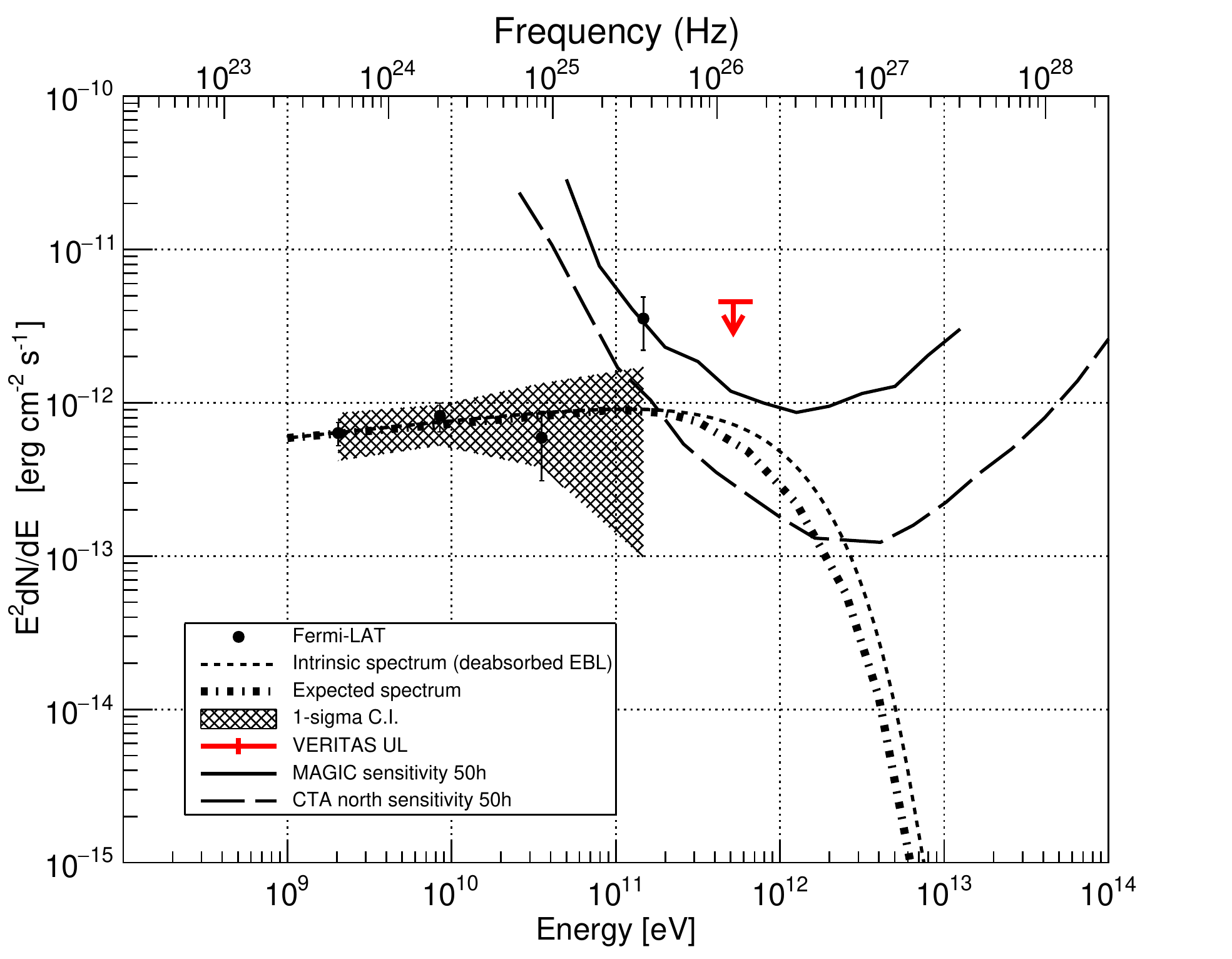}\label{fig:1RXSJ021417.8+514457-PWL-extrapolation} \includegraphics[width=0.52\textwidth]{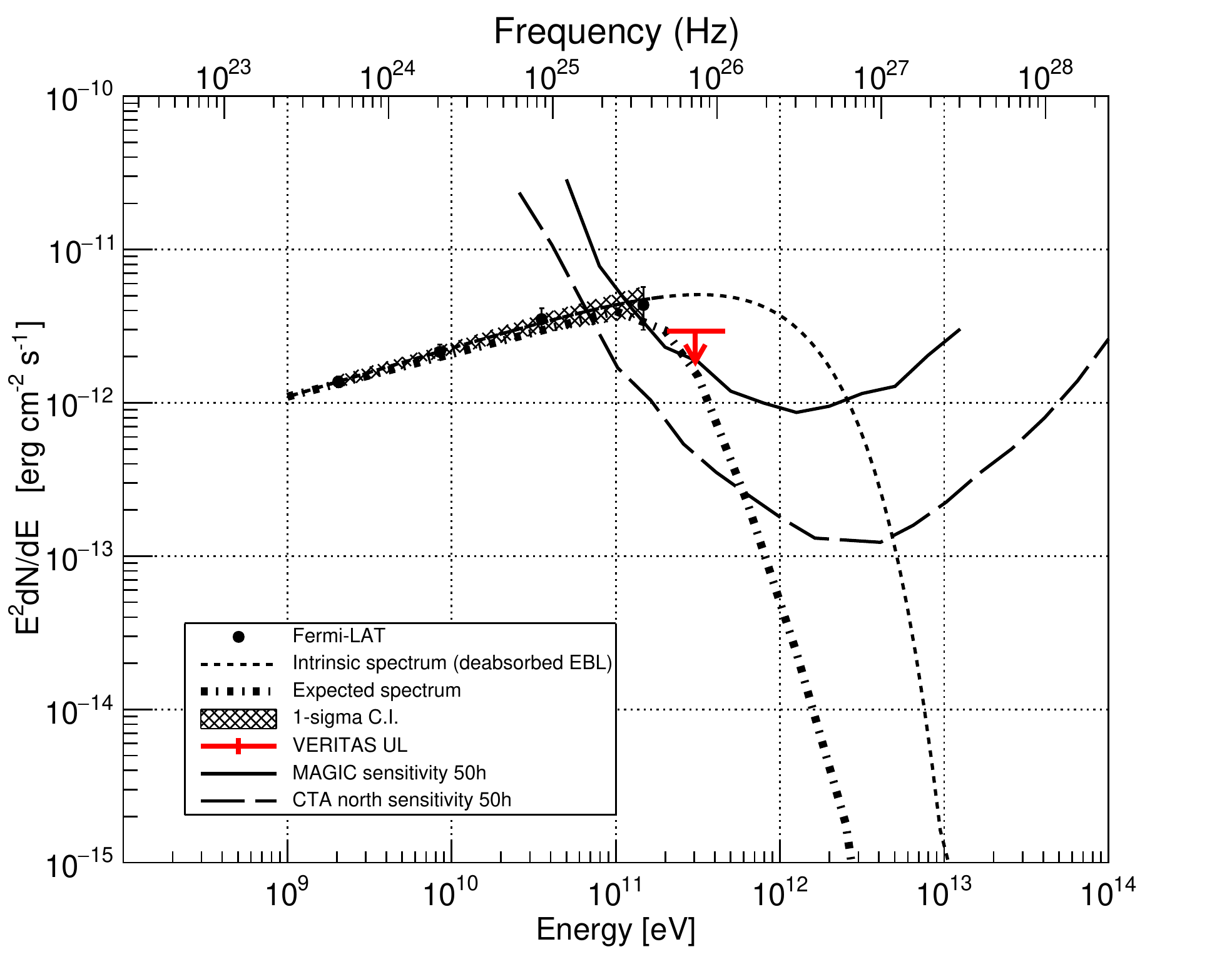}\label{fig:1ES1028+511-PWL-extrapolation}}\\

\end{figure*}

\begin{figure*}
\centering
\ContinuedFloat
\subfloat[][BZB J0244-5819 with $z=0.265$ (left) and 1RXS J032521.8-56354 with $z=0.06$ (right).] {\includegraphics[width=0.52\textwidth]{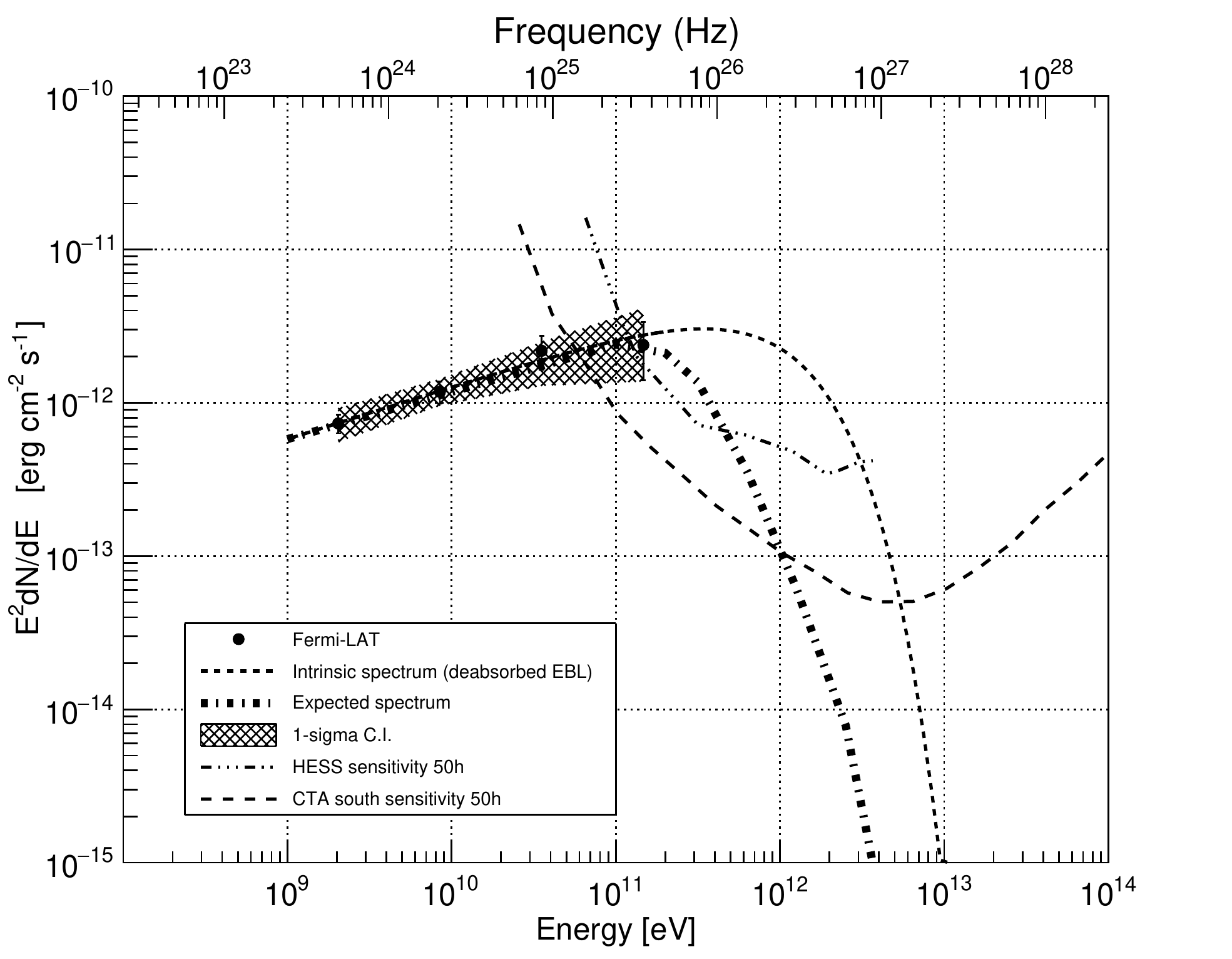}\label{fig:BZBJ0244-5819-PWL-extrapolation} \includegraphics[width=0.52\textwidth]{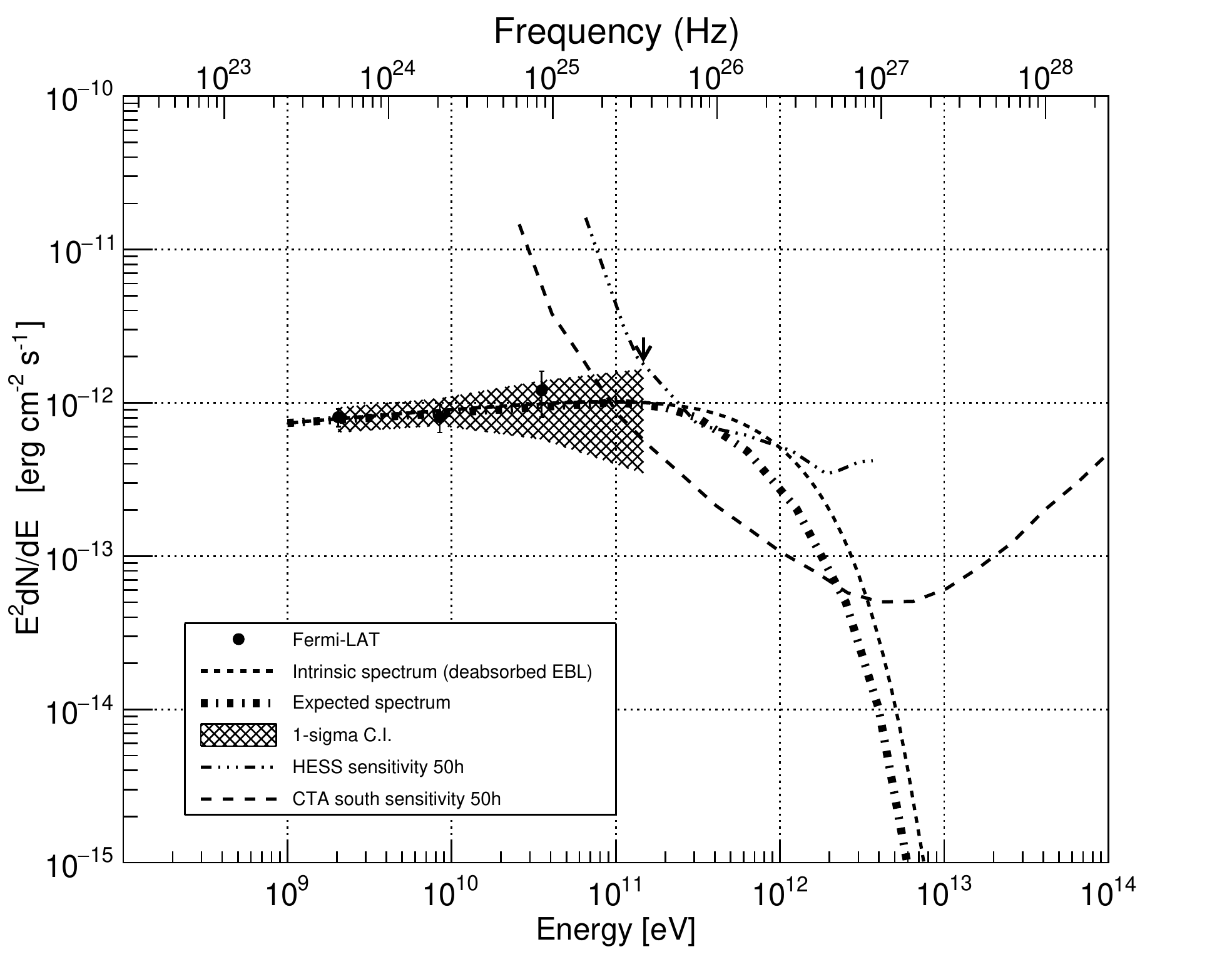}\label{fig:2MASXJ03252346-5635443-PWL-extrapolation}}\\

\addtocounter{figure}{2}
\caption{Power-law extrapolation above 1 GeV of  the \emph{Fermi}-LAT data after ten years of operation (not EBL de-absorbed) of all 14 TeV gamma-ray undetected sources in our final sample in \Cref{tab:tevundetected}. The thicker dashed line is the power-law extrapolation absorbed for EBL using the model by \citet{Franceschini17}. CTA, MAGIC \citep{MAGICsens}, and H.E.S.S. \citep{HESSsensitivityNEW} sensitivities for 50h of observations are also reported in the plots. The available upper-limits on the already observed sources by the VERITAS telescopes \citep{Veritas-2masx} and by the H.E.S.S. telescopes \citep{hess-ul} are reported. }
\label{fig:PWL-extrapolation}
\end{figure*}

\end{document}

%% file: table_big.tex
\newpage

\thispagestyle{empty}
\newgeometry{margin=01.5cm}

\begin{landscape}
\begin{table}
\renewcommand{\arraystretch}{1.37} 
 \centering  
 \vspace*{-15pt}
\scalebox{0.98}{
\hspace*{-15pt}
\begin{tabular}{ccccccccccccccccc}
\hline
\hline
\# &	\emph{Swift}-BAT name & Counterpart	&	RA	&	DEC	&	Redshift	&	 \emph{Swift}-BAT flux & $\Delta\,\text{flux}_\text{BAT}$ & \emph{Fermi}& \emph{Fermi} flux&\emph{Fermi}  &\emph{Fermi}  &Log $\nu_{\text{peak}}^{\text{sync}}$ &TeV  	\\
 & &	&	(deg) 	&	 (deg)	& & $(10^{-12}$ erg/s/cm$^2)$	& 	&  variab. &$(10^{-10}$ ph/s/cm$^2)$ & spectral index  & TS & (Hz)  &detect.\\
\hline
\hline
1&SWIFT J0232.8+2020&1ES 0229+200&38.188&20.29&0.140&23.46$^{+2.62}_{-2.27}$&4.9&130&2.7 $^+_-$ 0.5&-1.74 $^+_-$ 0.13&49&18.5 $^+_-$ 0.2&D\\
2&SWIFT J2251.8-3210&1RXS J225146.9-320614&342.944&-32.096&0.246&13.98$^{+2.47}_{-1.92}$  (0.6) &4.4&47&1.1 $^+_-$ 0.4 (0.4) &-1.55 $^+_-$ 0.26 (0.7) &37&18.3 $^+_-$ 0.3&N\\
3&SWIFT J0733.9+5156&3FGL J0733.5+5153 &113.404&51.931&&8.17$^{+2.27}_{-2.17}$  (0.3) &4.4&107&2 $^+_-$ 0.4 (0.7) &-1.69 $^+_-$ 0.13 (0.3) &41&18.3 $^+_-$ 0.2&N \\
4&SWIFT J0244.8-5829&BZB J0244-5819&41.188&-58.299&0.265&10.13$^{+2.46}_{-1.13}$  (0.4) &3.6&429&5.1 $^+_-$ 0.5 (1.9) &-1.65 $^+_-$ 0.08 (0.6) &37&18.2 $^+_-$ 0.3&N\\
5&SWIFT J1136.7+6738&RX J1136.5+6737 &174.104&67.645&0.134&12.73$^{+2.24}_{-1.53}$  (0.5) &3.8&471&4.4 $^+_-$ 0.4 (1.6) &-1.67 $^+_-$ 0.07 (0.5) &51&18.2 $^+_-$ 0.6&D\\
6&SWIFT J0709.3-1527&PKS 0706-15&107.329&-15.437&&7.37$^{+1.99}_{-1.63}$  (0.3) &3.6&82&3.3 $^+_-$ 0.8 (1.2) &-1.77 $^+_-$ 0.15 (0.2) &39&18.0 $^+_-$ 0.2&N\\
7&SWIFT J0156.5-5303&RBS 259&29.13&-53.036&&7.32$^{+1.56}_{-2.46}$  (0.3) &4&361&5.5 $^+_-$ 0.5 (2) &-1.89 $^+_-$ 0.08 (1.0) &56&18.0 $^+_-$ 0.2&N\\
8&SWIFT J1428.7+4234&1ES 1426+428&217.149&42.655&0.129&20.85$^{+1.5}_{-1.04}$  (0.9) &2.5&731&6.1 $^+_-$ 0.5 (2.3) &-1.55 $^+_-$ 0.06 (1.3) &59&18.0 $^+_-$ 0.2&D\\
9&SWIFT J0353.4-6830&PKS 0352-686&58.28&-68.532&0.087&12.24$^{+1.67}_{-1.44}$  (0.5) &3.1&165&2.3 $^+_-$ 0.4 (0.9) &-1.46 $^+_-$ 0.11 (1.6) &42&18.0 $^+_-$ 0.2&O\\
10&SWIFT J0710.3+5908&3FGL J0710.3+5908 &107.635&59.14&0.125&24.06$^{+2.77}_{-2.3}$  (1.0) &5.1&445&4.4 $^+_-$ 0.4 (1.6) &-1.57 $^+_-$ 0.07 (1.2) &78&17.8 $^+_-$ 0.2&D\\
11&SWIFT J0036.0+5951&1ES 0033+595&8.989&59.841&0.086&25.96$^{+1.13}_{-1.57}$  (1.1) &2.7&3138&32.2 $^+_-$ 1.2 (11.9) &-1.7 $^+_-$ 0.03 (0.3) &70&17.9 $^+_-$ 0.2&D\\
12&SWIFT J0550.7-3212A&PKS 0548-322&87.689&-32.273&0.069&18.21$^{+6.16}_{-5.29}$  (0.8) &11.5&157&2.8 $^+_-$ 0.5 (1) &-1.73 $^+_-$ 0.12 (0.1) &48&17.8 $^+_-$ 0.2&D\\
13&SWIFT J0122.9+3420&1ES 0120+340&20.776&34.371&0.272&11.27$^{+2.07}_{-1.62}$  (0.5) &3.7&132&1.9 $^+_-$ 0.4 (0.7) &-1.49 $^+_-$ 0.15 (1.3) & 128&17.7 $^+_-$ 0.2&O\\
14&SWIFT J1654.0+3946&Mrk 501&253.472&39.76&0.034&71.58$^{+2.23}_{-2.29}$  (3.1) &4.5&24696&107.8 $^+_-$ 1.7 (39.9) &-1.76 $^+_-$ 0.01 (0.2) &101&17.7 $^+_-$ 0.2&D\\
15&SWIFT J2346.8+5143&1ES 2344+514&356.764&51.692&0.044&10$^{+2.13}_{-1.51}$  (0.4) &3.6&2986&28.9 $^+_-$ 1.1 (10.7) &-1.88 $^+_-$ 0.03 (1.0) &2935&17.7 $^+_-$ 0.4&D\\
16&SWIFT J1417.7+2539&BZB J1417+2543&214.449&25.72&0.237&5.93$^{+1.77}_{-1.73}$  (0.3) &3.5&211&2.8 $^+_-$ 0.4 (1) &-1.52 $^+_-$ 0.1 (1.3) &43&17.7 $^+_-$ 0.2&O\\
17&SWIFT J0640.3-1286&TXS 0637-128 &100.07&-12.866&&7.87$^{+2.51}_{-2.08}$  (0.3) &4.6&186&3 $^+_-$ 0.6 (1.1) &-1.44 $^+_-$ 0.1 (1.8) &51&17.7 $^+_-$ 0.2&N\\
18&SWIFT J2246.7-5208&RBS 1895&341.661&-52.12&0.194&6.97$^{+1.76}_{-1.57}$  (0.3) &3.3&81&1.8 $^+_-$ 0.4 (0.7) &-1.67 $^+_-$ 0.15 (0.4) &37&17.6 $^+_-$ 0.2&N\\
19&SWIFT J0213.7+5147&1RXS J021417.8+514457&33.553&51.772&0.049&13.65$^{+1.67}_{-1.81}$  (0.6) &3.5&157&4.1 $^+_-$ 0.6 (1.5) &-1.88 $^+_-$ 0.1 (0.9) &41&17.6 $^+_-$ 0.2&O\\
20&SWIFT J2359.0-3038&H 2356-309&359.777&-30.579&0.165&14.88$^{+1.97}_{-1.85}$  (0.6) &3.8&292&4.6 $^+_-$ 0.5 (1.7) &-1.8 $^+_-$ 0.09 (0.4) &47&17.6 $^+_-$ 0.2&D\\
21&SWIFT J1031.5+5051&1ES 1028+511&157.854&50.903&0.360&7.85$^{+1.68}_{-1.31}$  (0.3) &3&1130&9.4 $^+_-$ 0.6 (3.5) &-1.72 $^+_-$ 0.06 (0.1) &131&17.5 $^+_-$ 0.2&O\\
22&SWIFT J1221.3+3012&1ES 1218+304&185.343&30.16&0.184&10.62$^{+1.29}_{-1.68}$  (0.5) &3&3971&33.3 $^+_-$ 1.2 (12.3) &-1.68 $^+_-$ 0.03 (0.4) &45&17.9 $^+_-$ 0.2&D\\
23&SWIFT J0349.2-1159&1ES 0347-121&57.368&-11.983&0.180&15.68$^{+3.31}_{-2.25}$  (0.7) &5.6&190&2.9 $^+_-$ 0.4 (1.1) &-1.64 $^+_-$ 0.1 (0.6) &39&17.7 $^+_-$ 0.4&D\\
24&SWIFT J1103.5-2329&1ES 1101-232&165.868&-23.471&0.186&10.8$^{+1.66}_{-3.01}$  (0.5) &4.7&241&4 $^+_-$ 0.5 (1.5) &-1.6 $^+_-$ 0.09 (0.9) &92&17.7 $^+_-$ 0.2&D\\
25&SWIFT J0507.7+6732&1ES 0502+675 &76.92&67.533&0.314&9.03$^{+2.18}_{-2.28}$  (0.4) &4.5&2292&15.4 $^+_-$ 0.3 (5.7) &-1.47 $^+_-$ 0.01 (2.1) &251&17.5 $^+_-$ 0.2&D\\
26&SWIFT J0930.1+4987&1ES 0927+500 &142.523&49.878&0.187&7.44$^{+2.22}_{-1.71}$  (0.3) &3.9&50&1.3 $^+_-$ 0.4 (0.5) &-1.61 $^+_-$ 0.2 (0.5) &44&17.4 $^+_-$ 0.4&O\\
27&SWIFT J1959.6+6507&1ES 1959+650&299.973&65.158&0.047&29.03$^{+1.72}_{-1.83}$  (1.2) &3.6&17681&95.2 $^+_-$ 1.7 (35.3) &-1.84 $^+_-$ 0.02 (0.8) &158&17.4 $^+_-$ 0.2&D\\
28&SWIFT J0326.0-5633&1RXS J032521.8-56354&51.47&-56.526&0.060&8.49$^{+2.35}_{-2.21}$  (0.4) &4.6&251&4.9 $^+_-$ 0.5 (1.8) &-1.97 $^+_-$ 0.1 (1.4) &755&17.4 $^+_-$ 0.2&N\\
29&SWIFT J0136.5+3906&B3 0133+388&24.126&39.05&&7.91$^{+2.28}_{-1.84}$  (0.3) &4.1&6511&42 $^+_-$ 1.2 (15.6) &-1.72 $^+_-$ 0.02 (0.2) &44&17.4 $^+_-$ 0.2&D\\
30&SWIFT J2009.6-4851&PKS 2005-489&302.477&-48.866&0.071&6.03$^{+2.61}_{-3.07}$  (0.3) &5.7&2943&28.7 $^+_-$ 1.1 (10.6) &-1.84 $^+_-$ 0.03 (0.7) &55&17.0 $^+_-$ 0.9&D\\
31&SWIFT J2056.8+4939&RX J2056.6+4940&314.202&49.665&&13.01$^{+1.7}_{-2.02}$  (0.6) &3.7&610&16.3 $^+_-$ 0.9 (6) &-1.81 $^+_-$ 0.05 (0.5) &178&17.0 $^+_-$ 0.9&D\\
32&SWIFT J1104.4+3812&Mrk 421&166.103&38.214&0.033&141$^{+1.11}_{-2.01}$  (6.0) &3.1&101631&344.7 $^+_-$ 3.1 (127.7) &-1.77 $^+_-$ 0.01 (0.2) &62&17.0 $^+_-$ 0.3&D\\

\hline
\hline
\end{tabular}
 }
\caption{List of all the 32 sources of our final sample selected in the \emph{Swift}-BAT 105-months catalog \citep{BATcatalog105} (see \Cref{sec:2} and \Cref{appendix:selection} for details). All of them are detected in the \emph{Fermi}-LAT 3LAC catalog \citep{3LACcatalog}, except for sources numbers 2 and 30 that have been selected using the 2WHSP catalog \citep{2whsp}. In this table we report from the \emph{Swift}-BAT catalog the name of each source, the counterpart, the galactic coordinates (in degrees), and the redshift (when known). Then we report the \emph{Swift}-BAT flux and the $\Delta \,\text{flux}_{\text{BAT}}$, that is a measure of variability computed as difference between the lowest flux and highest flux in the \emph{Swift}-BAT catalog. After that, we present the updated ten-years \emph{Fermi}-LAT TS, the 1-300 GeV flux, the spectral index and its compatibility with the one of 1ES~0229+200 (see \Cref{appendix:fermi} for details), and the \emph{Fermi}-LAT 3FGL \citep{fermi3fgl} variability index. 
In the last columns we present also the logarithm of the synchrotron peak frequency, that we estimated with a log-parabolic fit through X-ray and hard X-ray data. Finally, we indicate the TeV detection (``D'' if already detected, ``N'' if not detected yet) reported in the TeGeV \citep{tegev} and TeVCat catalogs (\url{http://tevcat.uchicago.edu}) association. If the source has been observed by the current Cherenkov telescopes but without detection, we indicate it with ``O''.}
\label{tab:sourcelist}
\end{table}
\end{landscape}

\restoregeometry
\newpage

%% file: tev_table.tex
\begin{table*}
\renewcommand{\arraystretch}{1.35} % Default value: 1
 \centering  
 \vspace*{-2cm}

% \vspace*{-5pt}
\scalebox{1.0}{
%  \resizebox{\columnwidth}{!}{%
\hspace*{-20pt}
\begin{tabular}{ccccccccccccccccc}
\hline
\hline
\# &	\emph{Swift}-BAT name & Counterpart	&	Log $\nu_{\text{peak}}^{\text{sync}}$ &HE & HE-VHE &HE-VHE& IC  &TeV  & TeV	\\
 & &	& (Hz)  & $\chi^2$ &	slope& $\chi^2$ &peak (TeV)&slope & $\chi^2$\\
\hline
\hline
1&SWIFT J0232.8+2020&1ES 0229+200&18.5 $^+_-$  0.2&1.1&0.43 $^+_-$  0.04&0.4&>10&0.49 $^+_-$  0.19&0.4\\
2&SWIFT J1136.7+6738&RX J1136.5+6737 &18.2 $^+_-$  0.6&2.4&&&&&\\
3&SWIFT J1428.7+4234&1ES 1426+428&18.0 $^+_-$  0.2&0.7&0.47 $^+_-$  0.04&0.5&>10&0.58 $^+_-$  0.15&0.4\\
4&SWIFT J1221.3+3012&1ES 1218+304&17.9 $^+_-$  0.2&1.4&0.16 $^+_-$  0.02&2.1&0.2 $^+_-$  0.3&-0.39 $^+_-$  0.22&0.6\\
5&SWIFT J0036.0+5951&1ES 0033+595&17.9 $^+_-$  0.2&0.5&&&&&\\
6&SWIFT J0710.3+5908&3FGL J0710.3+5908 &17.8 $^+_-$  0.2&1.5&0.33 $^+_-$  0.03&0.4&>10&0.22 $^+_-$  0.16&0.3\\
7&SWIFT J0550.7-3212A&PKS 0548-322&17.8 $^+_-$  0.2&2.8&0.15 $^+_-$  0.04&1.9&0.3 $^+_-$  0.2&-0.47 $^+_-$  0.23&0.0\\
8&SWIFT J2346.8+5143&1ES 2344+514&17.7 $^+_-$  0.4&1.4&0.03 $^+_-$  0.01&6.7&0.3 $^+_-$  0.3&-0.41 $^+_-$  0.03&1.8\\
9&SWIFT J1654.0+3946&Mrk 501&17.7 $^+_-$  0.2&0.9&0.13 $^+_-$  0.01&13.9&0.2 $^+_-$  0.1&-0.52 $^+_-$  0.11&0.1\\
10&SWIFT J0349.2-1159&1ES 0347-121&17.7 $^+_-$  0.4&0.1&0.42 $^+_-$  0.04&0.6&>10&0.32 $^+_-$  0.16&0.6\\
11&SWIFT J1103.5-2329&1ES 1101-232&17.7 $^+_-$  0.2&1.0&0.40 $^+_-$  0.03&0.5&>10&0.34 $^+_-$  0.17&0.5\\
12&SWIFT J2359.0-3038&H 2356-309&17.6 $^+_-$  0.2&0.5&0.20 $^+_-$  0.04&0.3&>1&-0.14 $^+_-$  0.32&0.1\\
13&SWIFT J0507.7+6732&1ES 0502+675 &17.5 $^+_-$  0.2&0.3&&&&&\\
14&SWIFT J1959.6+6507&1ES 1959+650&17.4 $^+_-$  0.2&1.4&0.03 $^+_-$  0.01&5.0&0.2 $^+_-$  0.2&-0.31 $^+_-$  0.06&0.8\\
15&SWIFT J0136.5+3906&B 30133+388&17.4 $^+_-$  0.2&61.7&&&&&\\
16&SWIFT J2009.6-4851&PKS 2005-489&17.0 $^+_-$  0.9&0.4&-0.11 $^+_-$  0.01&3.8&0.1 $^+_-$  0.1&-0.82 $^+_-$  0.12&0.9\\
17&SWIFT J2056.8+4939&RX J2056.6+4940&17.0 $^+_-$  0.9&6.0&&&&&\\
18&SWIFT J1104.4+3812&Mrk 421&17.0 $^+_-$  0.3&8.1&&173.1&0.7 $^+_-$  0.1&-0.35 $^+_-$  0.01&85.3\\
\hline
\hline
\end{tabular}
 }
\caption{List of all the 18 TeV gamma-ray detected sources of our final sample. We report here for convenience the synchrotron peak frequencies of \Cref{tab:sourcelist}. In the \texttt{HE $\chi^2$} column, we present the \emph{Fermi}-LAT curvature test estimated with a reduced $\chi^2$ on the ten-years \emph{Fermi}-LAT data. Then we show the \texttt{HE-VHE slope} and \texttt{HE-VHE} reduced-$\chi^2$ test in the 100 MeV - 10 TeV range, and the \texttt{TeV slope} and \texttt{TeV} $\chi^2$ test in the 100 GeV - 10 TeV range. The \texttt{TeV slope} $\Gamma$ is related to the power-law spectral index $S$ in the same energy band by  $S = 2.0 - \Gamma$.}
%\captionsetup{width=10cm, margin=0cm}
\label{tab:tevdetected}
\end{table*}

%% file: tevundetected_table.tex
\begin{table*}
\renewcommand{\arraystretch}{1.35}
 \centering  
\scalebox{1.0}{
\hspace*{-20pt}
\begin{tabular}{ccccccccccccccccc}
\hline
\hline
\# &	\emph{Swift}-BAT name & Counterpart	&	Log $\nu_{\text{peak}}^{\text{sync}}$ &HE & \multicolumn{3}{c}{Cherenkov telescopes observations}	\\
 & &	&	(Hz)  & $\chi^2$ & obs time (h)& signif.& UL  ($10^{-12}$ erg/s/cm$^2$)\\
\hline
\hline
1&SWIFT J2251.8-3210&1RXS J225146.9-320614&18.3 $^+_-$ 0.3&0.2&&&\\
2&SWIFT J0733.9+5156&3FGL J0733.5+5153 &18.3 $^+_-$ 0.2&0.1&&&\\
3&SWIFT J0244.8-5829&BZB J0244-5819&18.2 $^+_-$ 0.3&0.2&&&\\
4&SWIFT J0709.3-1527&PKS 0706-15&18.0 $^+_-$ 0.2&0.2&&&\\
5&SWIFT J0156.5-5303&RBS 259&18.0 $^+_-$ 0.2&7.8&&&\\
6&SWIFT J0353.4-6830&PKS 0352-686&18.0 $^+_-$ 0.2&0.6&15.0&0.1 $\sigma$&1.2\text{ at }710\text{ GeV}\\
7&SWIFT J0122.9+3420&1ES 0120+340&17.7 $^+_-$ 0.2&0.6&5.9&1.5 $\sigma$&3.2\text{ at }283\text{ GeV}\\
8&SWIFT J1417.7+2539&BZB J1417+2543&17.7 $^+_-$ 0.2&1.4&10.0&1.9  $\sigma$&2.9\text{ at }327\text{ GeV}\\
9&SWIFT J0640.3-1286&TXS 0637-128 &17.7 $^+_-$ 0.2&0.1&&&\\
10&SWIFT J2246.7-5208&RBS 1895&17.6 $^+_-$ 0.2&1.0&&&\\
11&SWIFT J0213.7+5147&1RXS J021417.8+514457&17.6 $^+_-$ 0.2&2.4&5.1&0.3  $\sigma$&2.9\text{ at }336\text{ GeV}\\
12&SWIFT J1031.5+5051&1ES 1028+511&17.5 $^+_-$ 0.2&6.8&24.1&1.2  $\sigma$&1.9\text{ at }305\text{ GeV}\\
13&SWIFT J0930.1+4987&1ES 0927+500 &17.4 $^+_-$ 0.4&0.1&11.7&-0.2 $\sigma$&2.1\text{ at }346\text{ GeV}\\
14&SWIFT J0326.0-5633&1RXS J032521.8-56354&17.4 $^+_-$ 0.2&1.4&&&\\
\hline
\hline
\end{tabular}
}
\caption{List of all the 14 currently TeV gamma-ray undetected sources of our final sample. We report here for convenience the synchrotron peak frequencies of \Cref{tab:sourcelist}. In the \texttt{HE $\chi^2$} column, we show here the \emph{Fermi}-LAT curvature test estimated with a reduced-$\chi^2$ test on the ten-years \emph{Fermi}-LAT data. The available upper-limits on the already observed sources by the VERITAS telescopes \citep{Veritas-2masx} and by the H.E.S.S. telescopes \citep{hess-ul} are reported.}
\label{tab:tevundetected}
\end{table*}

%% file: mwlplots.tex
%\onecolumn

%\vspace{-30cm}

\begin{figure*}
\centering
\vspace*{-2cm}
\subfloat[][1ES 0033+595 (left) and 3FGL J0733+5153 (right).] {\includegraphics[width=0.52\textwidth]{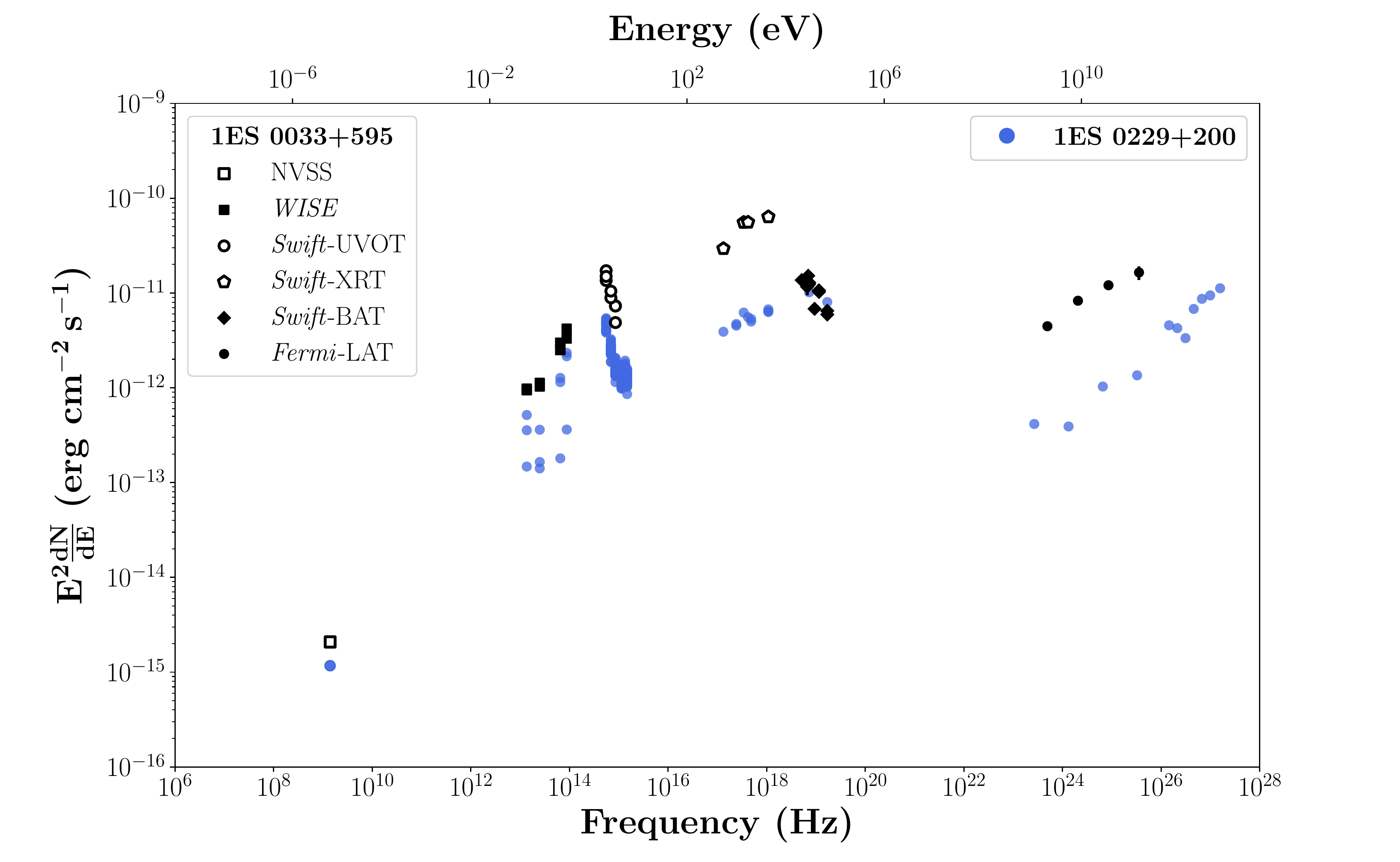} \includegraphics[width=0.52\textwidth]{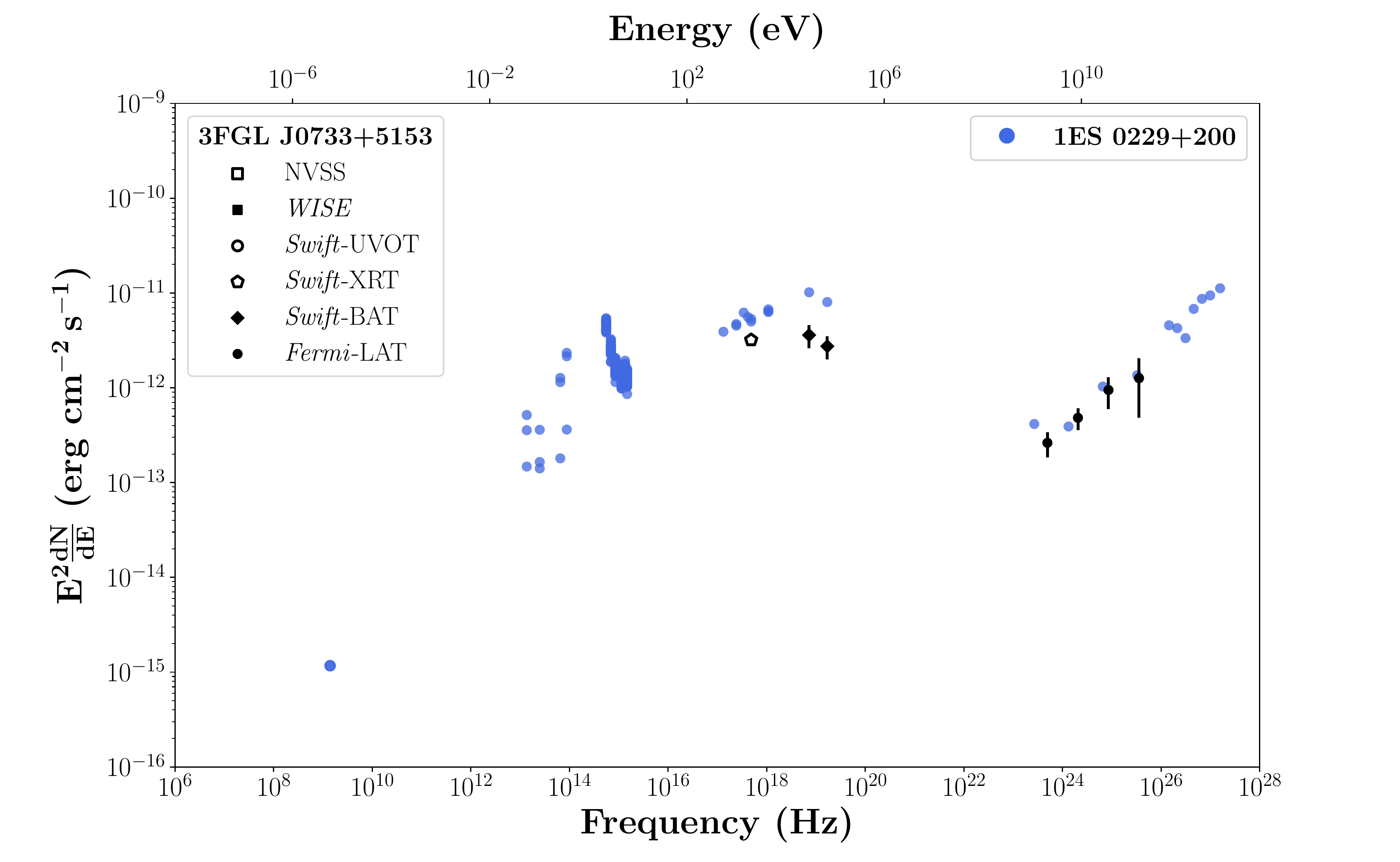}}\\
\subfloat[][1ES 0120+340 (left) and B3~0133+388 (right).] {\includegraphics[width=0.52\textwidth]{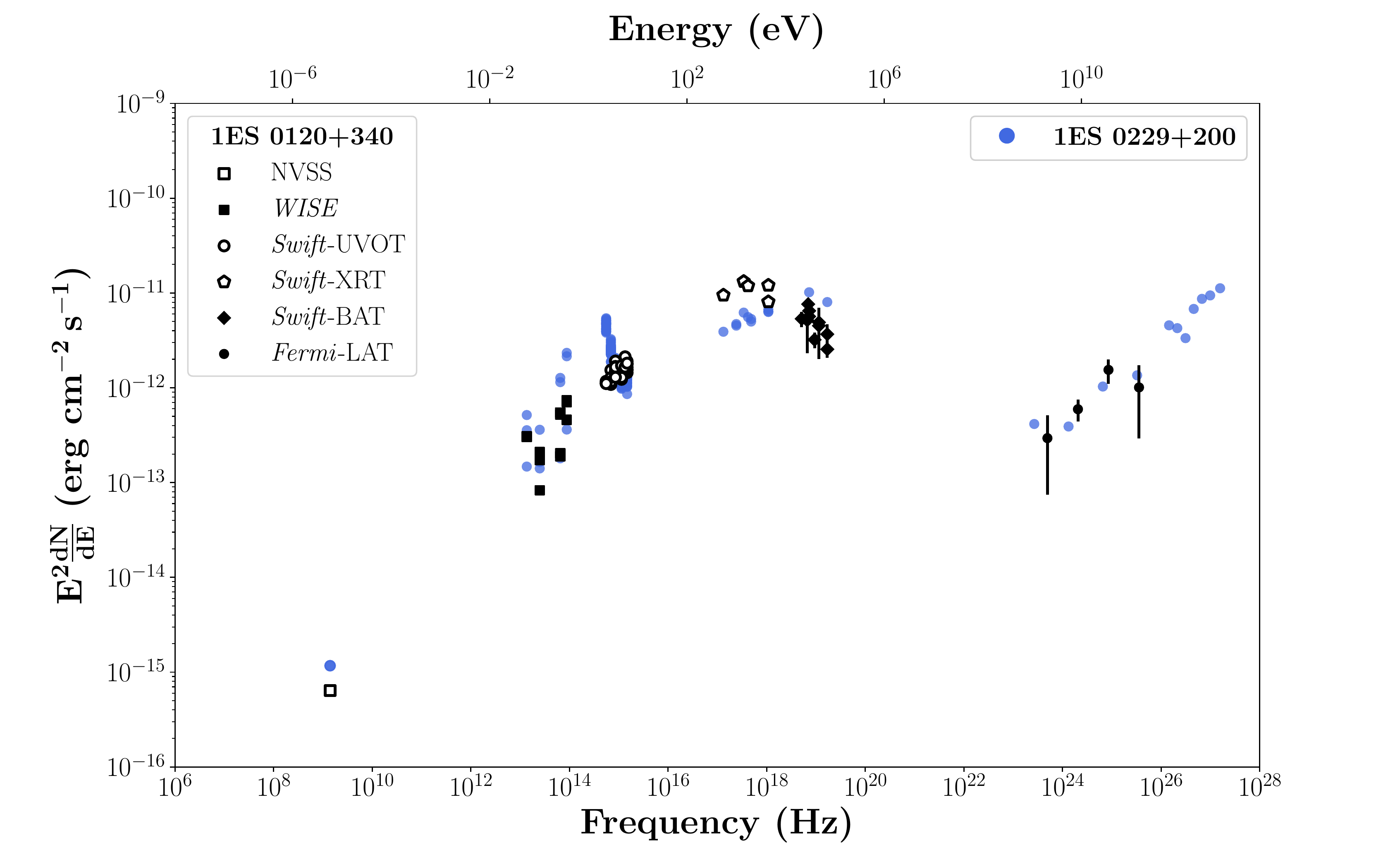} \includegraphics[width=0.52\textwidth]{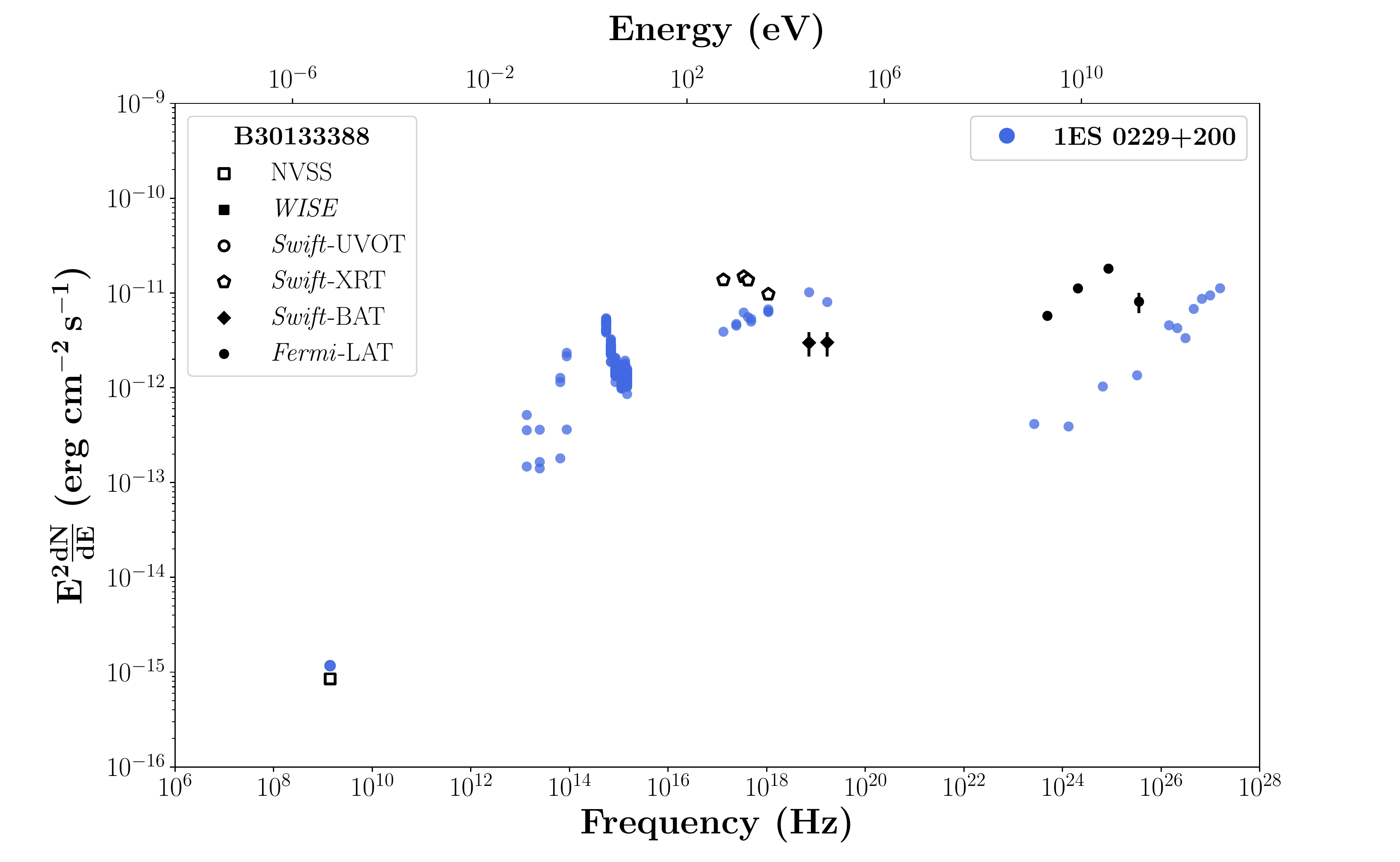}}\\
\subfloat[][1RXS J032521-56354 (left) and BZB J0244-5819 (right).]  {\includegraphics[width=0.52\textwidth]{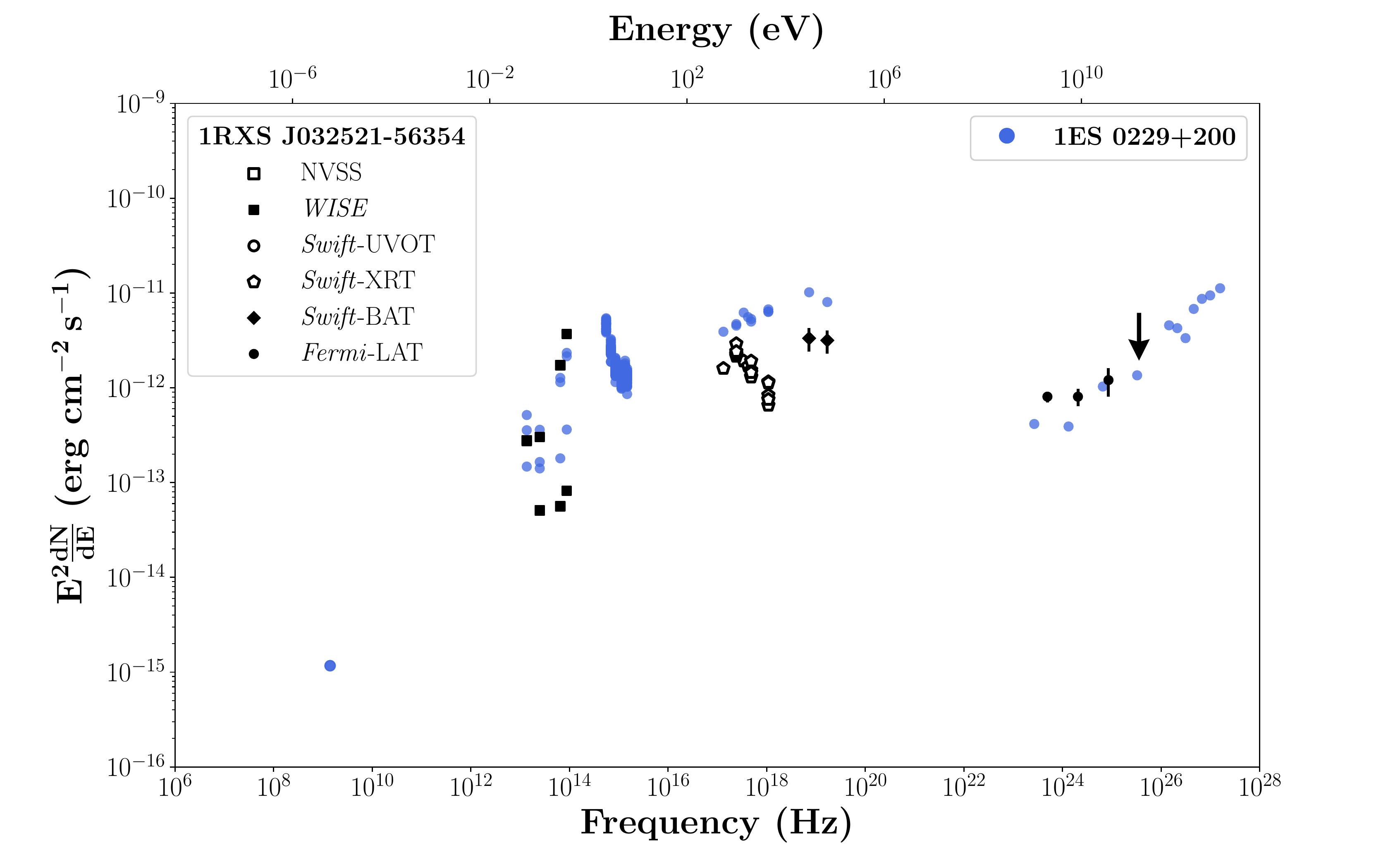} \includegraphics[width=0.52\textwidth]{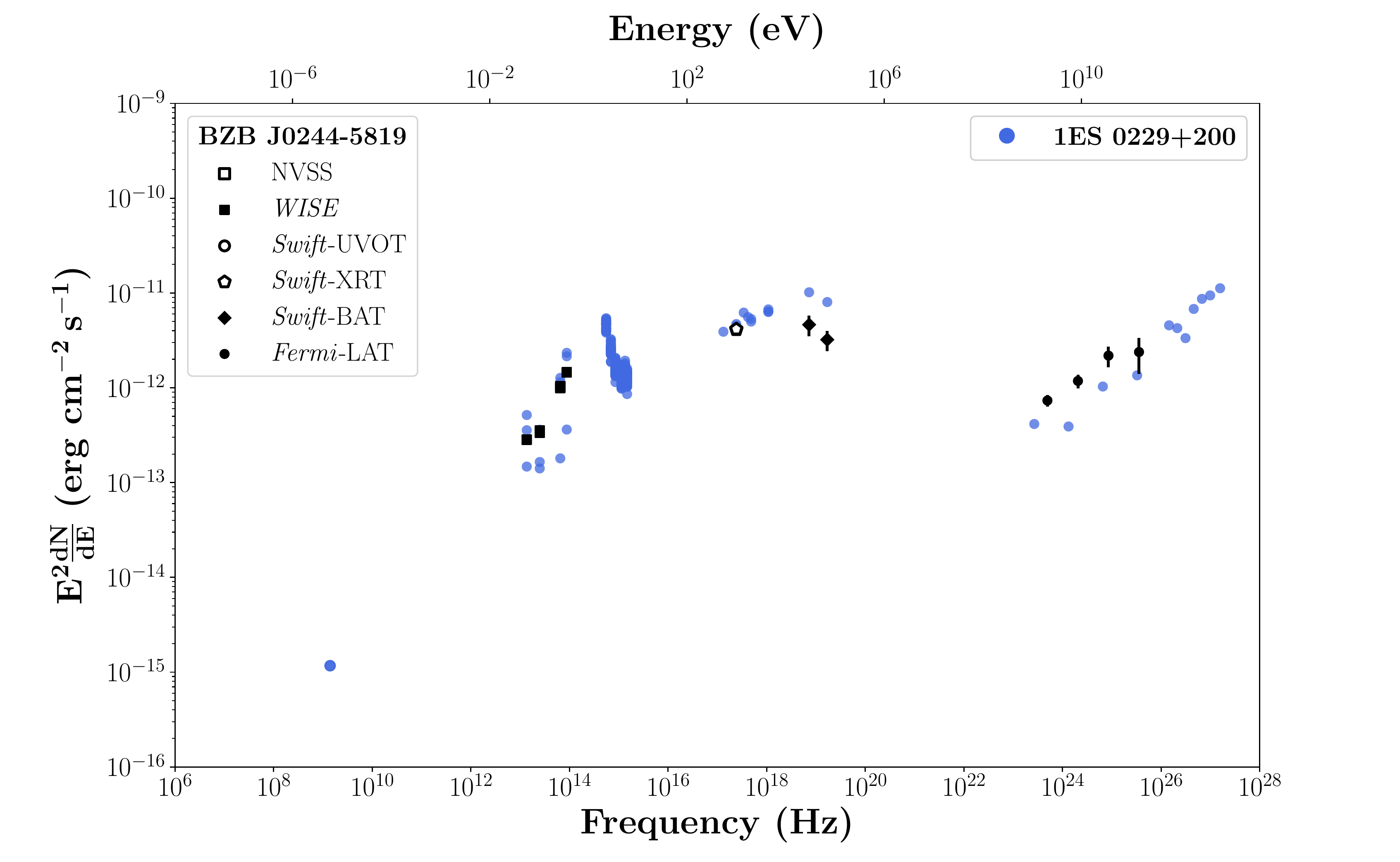}}\\
\subfloat[][BZB J1417+2543 (left) and 1ES 0347-121 (right).] {\includegraphics[width=0.52\textwidth]{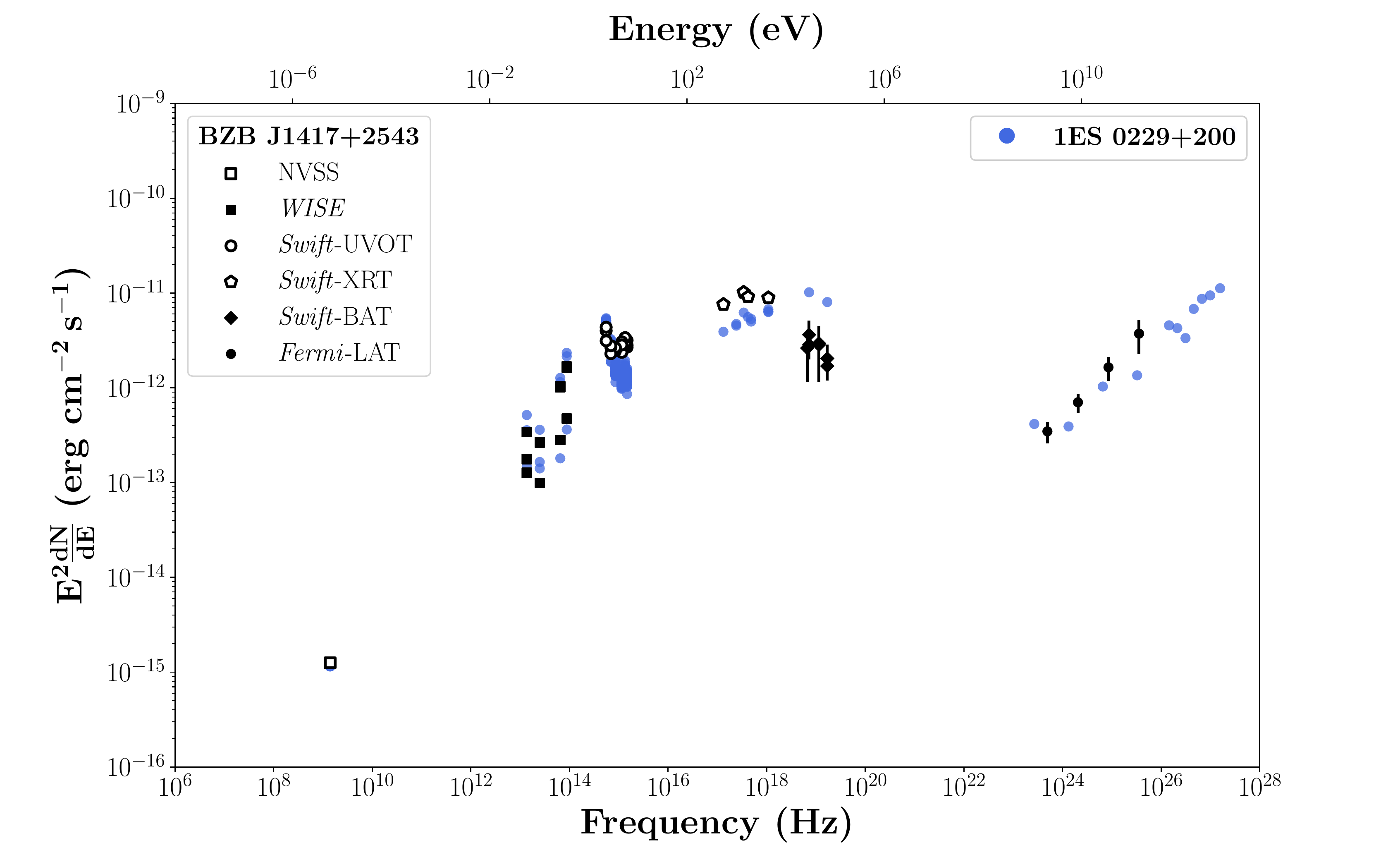} \includegraphics[width=0.52\textwidth]{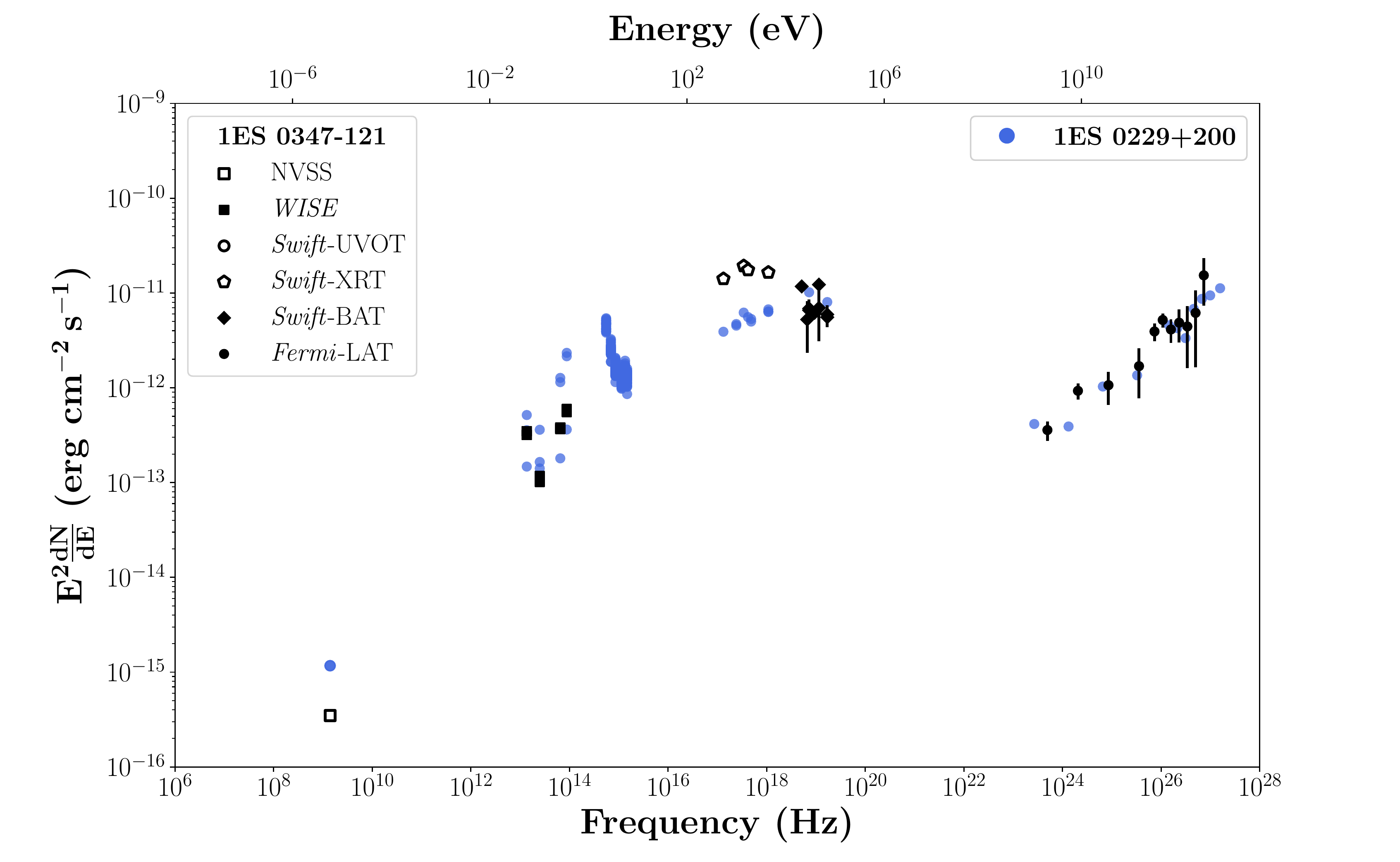}}\\
\addtocounter{figure}{0}
\caption{ }
\end{figure*}

\newpage

\begin{figure*}
\centering
\ContinuedFloat
\vspace*{-2cm}
\subfloat[][1ES 0502+675 (left) and H 2356-309 (right).] {\includegraphics[width=0.52\textwidth]{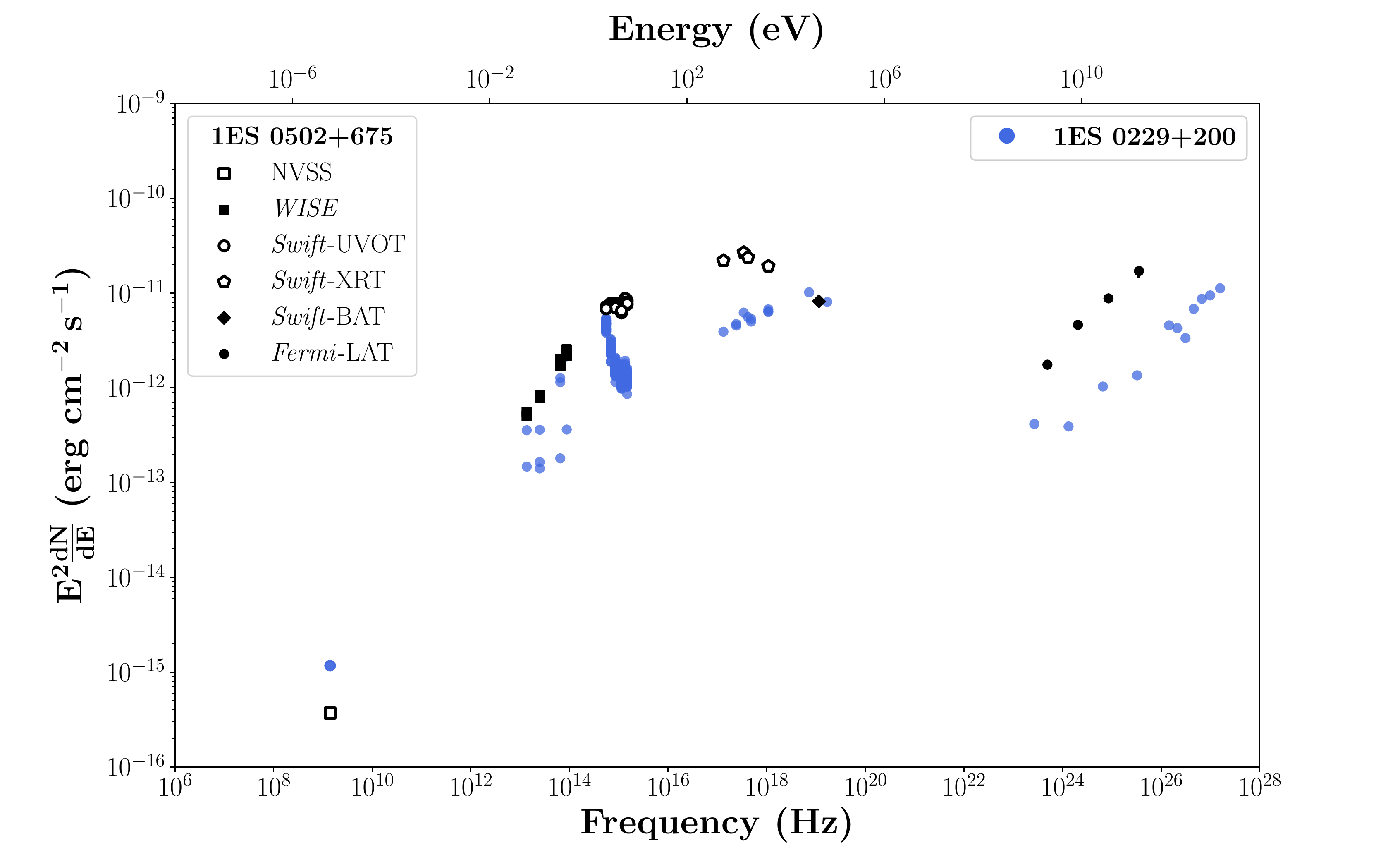} \includegraphics[width=0.52\textwidth]{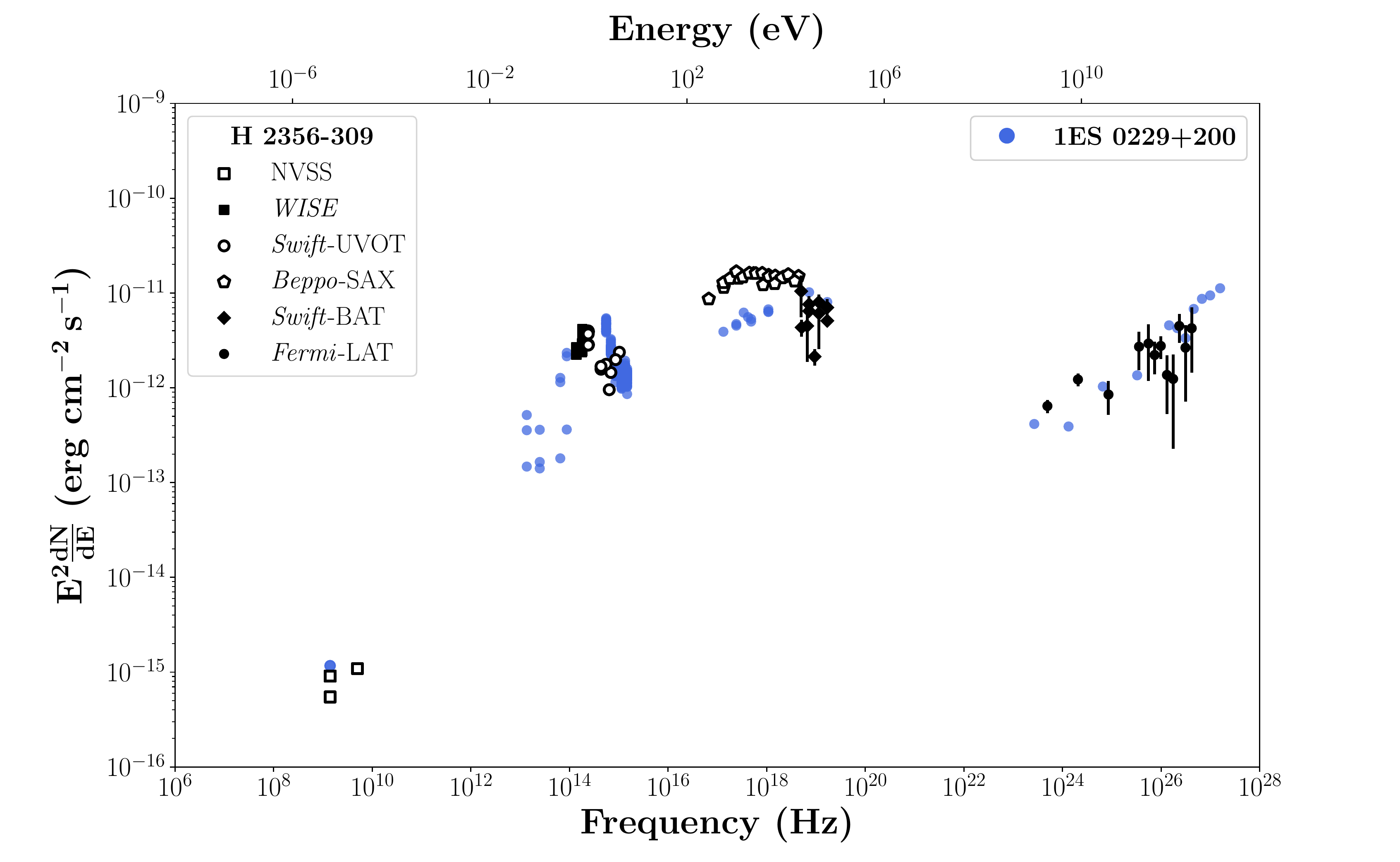}}\\
\subfloat[][1ES 0927+500 (left) and Mrk 421 (right).] {\includegraphics[width=0.52\textwidth]{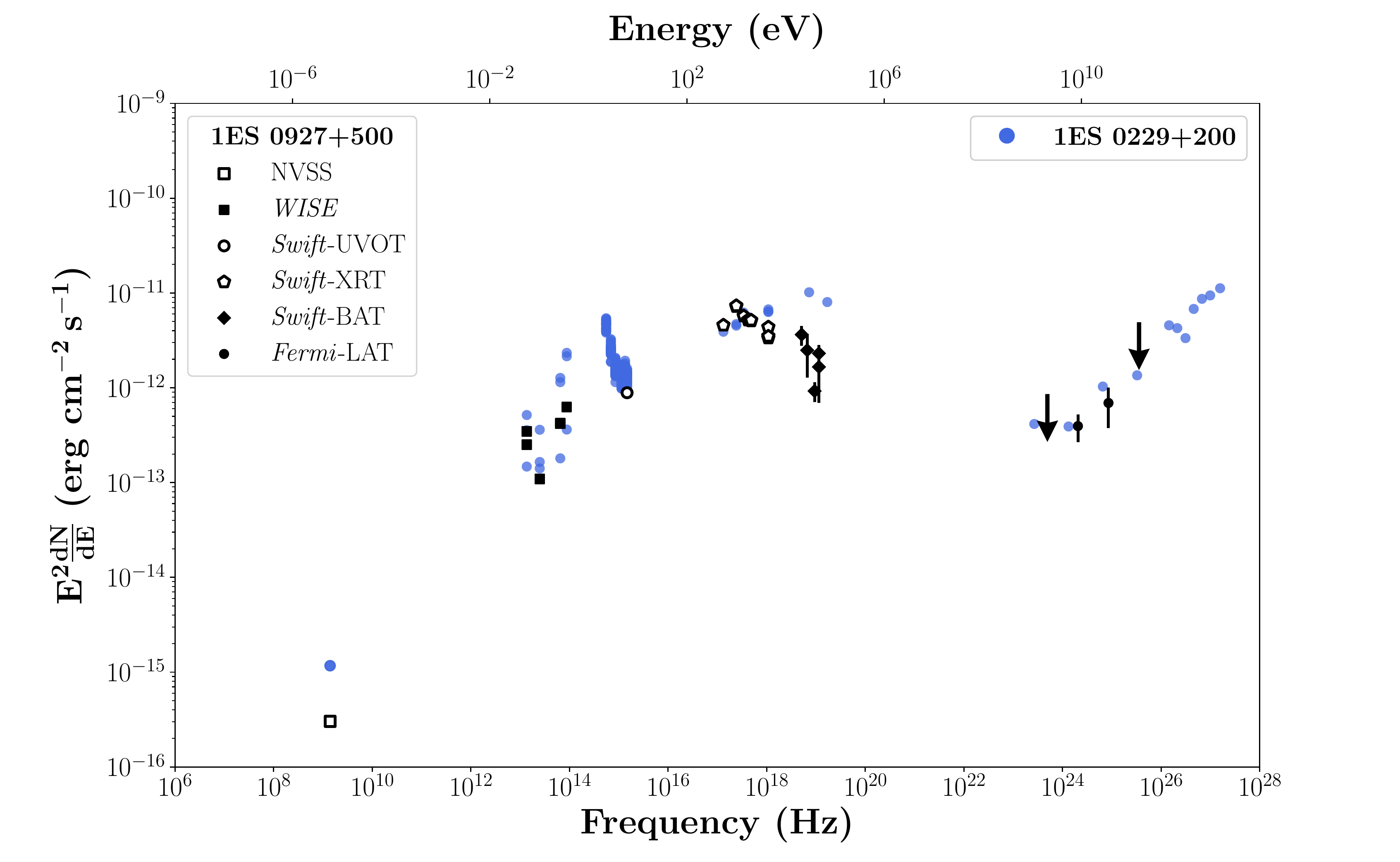} \includegraphics[width=0.52\textwidth]{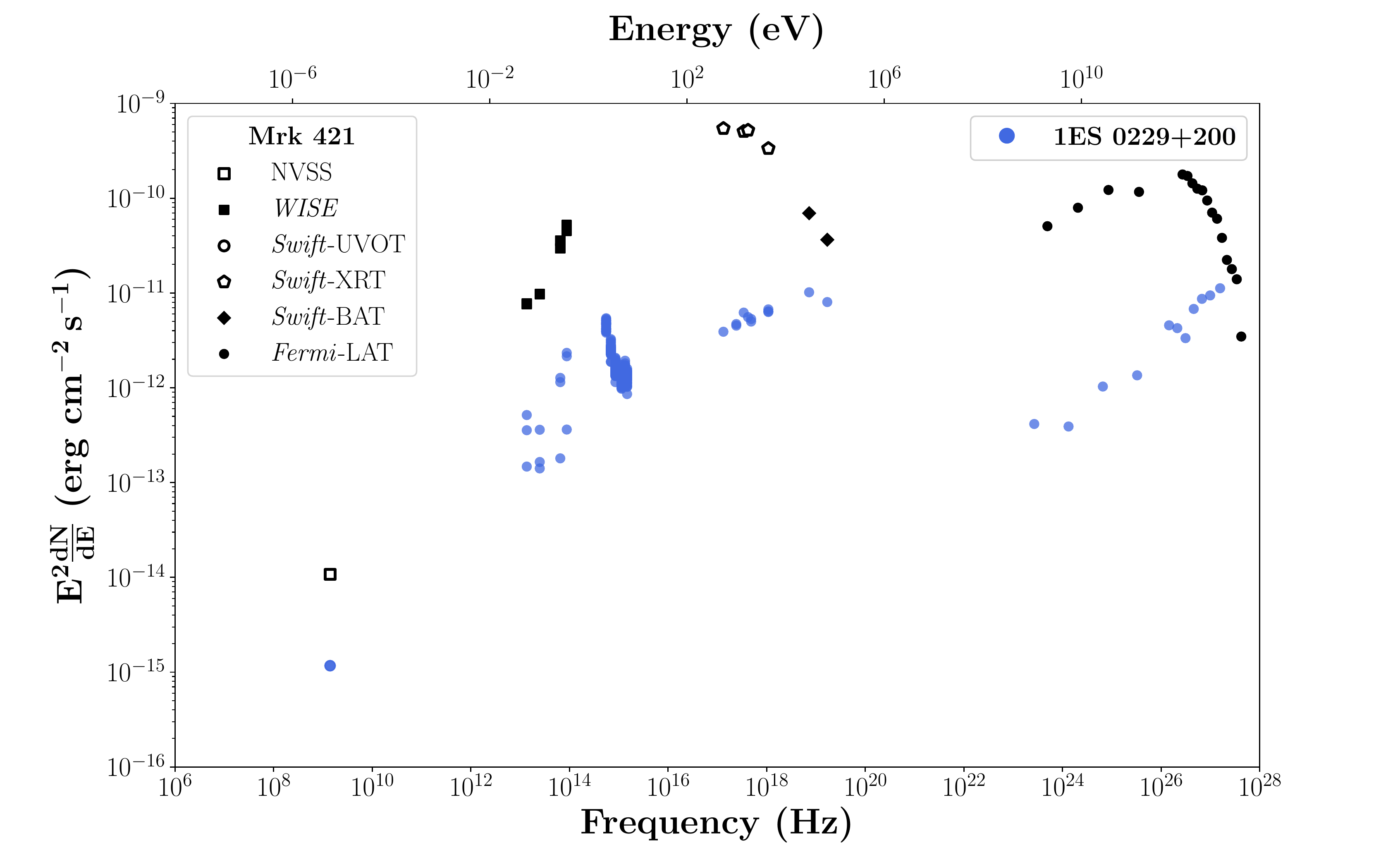}}\\
\subfloat[][1ES 1028+511 (left) and Mrk 501 (right).] {\includegraphics[width=0.52\textwidth]{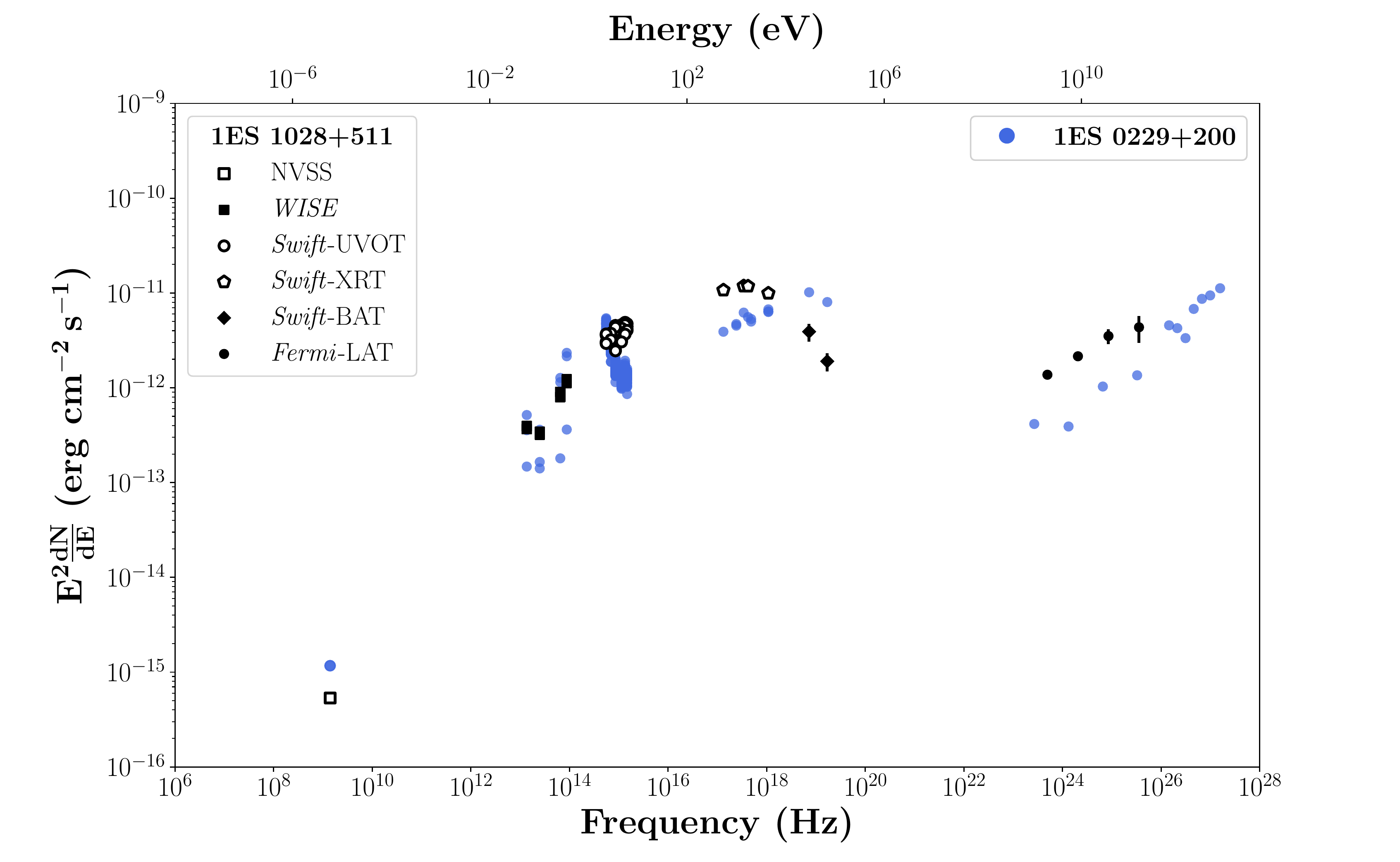} \includegraphics[width=0.52\textwidth]{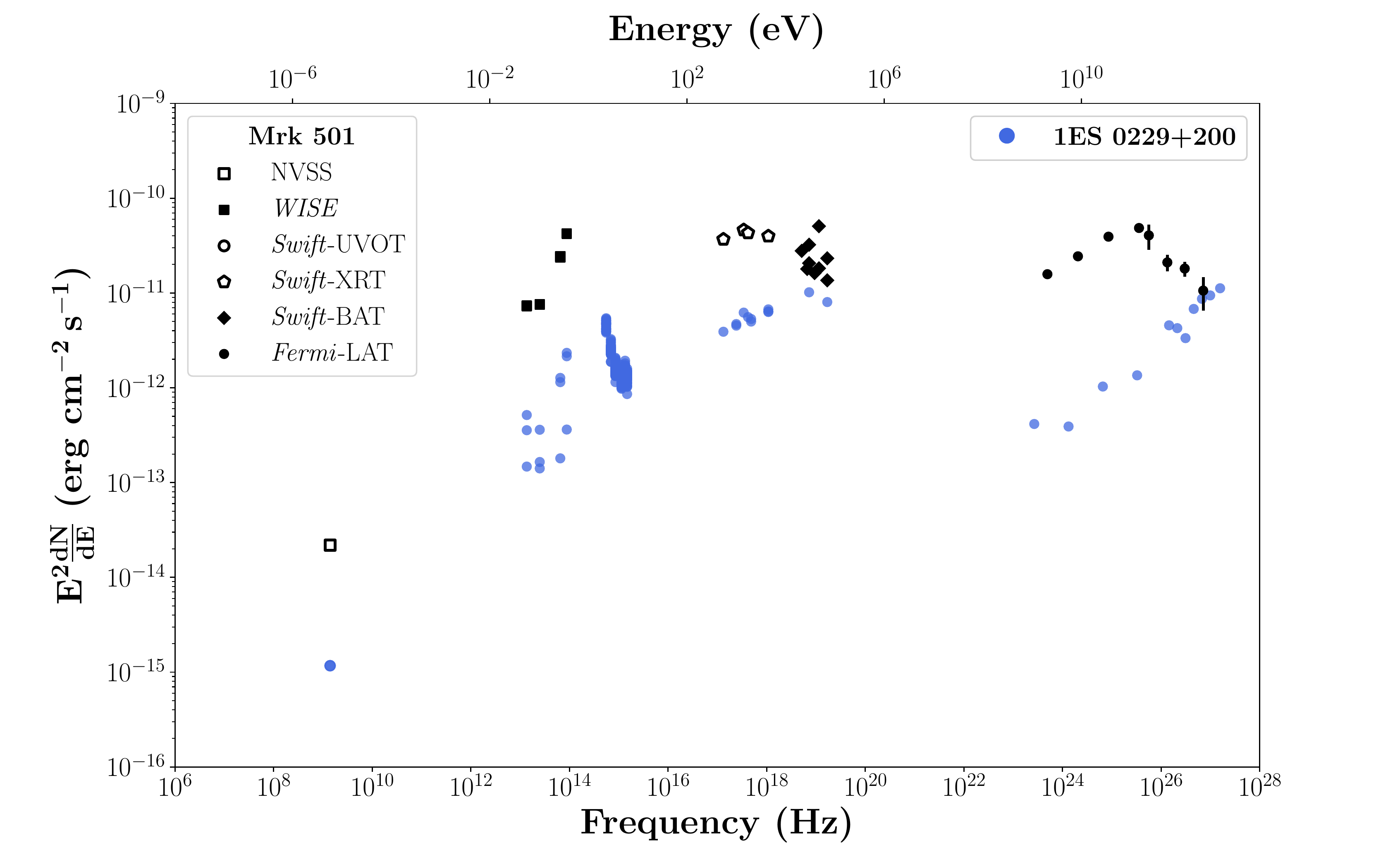}}\\
\subfloat[][1ES 1101-232 (left) and PKS 0352-686 (right).] {\includegraphics[width=0.52\textwidth]{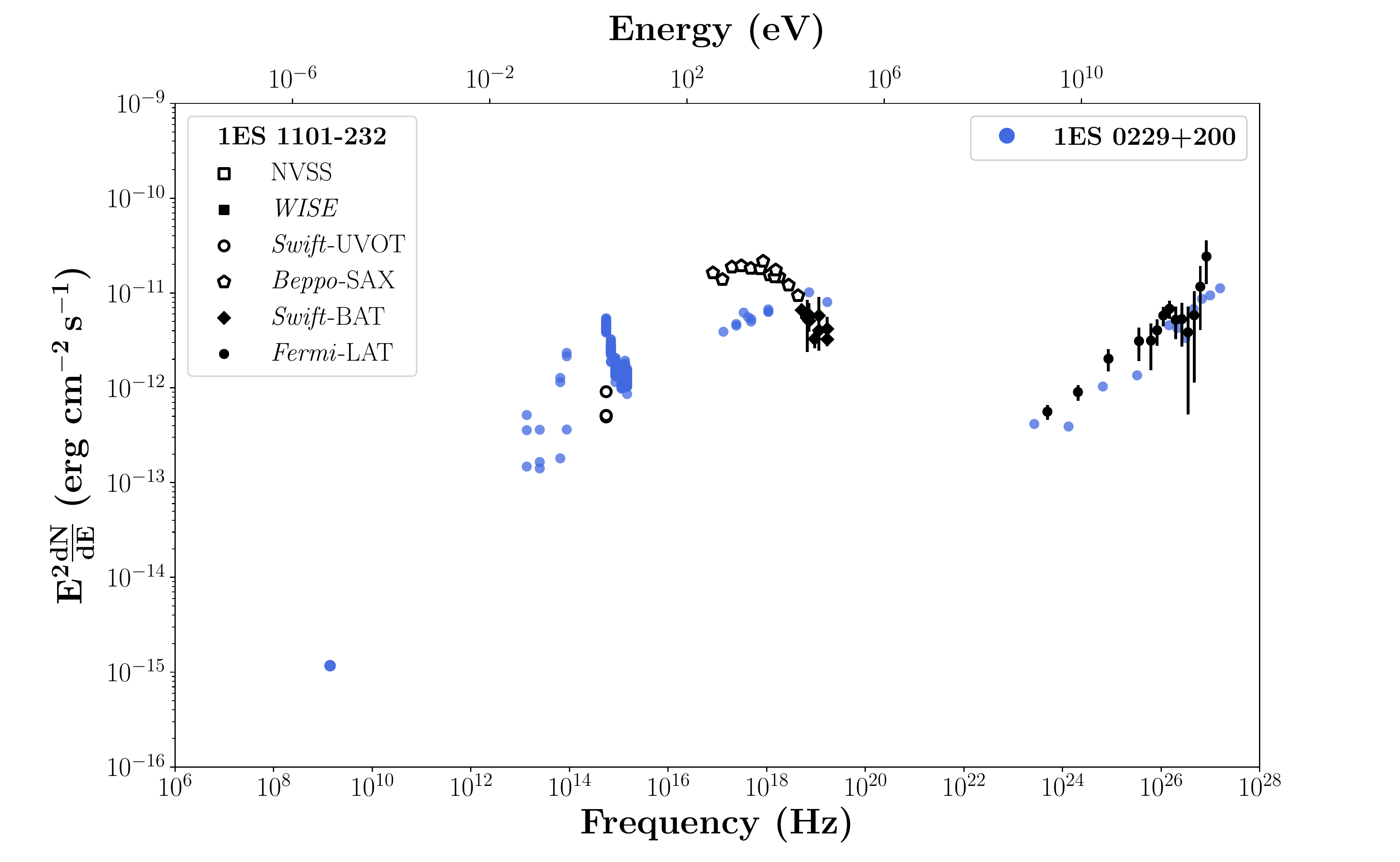} \includegraphics[width=0.52\textwidth]{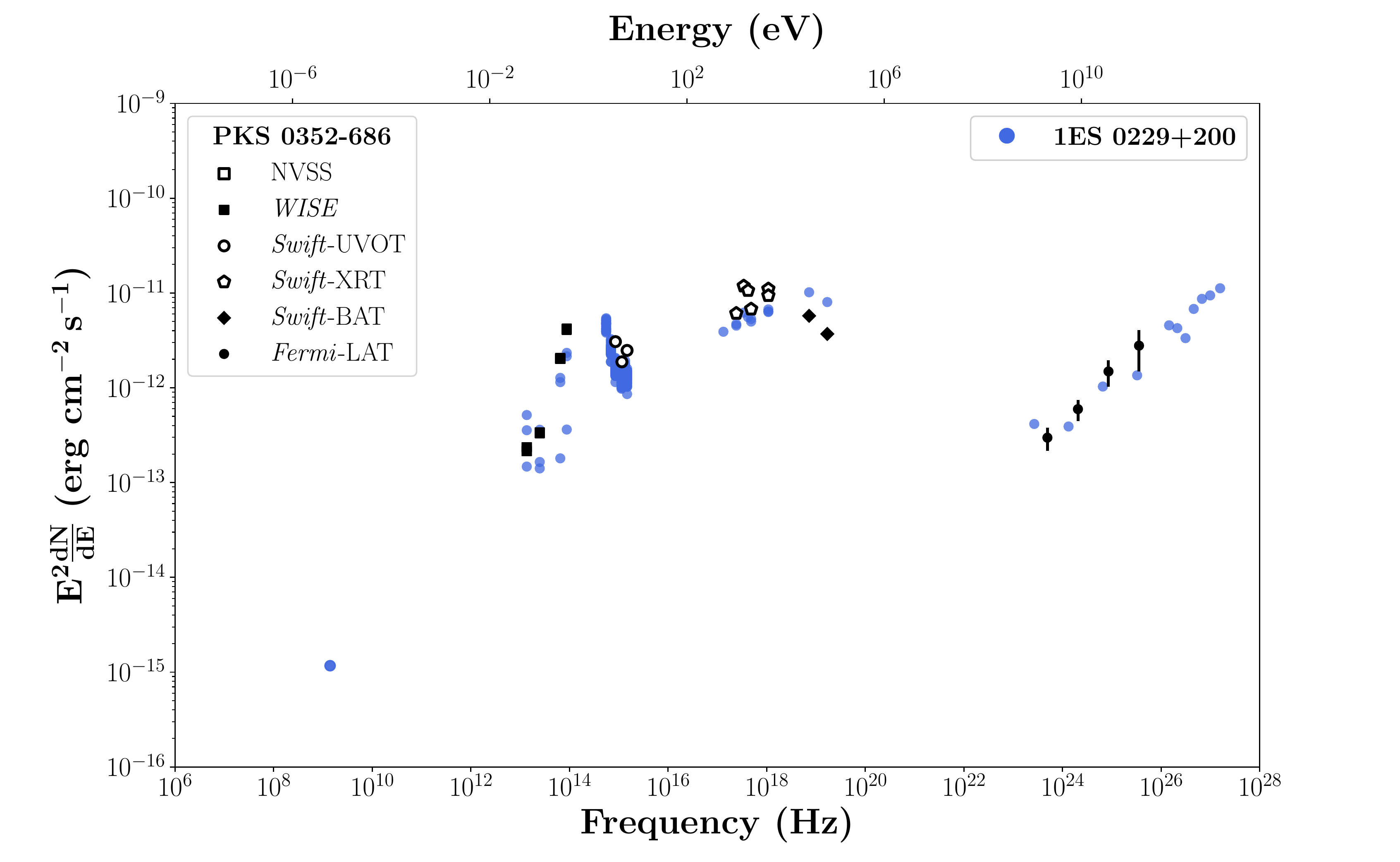}}\\
\addtocounter{figure}{0}

\caption{ }
\end{figure*}

\begin{figure*}
\centering
\ContinuedFloat
\vspace*{-2cm}
\subfloat[][1ES 1218+304 (left) and PKS 0548-322 (right).] {\includegraphics[width=0.52\textwidth]{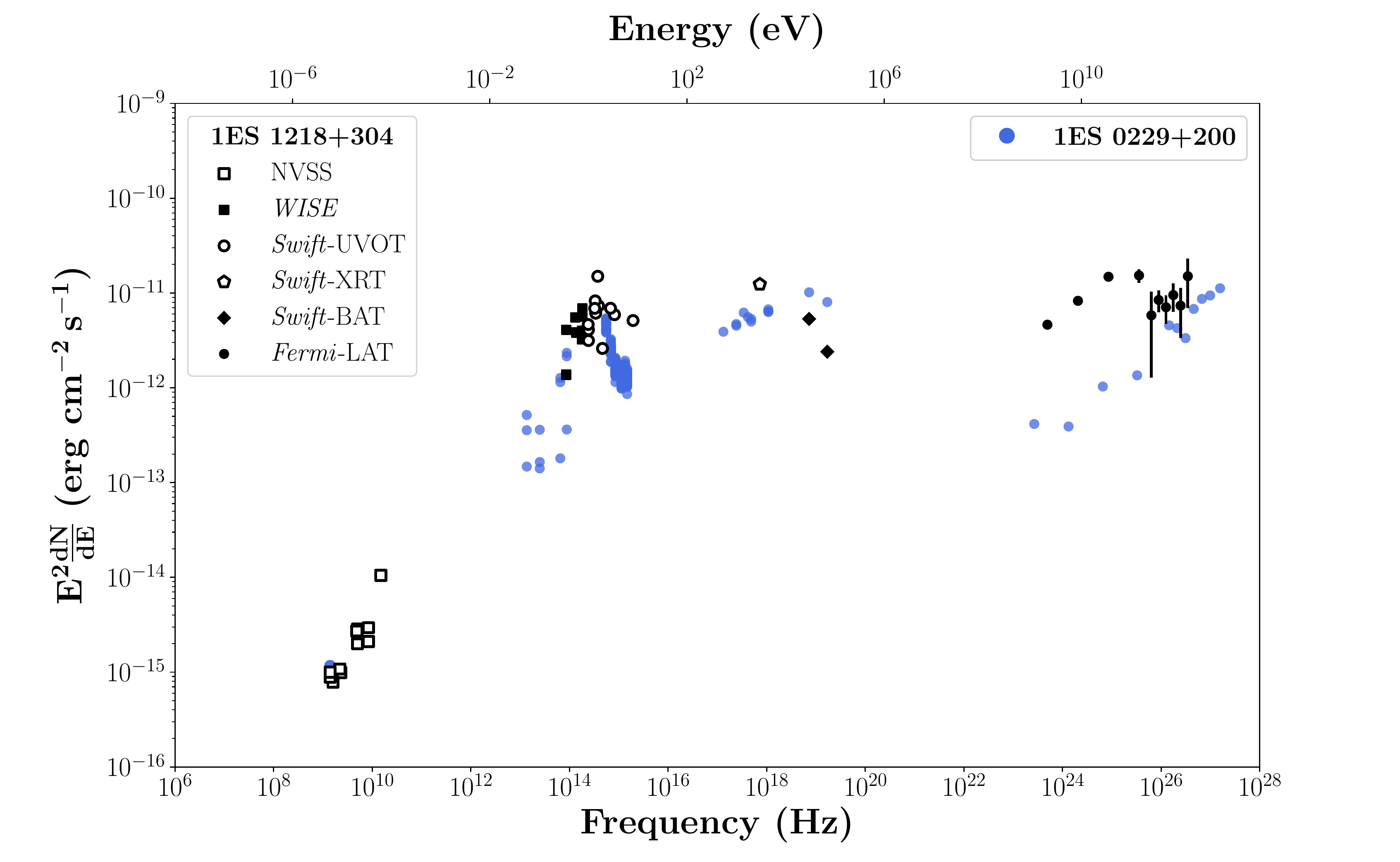} \includegraphics[width=0.52\textwidth]{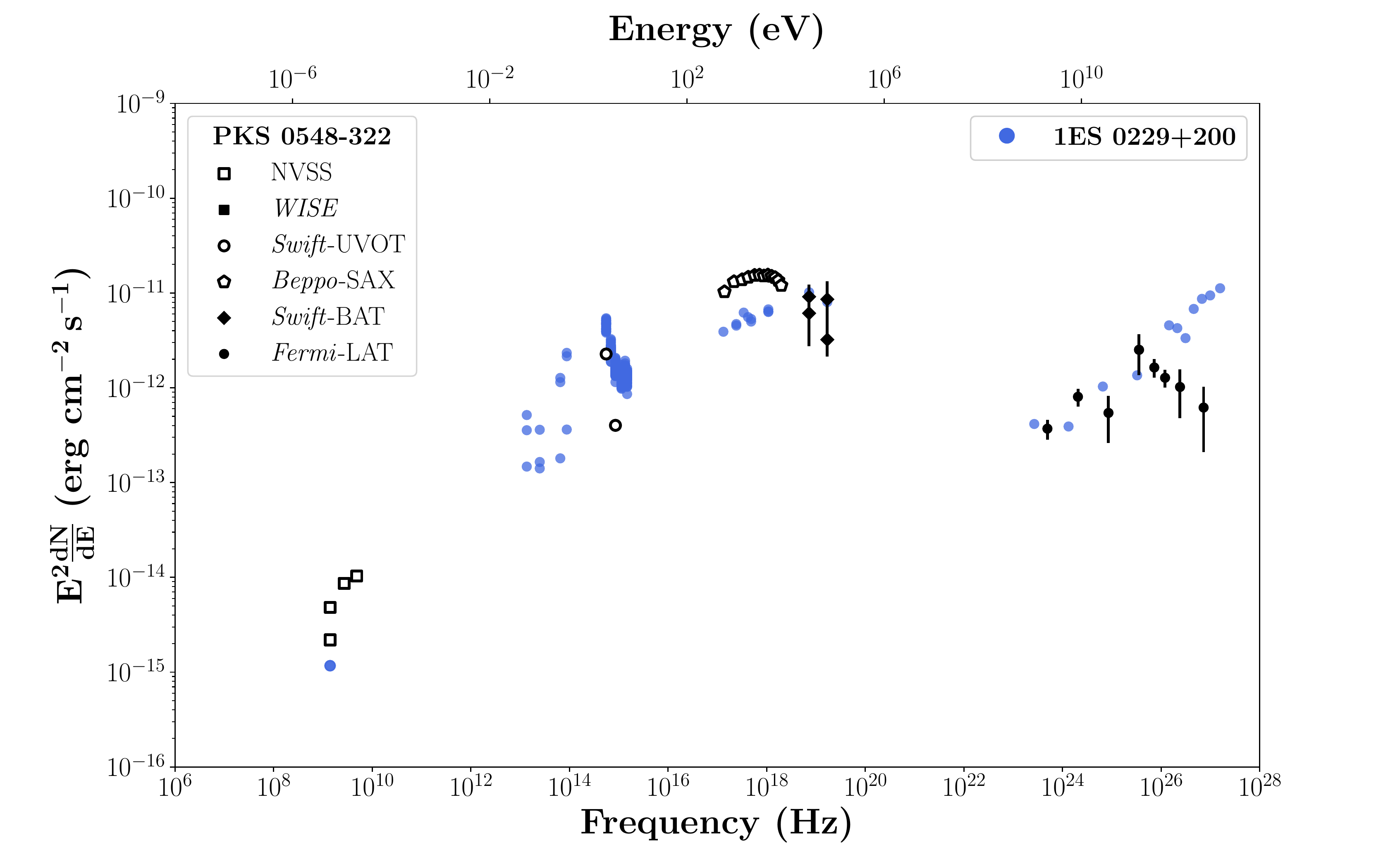}}\\
\subfloat[][1ES 1426+428 (left) and PKS 0706-15 (right).] {\includegraphics[width=0.52\textwidth]{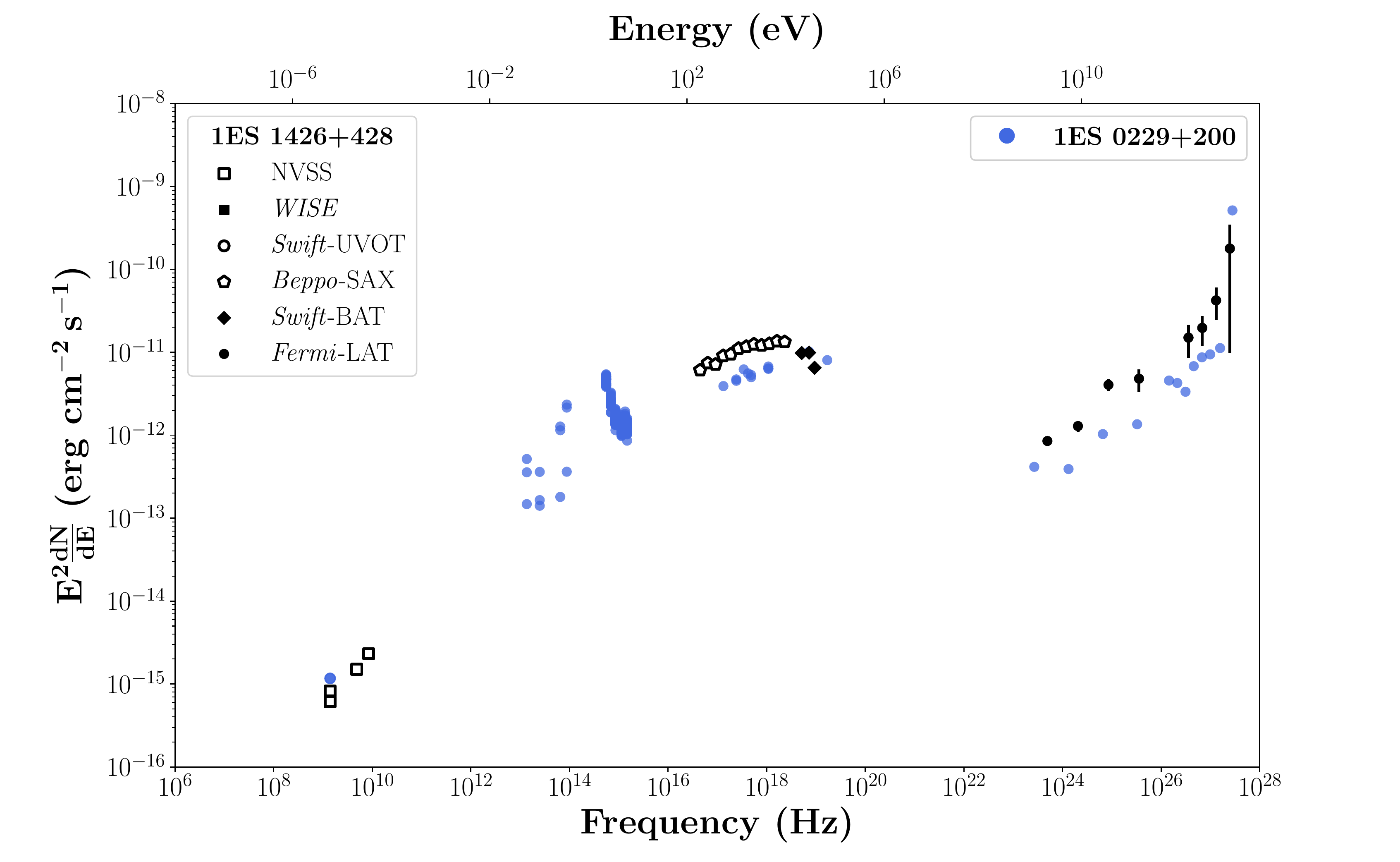} \includegraphics[width=0.52\textwidth]{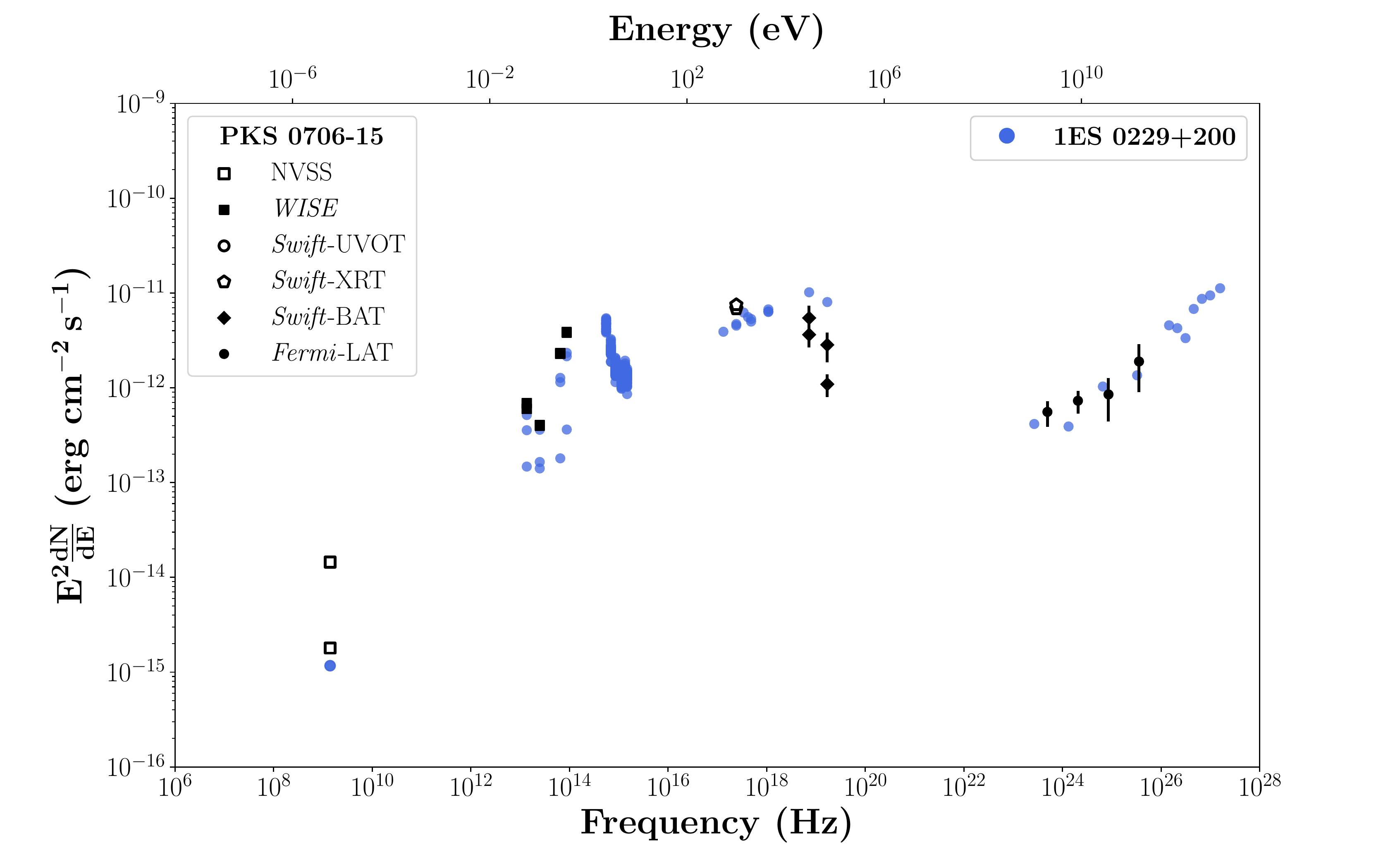}}\\
\subfloat[][1ES 1959+650 (left) and PKS 2005-489 (right).] {\includegraphics[width=0.52\textwidth]{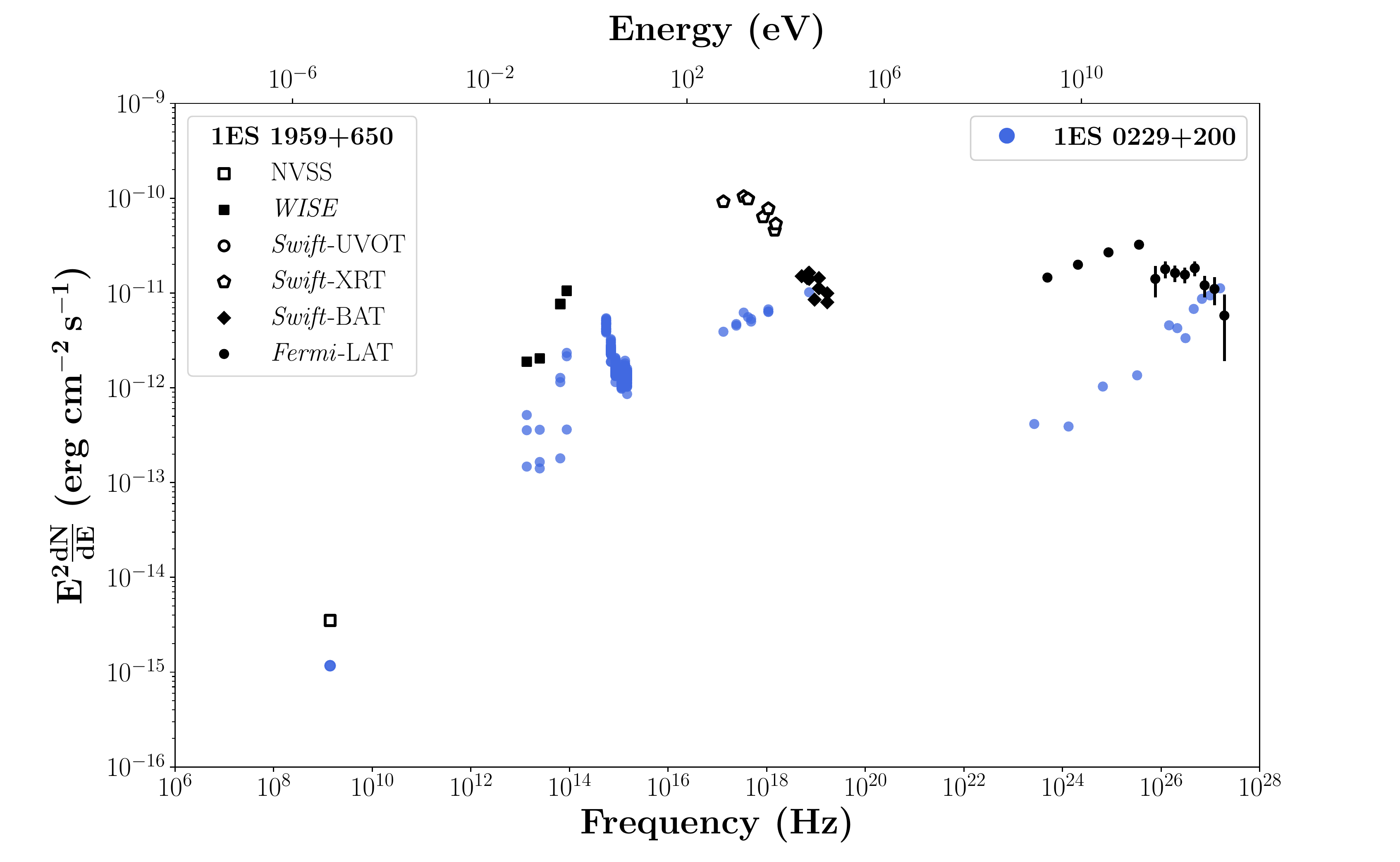} \includegraphics[width=0.52\textwidth]{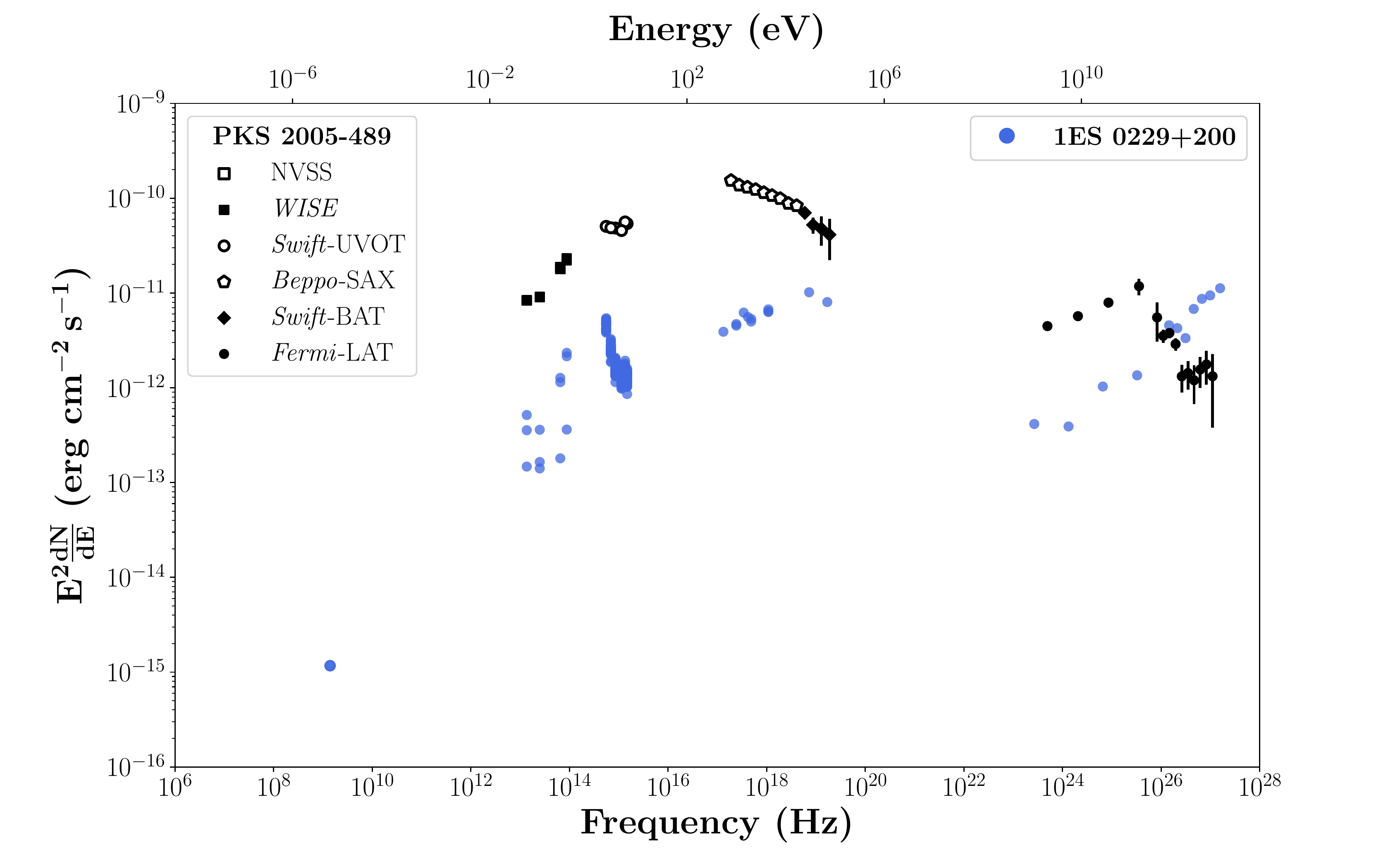}}\\
\subfloat[][1ES 2344+514 (left) and RBS 1895 (right).] {\includegraphics[width=0.52\textwidth]{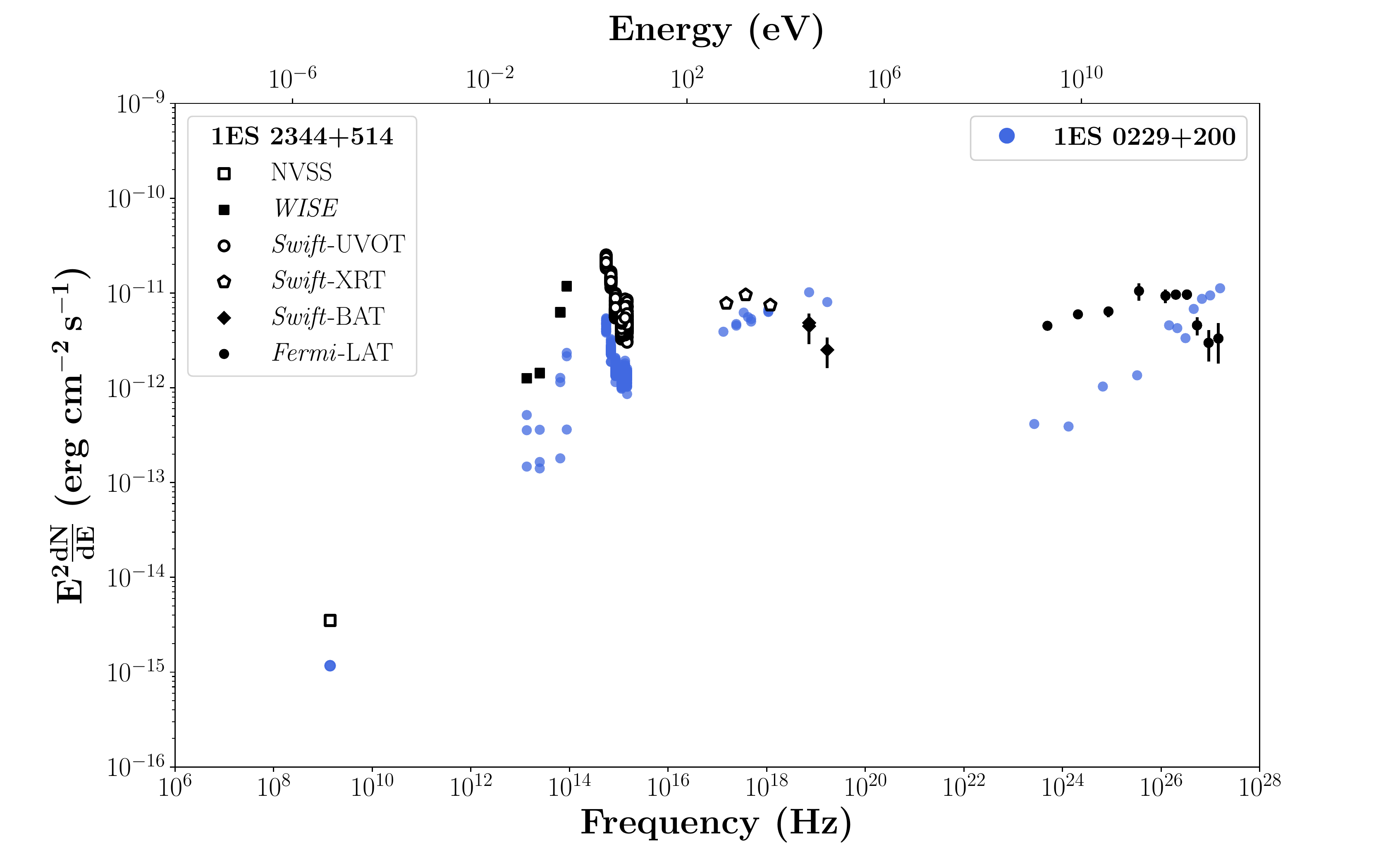} \includegraphics[width=0.52\textwidth]{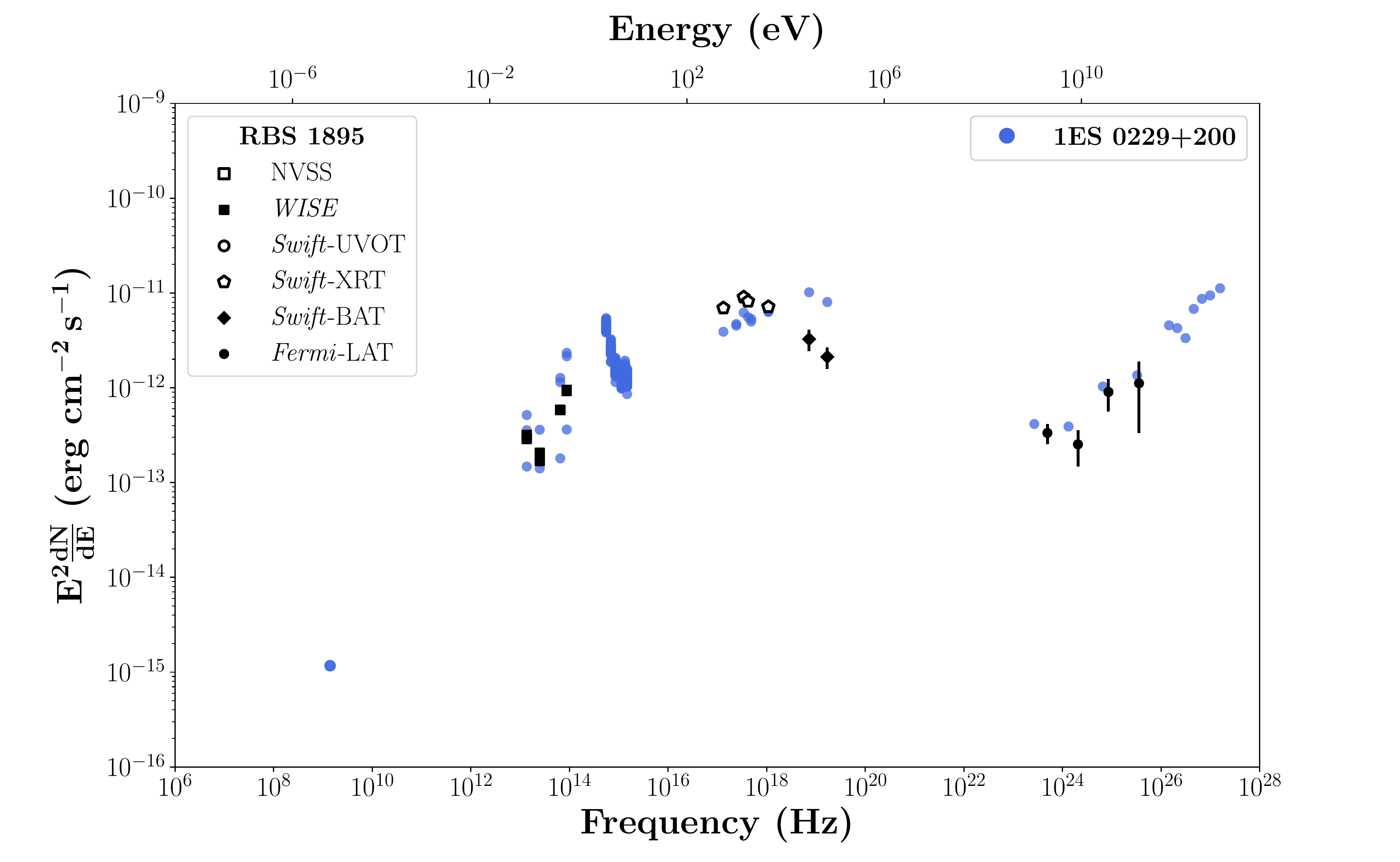}}\\
\addtocounter{figure}{0}
\caption{ }
\end{figure*}

\begin{figure*}
\centering
\ContinuedFloat
\vspace*{-2.8cm}
\subfloat[][1RXS J0214178+514457 (left) and RBS 259 (right).] {\includegraphics[width=0.52\textwidth]{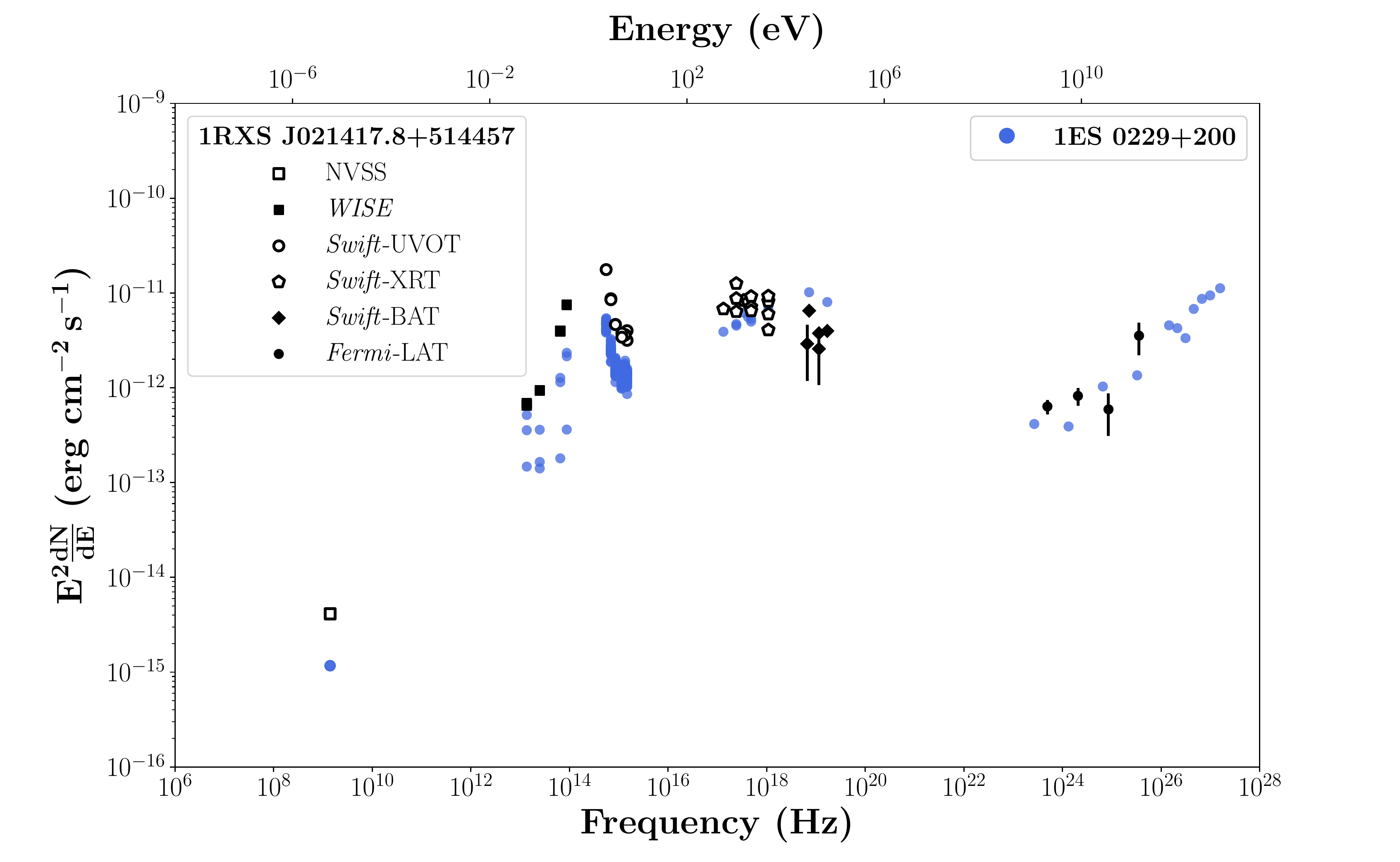} \includegraphics[width=0.52\textwidth]{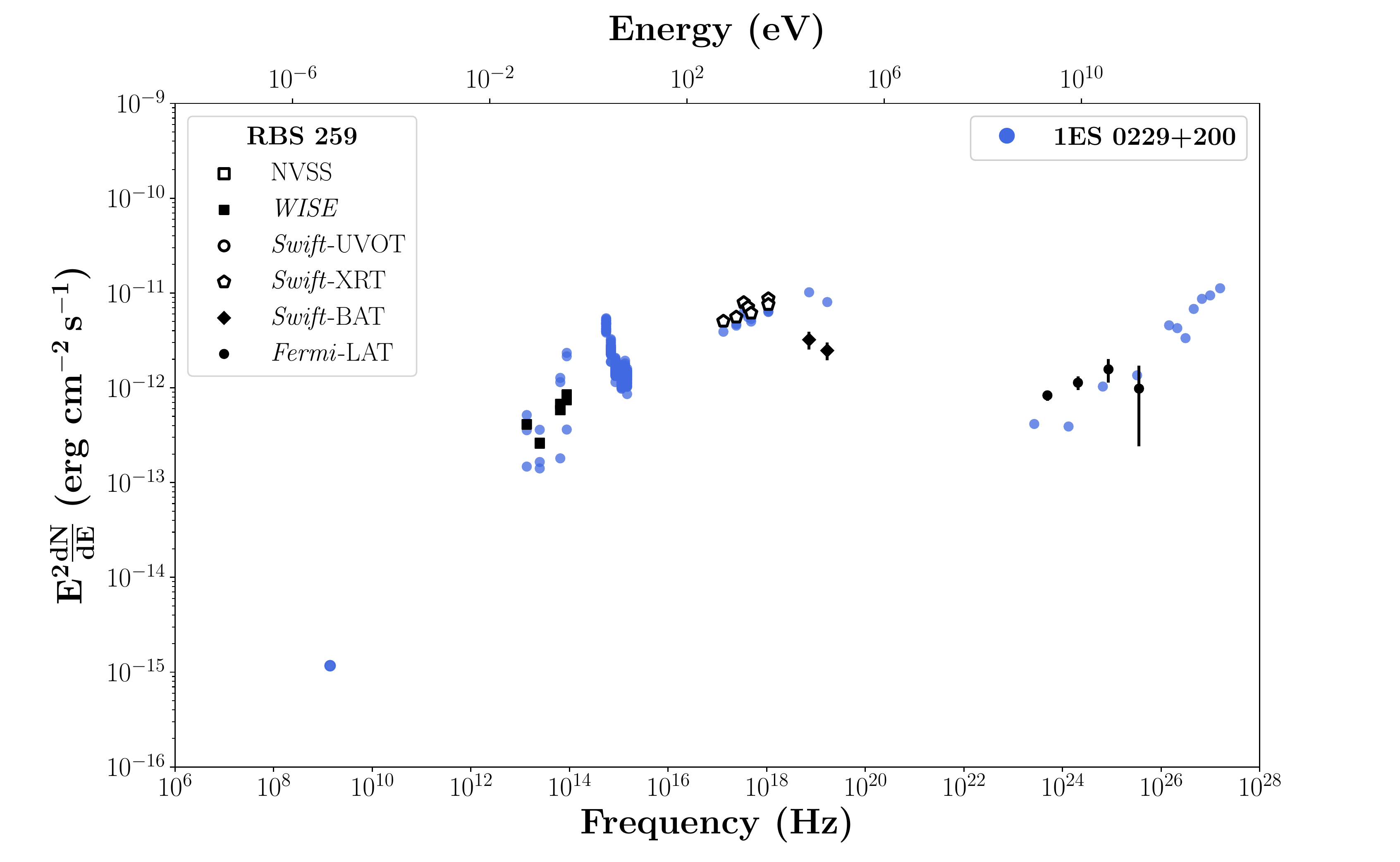}}\\
\subfloat[][1RXS J225146.9-320614 (left) and RX J11365+6737 (right).] {\includegraphics[width=0.52\textwidth]{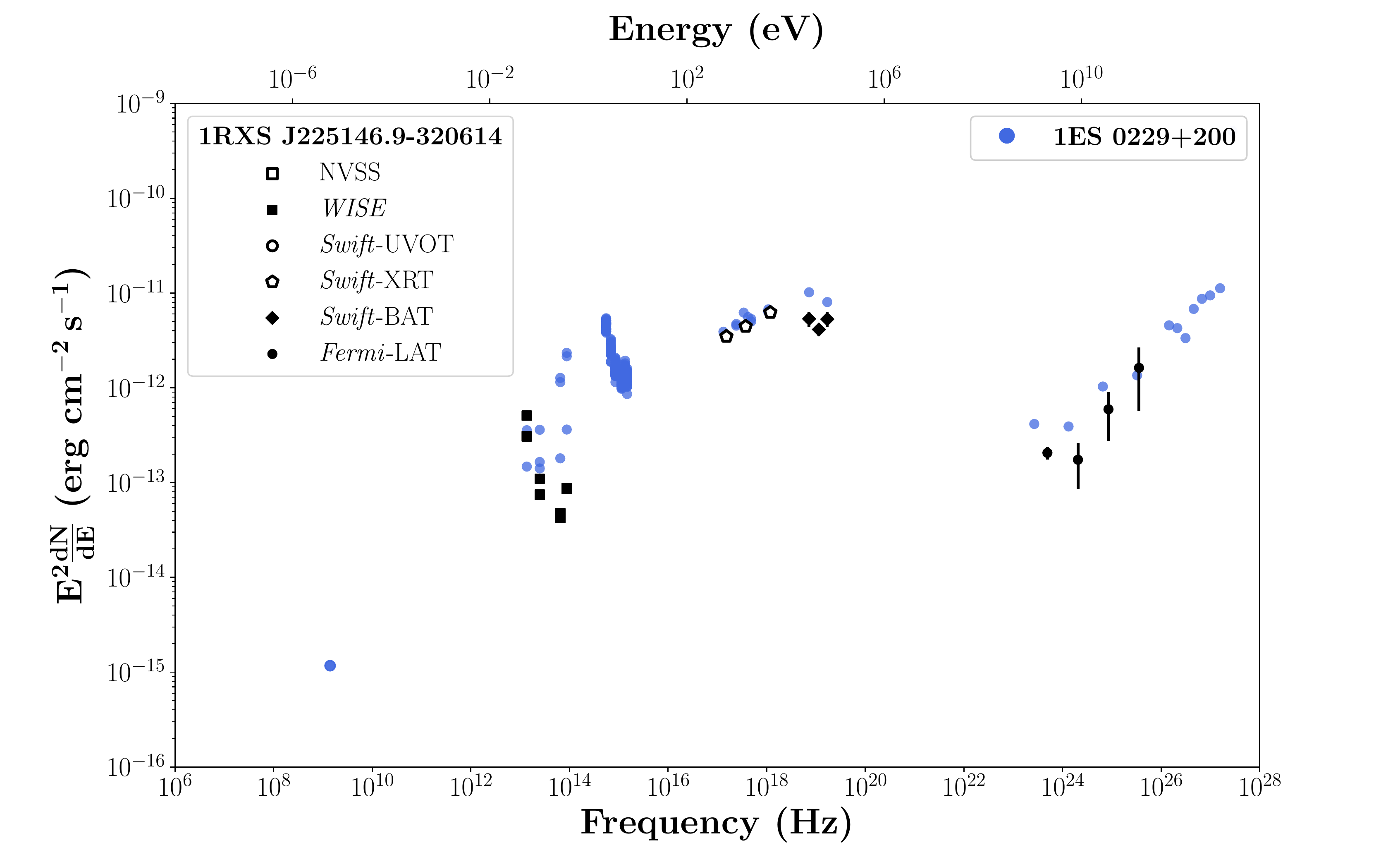} \includegraphics[width=0.52\textwidth]{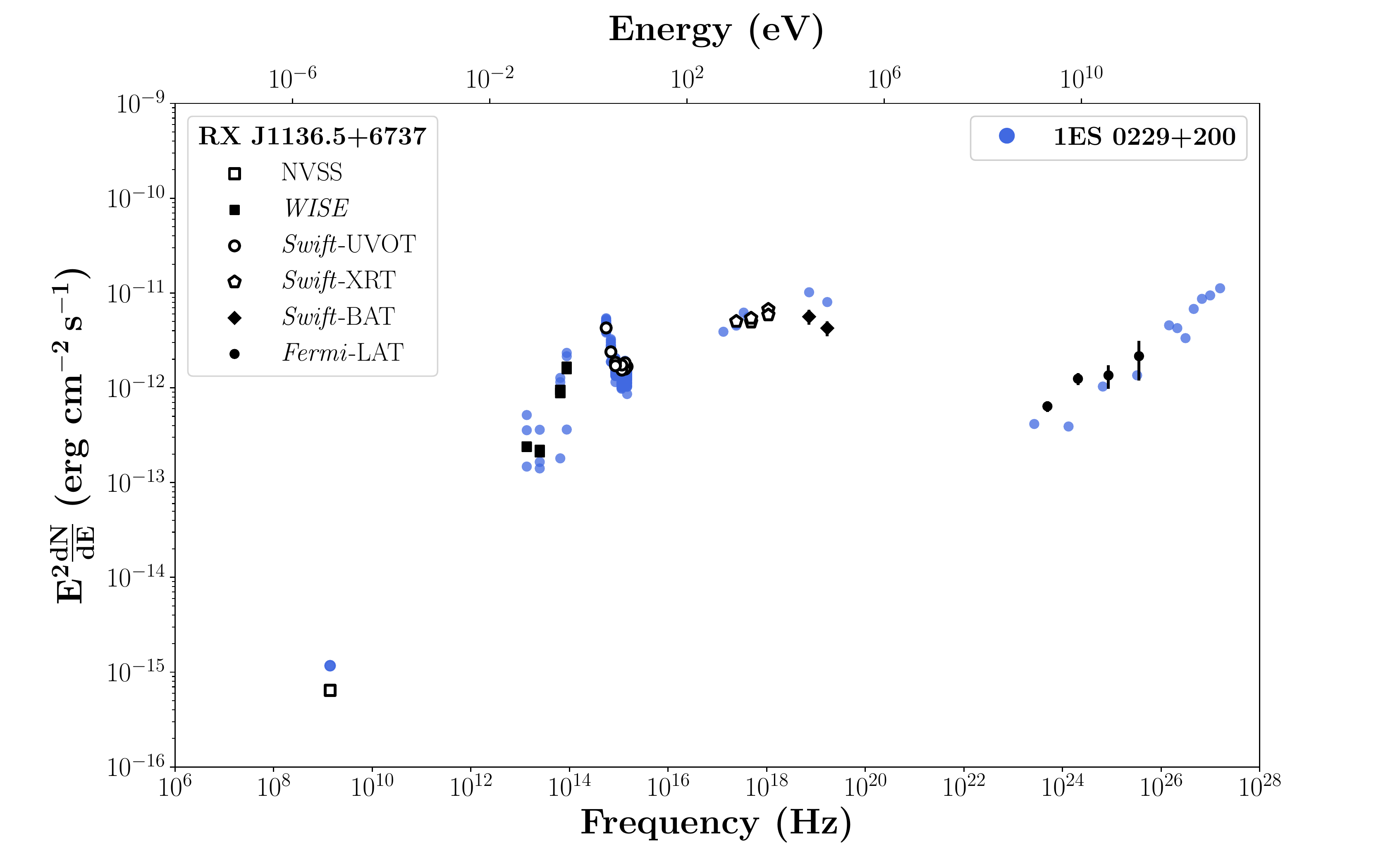}}\\
\subfloat[][RX J2056.6+4940 (left) and 3FGL J0710.3+5908  (right).] {\includegraphics[width=0.52\textwidth]{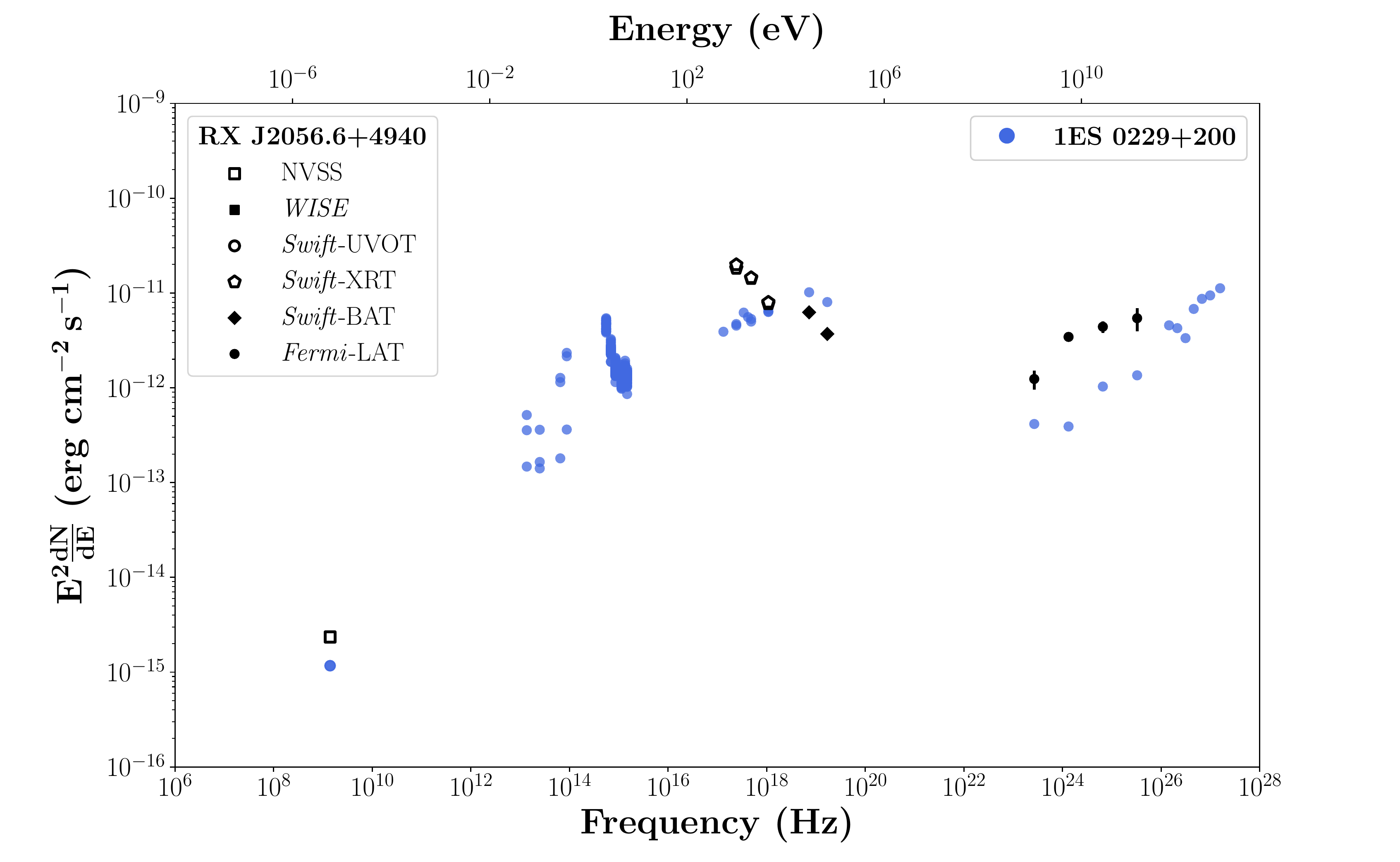} \includegraphics[width=0.52\textwidth]{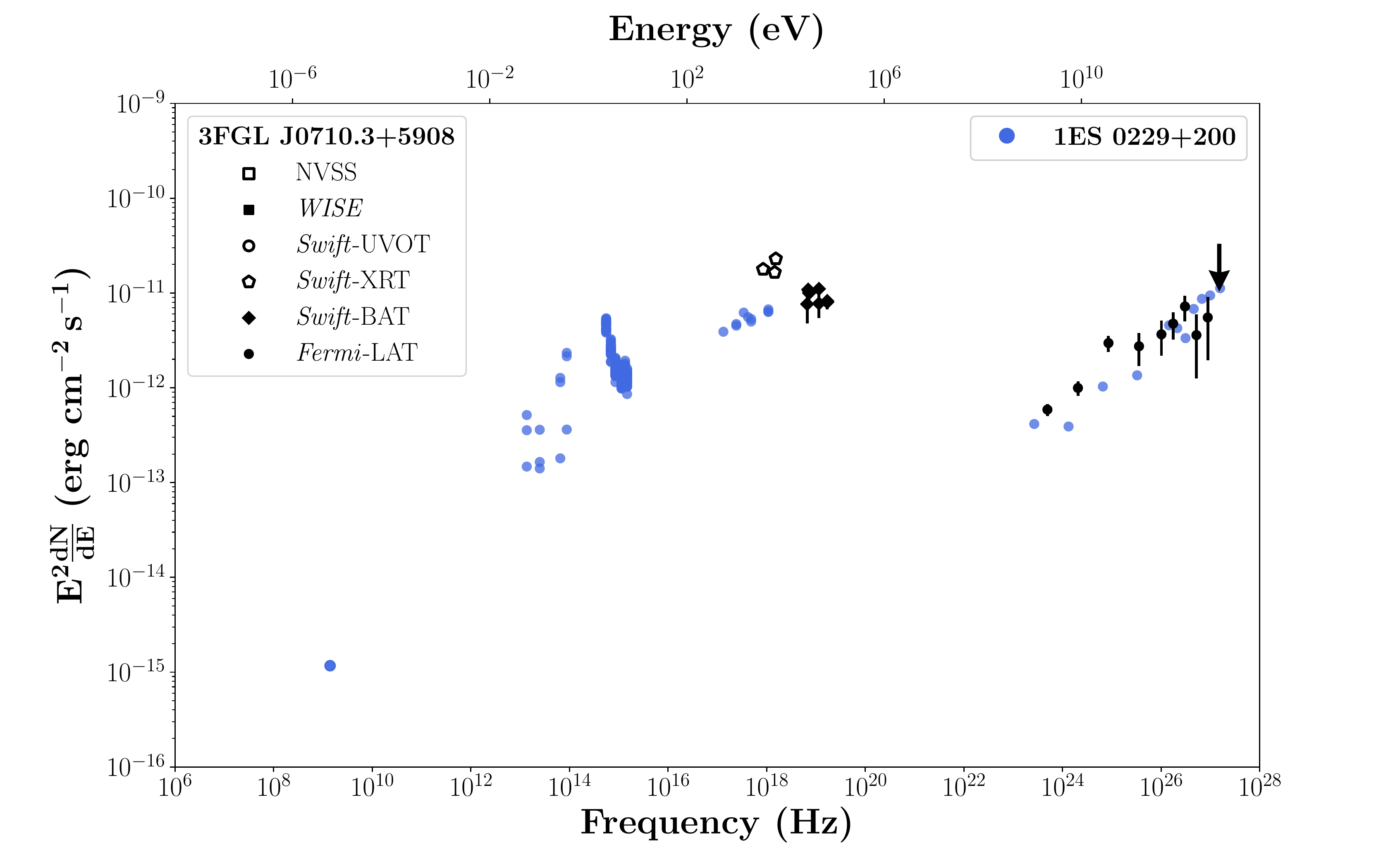}}\\
\addtocounter{figure}{0}
\caption{SED superimposition of SSDC (not simultaneous) archival data of EHBL objects in our sample (black symbols) and 1ES~0229+200 (blue circles), where data of 1ES~0229+200 are drawn without error bars to be better readable. We show NVSS \citep{NVSS} points for radio band, \emph{WISE} points for the infrared band, \emph{Swift}-UVOT \citep{Swift-UVOT} points for the optical-UV band, \emph{Swift}-XRT \citep{Swift-XRT} points or the \emph{Beppo}-SAX \citep{bepposax} data (when available) for the soft-X-ray band, and \emph{Swift}-BAT 105-months points for the hard-X-ray band. Arrows represent upper limits. Data in the HE and VHE gamma-ray bands are de-absorbed using \citet{Franceschini17} model with the source redshift $z$ to show the intrinsic spectra. In the HE gamma-ray band the ten-years \emph{Fermi}-LAT data are reported. Please see \Cref{appendix:tevdata} for further information about TeV data.
}
\label{fig:sed_superposition}
\end{figure*}